\RequirePackage{fix-cm}
\documentclass[smallcondensed]{svjour3}     
\smartqed  
\usepackage[T1]{fontenc}
\usepackage[utf8]{inputenc}
\usepackage[english]{babel}
\usepackage{amssymb}
\usepackage{amsmath}
\usepackage{latexsym}
\usepackage{mathtools}
\usepackage{graphicx}
\usepackage{bm}
\usepackage{amsbsy}
\usepackage{wasysym}
\usepackage{booktabs}
\usepackage{subcaption}
\usepackage{caption}
\usepackage{tabularx}
\usepackage{epsfig}
\usepackage{natbib}
\usepackage[pagewise]{lineno}
\usepackage{placeins}
\usepackage{hyperref}
\hypersetup{
    colorlinks=true,
    linkcolor=blue,
    filecolor=blue,      
    urlcolor=blue,
    citecolor=blue,
}

\journalname{Experimental Astronomy}

\def\arxivprefixesep{:}

\newcommand{\eprint}[2][]{%
{\tt\if!#1!#2\else#1\arxivprefixesep\ignorespaces#2\fi}%
}

\begin{document}

\title{The Scientific Performance of the Microchannel X-ray Telescope on board the SVOM Mission}
\titlerunning{MXT Scientific Performance} 

\author{D.~G\"otz\textsuperscript{*}\textsuperscript{1} \and
        M.~Boutelier\textsuperscript{2} \and
        V.~Burwitz\textsuperscript{3} \and
        R.~Chipaux\textsuperscript{1} \and
        B.~Cordier\textsuperscript{1} \and
        C.~Feldman\textsuperscript{4} \and
        P.~Ferrando\textsuperscript{1} \and
        A.~Fort\textsuperscript{2} \and
        F.~Gonzalez\textsuperscript{2} \and
        A.~Gros\textsuperscript{1} \and
        S.~Hussein\textsuperscript{5} \and
        J.-M.~Le~Duigou\textsuperscript{2} \and
        N.~Meidinger\textsuperscript{3} \and
        K.~Mercier\textsuperscript{2} \and
        A.~Meuris\textsuperscript{1} \and 
        J.~Pearson\textsuperscript{4} \and
        N.~Renault-Tinacci\textsuperscript{1}\and
        F.~Robinet\textsuperscript{5}\and
        B.~Schneider\textsuperscript{1}\and
        R.~Willingale\textsuperscript{4}}
\authorrunning{D.~G\"otz~et~al.} 

\institute{
\textsuperscript{1} Universit\'e Paris-Saclay, Universit\'e Paris Cit\'e, CEA, CNRS, AIM, 91191, Gif-sur-Yvette, France \\
\textsuperscript{2} Centre National d’Etudes Spatiales, Centre spatial de Toulouse, 18 avenue Edouard Belin,
31401 Toulouse Cedex 9, France\\
\textsuperscript{3} Max-Planck-Institut f\"ur Extraterrestrische Physik, 85748 Garching, Germany\\
\textsuperscript{4} School of Physics and Astronomy, University of Leicester, University road, LE1 7RH, United Kingdom\\
\textsuperscript{5} Universit\'e Paris-Saclay, CNRS/IN2P3, IJCLab, 91405 Orsay, France\\
\textsuperscript{*} \email{diego.gotz@cea.fr}
}

\date{Received: XXXX / Accepted: YYYY}

\maketitle

\begin{abstract}
    The Microchannel X-ray Telescope (MXT) will be the first focusing X-ray telescope based on a "Lobster-Eye" optical design to be flown on Sino-French mission SVOM. SVOM will be dedicated to the study of Gamma-Ray Bursts and more generally time-domain astrophysics. The MXT telescope is a compact (focal length $\sim$ 1.15 m) and light (< 42 kg) instrument, sensitive in the 0.2--10 keV energy range. It is composed of an optical system, based on micro-pore optics (MPOs) of 40 $\mu$m pore size, coupled to a low-noise pnCDD X-ray detector. In this paper we describe the expected scientific performance of the MXT telescope, based on the End-to-End calibration campaign performed in fall 2021, before the integration of the SVOM payload on the satellite. 
\end{abstract}

\section{Introduction}
The Space-based Variable astronomical Object Monitor (SVOM; \citealt{svom}) is a Sino-French mission developed, in cooperation by the Chinese National Space Agency (CNSA) and the French Space Agency (CNES). 
The SVOM mission is dedicated to the study of Time Domain Astrophysics (TDA) and in particular to Gamma-Ray Bursts (GRBs). GRBs are short flashes of gamma rays lasting from less than a second to a few hundreds of
seconds, appearing from unpredictable directions over the entire sky. They have been discovered in the late '60s of the last century \citep{klebesadel73}, and have remained a mystery up to the discovery in the '90s of the so-called \textit{afterglows}, i.e. the electromagnetic emission following the GRBs at other wavelength (X-ray, optical, IR, radio), lasting a few hours up to several months after the event. In particular, the observation of the afterglows in X-rays, performed for the first time by the Italian
satellite Beppo\textit{SAX}, allowed the astronomers to finely localize the GRB counterparts (at $\sim$ arcmin level), identify their optical counterparts and host galaxies, measure their distances, and finally confirm their cosmological origin  \citep{costa97, vanparadijs97, frail97, metzger97}.
It turned out that the beaming corrected energies implied by these explosions are huge, of the order of 10$^{50}$-10$^{52}$ erg \citep[e.g.][]{liang08}, making them extremely powerful sources that can be detected up to the early Universe \citep[z$\sim$9.4 for GRB 090429B;][]{cucchiara11}.

Since the launch of the Neil Gehrels Observatory (aka \textit{Swift}, \citealt{swift}) in 2004, GRBs are routinely discovered and followed-up from ground-based telescopes. The origin of long bursts (i.e. those lasting more than 2 s) has been firmly identified as the collapse of massive stars, especially thanks to the spectroscopic identification of some GRBs with peculiar Supernovae of type Ibc \citep[e.g.][]{pian06}. 
The fact that they can be detected up to very high redshifts, makes them the perfect candidate sources to be used as tracers of the early Universe star formation \citep[e.g.][]{chary16}, and of the cosmological chemical enrichment \citep[e.g.][]{perley16}. Finally they can also potentially pinpoint the first generation (pop III) of stars \citep[][]{toma16}.

On the other hand, the origin of short GRBs (lasting less than 2 s) is less clear. However, the recent simultaneous and co-located detection of gravitational waves (GWs) from a binary neutron star (BNS) merger and a short GRB strongly supports the hypothesis of compact binary mergers as the progenitors of short GRBs \citep{abbott17a} and gave birth to modern multi-messenger astrophysics \citep{abbott17b}. 

In this paper we will present the scientific performance of the Microchannel X-ray Telescope (MXT), 
and show how it is adapted to contribute to the investigation of the nature of GRBs especially in the context of multi-messenger astrophysics. We will first introduce briefly the SVOM mission as a whole (\S \ref{sec:sovm}), then present the MXT design (\S \ref{sec:mxt}), and finally the scientific performance (\S \ref{sec:perfo}), as derived in October/November
2021 during the End-to-End testing and calibration campaign performed at the MPE PANTER X-ray testing facility.

\section{The SVOM Mission}
\label{sec:sovm}
The SVOM mission, to be launched in 2023, will be composed of a space segment, as well as a few ground based
dedicated follow-up facilities. The space segment is composed of four co-aligned instruments.
Two instruments (ECLAIRs and GRM) are sensitive in the hard-X/soft gamma-ray energy range and have wide fields of view, in order to monitor vast regions of the sky and detect gamma-ray transients. Two narrow field of view instruments (MXT and VT) will be used to follow-up and characterize the afterglow emission.

ECLAIRs is a coded-mask telescope, composed of a 54$\times$54\,cm$^{2}$ pseudo-random coded mask made of a 
Ti-Ta-Ti sandwich (10/0.6/10\,mm) placed 45.8 cm above a pixellated detection plane made of 80$\times$80 CdTe crystals (4$\times$4$\times$1\,mm$^{3}$). Its field of view is about 2\,sr (89$^{\circ}\times$89$^{\circ}$) wide. ECLAIRs is sensitive in the 4\,keV -- 150\,keV energy range, and it comprises an on-board software to detect and localize (to better than 13\,arcmin) in near-real-time the GRBs that appear in its FOV. Once a new transient is detected ECLAIRs issues an alert and requests the platform to slew so that the error box can be observed by the narrow-field instruments. 

ECLAIRs is complemented by the Gamma-Ray Monitor (GRM), a set of three 1.5 cm thick NaI scintillators of 16 cm in diameter, each one offset by 120$^{\circ}$ w.r.t. each other and with a combined FOV of $\sim$\,2.6 sr. The GRM has poor localization capabilities, but it extends the SVOM spectral range up to about 5\,MeV, and increases the probability of simultaneous detection of short GRBs and GW alerts.

The Visible Telescope (VT) is a Ritchey-Chretien telescope with a 40\,cm diameter primary mirror. Its field of view is 26$\times$26 arcmin$^{2}$ wide, adapted to cover the ECLAIRs error box in most of the cases. It has two channels, a blue one (400--650 nm) and a red one (650--1000 nm), and a sensitivity limit of $M_{V}=22.5$ in 300 s, allowing the detection $\sim$\,80\,\% of the ECLAIRs GRBs.
The space segment is completed by the MXT, that is described in detail in \S \ref{sec:mxt}.  The main characteristics of the SVOM space segment are summarized in Tab. \ref{tab:svom}.

   \begin{table}[ht]
    \centering
    \caption{Summary of the characteristics of the SVOM space instruments}
    \renewcommand{\arraystretch}{1.25}
        \begin{tabular}{lllll}
            \hline \hline
            & ECLAIRs & GRM & MXT & VT\\
            Energy/Wavelength  & 4--150 keV & 15-5000 keV & 0.1-10 keV & 650-1000 nm\\
            Field of View &  2 sr & 2.6 sr (combined) &  58$^{\prime}$$\times $58$^{\prime}$ & 26$^{\prime}$$ \times$ 26$^{\prime}$ \\
            Localization accuracy  & $<$ 12$^{\prime} $& $<$20$^{\circ}$&$<$2$^{\prime}$ &$<$1$^{\prime\prime}$\\
            Expected GRBs year$^{-1}$ &60 & 90 & 50 & 40\\
            \hline
        \end{tabular}
    \label{tab:svom}
    \end{table}

The SVOM alerts, generated by ECLAIRs (and GRM) for the GRB prompt phase, will reach the French Science Centre (FSC) through a series of VHF antenna receivers placed below the track of the satellite. From the FSC the alerts will be dispatched to the scientific community for further follow-up in less than 30 s from the on-board detection time, through GCN notices and/or VO events. Indeed, the SVOM pointing strategy is optimized in order to provide alerts that are always on the night side of the Earth and hence promptly observable by ground based telescopes. This strategy has been chosen to increase the number of GRBs with measured redshifts\footnote{SVOM does not have on-board optical/NIR spectroscopic capabilities and relies on follow-up from the ground for distance determination.}. The alerts containing the information on the afterglow properties, including the refined positions, generated by the  MXT and the VT on board, will also reach FSC and the ground observatories using the VHF system.

The SVOM mission is also provided with a number of dedicated telescopes on ground. In particular here we mention:
\begin{itemize}
    \item the Ground-Based Wide Angle Cameras (GWACs), a set of 36 optical cameras with a combined FOV of 5400 deg$^{2}$, located in Ali (China), whose goal is to catch the prompt optical emission for the ECLAIRs GRBs
    \item the Chinese Ground Follow-up Telescope (C-GFT), a robotic 1-m class telescope, with a 21$\times$21 arc min$^{2}$ FOV, located in Xinglog (China) and sensitive in the 400-950 nm wavelength range
    \item the French Ground Follow-up Telescope (F-GFT, Colibri), a robotic 1-m class telescope, with a 26$\times$26 arc min$^{2}$ FOV, located in San Pedro Martir (Mexico), and with multi-band photometry capabilities over the 400-1700 nm wavelength range
\end{itemize}

Other robotic telescopes will be part of the SVOM follow-up system, but they will not be fully dedicated to SVOM.

\section{The Microchannel X-ray Telescope}
\label{sec:mxt}
The Microchannel X-ray Telescope has been developed under the CNES responsibility in close collaboration with CEA-Saclay/Irfu , the Univeristy of Leicester, the Max Planck Institut f\"ur Extraterrestrische Physik (MPE) in Munich, and the IJCLab in Orsay. It is a light ($<$\,42\,kg) and compact (focal length $\sim$\,1.15\,m) focusing X-ray telescope; its sensitivity below 1\,mCrab makes it the ideal instrument to detect, identify and localize down to the arc min level X-ray afterglows of the SVOM GRBs.

\begin{figure*}[ht]
\centering
\includegraphics[width=5.5cm]{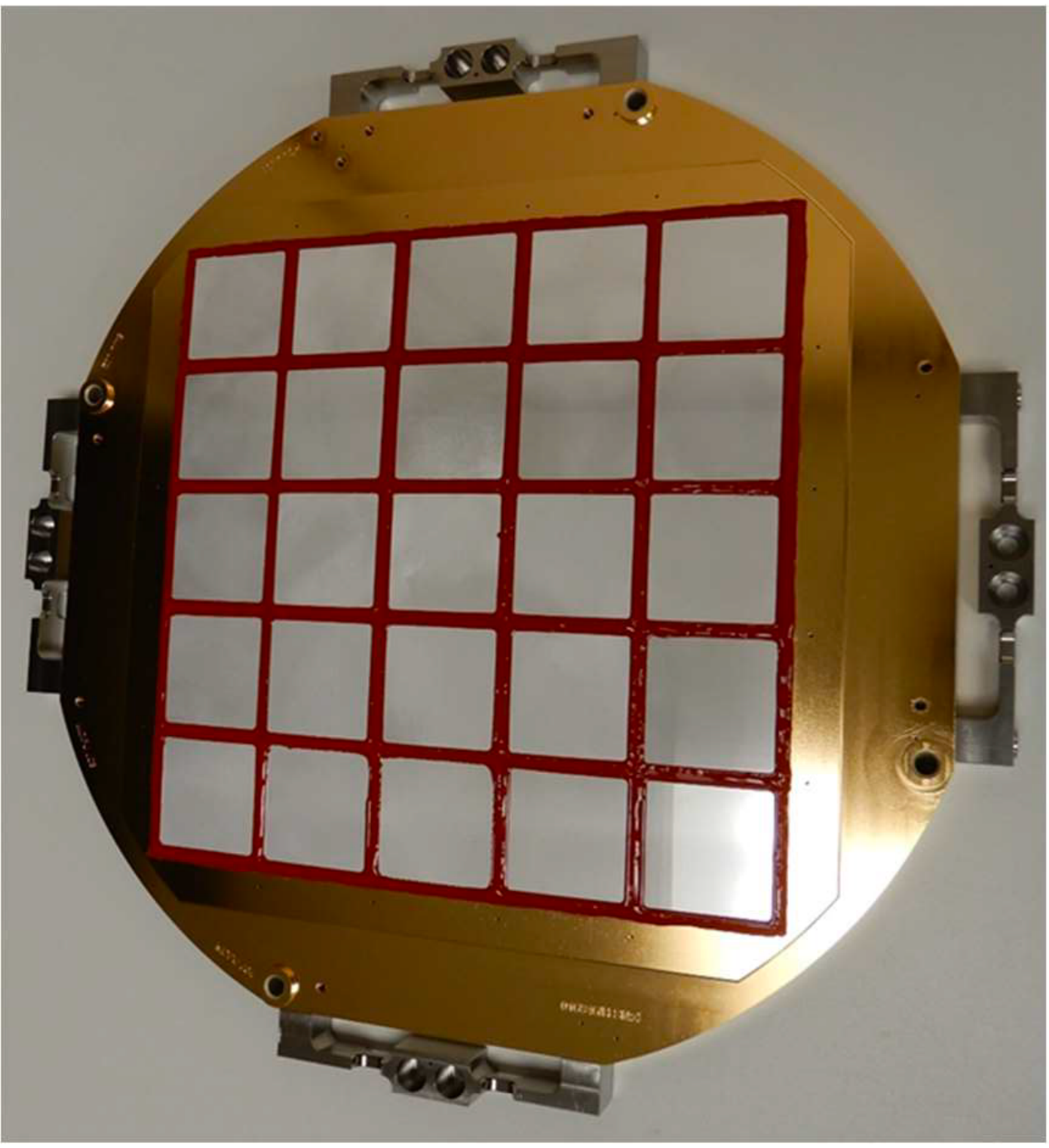}\includegraphics[width=6.cm]{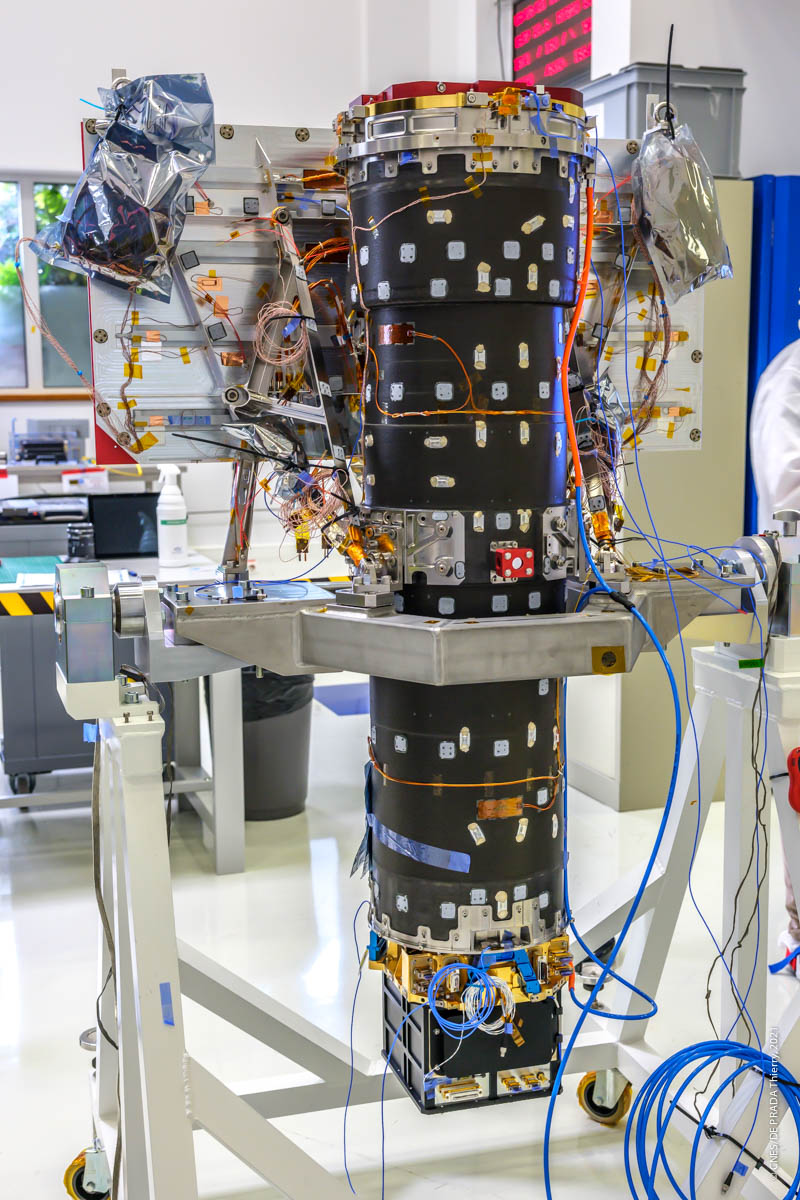}
\caption{Left: the MXT flight model optics. Right: the fully integrated (apart from MLI) MXT flight model in the CNES integration hall (MDPUs not shown).}
\label{fig:fmmop}
\end{figure*}

The MXT (see Fig. \ref{fig:fmmop}, \citealt{mercier22}) is composed of five main sub-systems:
\begin{itemize}
    \item the MOP: the MXT OPtical assembly, based on square Micropore Optics (MPO),
    \item the MCAM: the MXT CAMera hosting a pnCCD,
    \item the MST: the MXT carbon fiber STructure,
    \item the MDPU: the MXT Data Processing Units (in cold redundancy),
    \item the MXT radiator to dissipate the heat generated at focal plane level.
\end{itemize}

The MOP design is based on a ‘‘Lobster-Eye'' grazing incidence X-ray optics, first proposed by \citet{angel79}, and inspired by the vision of some crustacean decapods. It is composed of 25 square MPO plates of 40 mm each arranged in a 5$\times$5 configuration, see Fig.\ref{fig:fmmop}.

\begin{figure*}[ht]
\centering
\includegraphics[width=10cm]{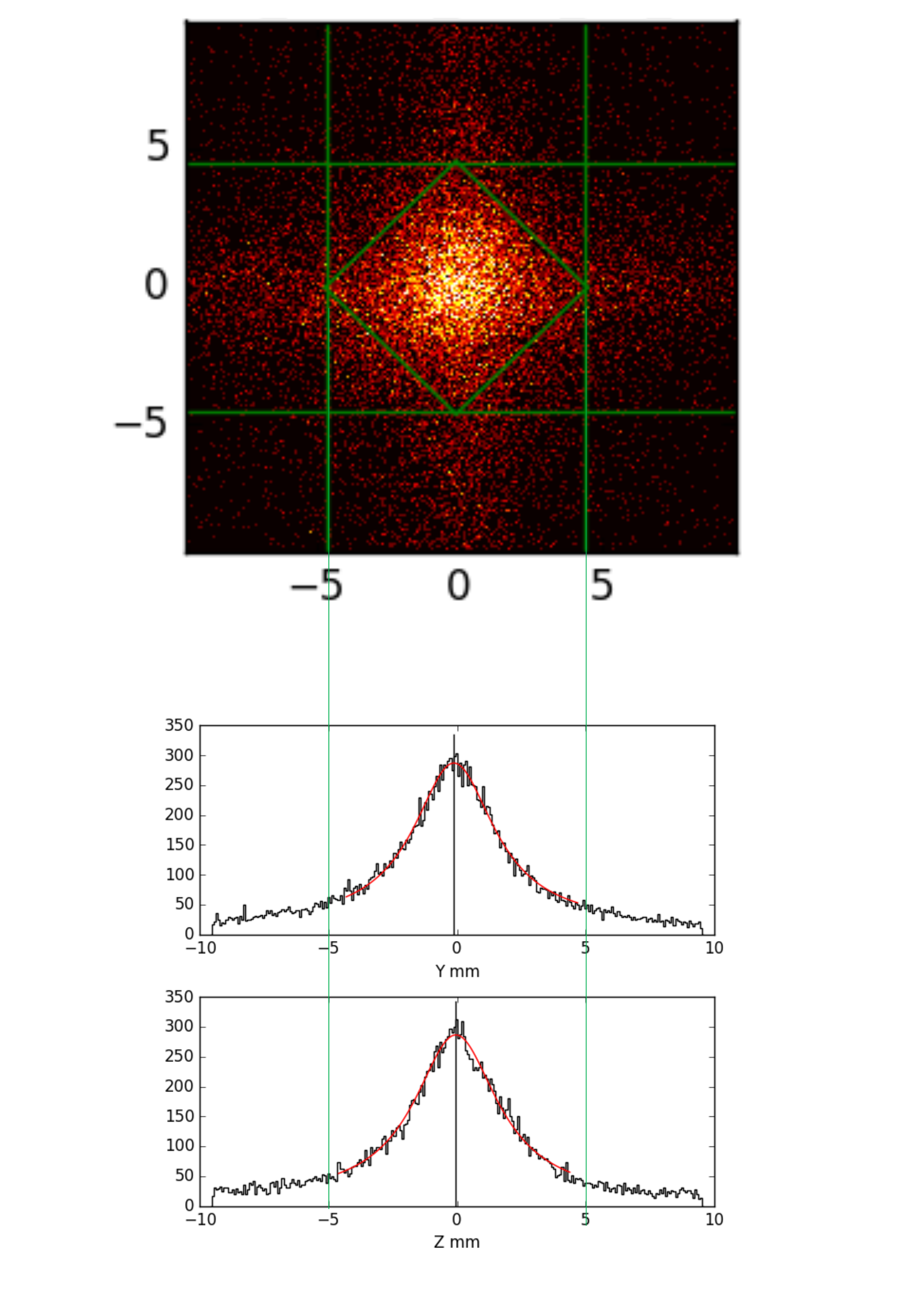}
\caption{Upper Panel: MXT PSF as measured at PANTER at Al-K. Middle Panel: Y profile of the measured PSF; the black line represents the data while the red line is the fitted model, see text. Lower Panel: Z Profile of the measured PSF.}
\label{fig:psf}
\end{figure*}

Although Lobster-Eye optics have been originally developed for large field of view telescopes (several tens of square degrees), the MXT optical design is optimized for a (relatively) small field of view\footnote {The FOV of the whole array is 6$^{\circ}\times$6$^{\circ}$, and the effective FOV is limited by the detector size. of 58$\times$58 arcmin$^{2}$} by making use of a combination of 1.2 and 2.4 mm thick plates (with a pore size of 40 $\mu$m), whose inner walls are coated with Ir to enhance reflectivity. The  entrance of the MPOs pores is covered with a 70 nm thick Al film, and the MPOs are then slumped to match a spherical surface which provides the requested X-ray focusing. This technique results in a peculiar point spread function (PSF), made by a central peak and two cross arms, see Fig. \ref{fig:psf}. The central peak is due to photons that are reflected twice on adjacent walls, while the cross arms are due to photons being reflected only once. Finally, a small part of the photons do not interact at all with the optics material and produce a diffuse background.

Despite the relative imaging complexity, the MXT MOP is very well adapted to GRB studies, for which the X-ray afterglow will be most of the time the only bright source in the MXT FOV.  Although less performing than classical X-ray Wolter-I type optics, that could be produced to match a 1$^{\circ}$ FOV, the MOP is very light ($<$ 2 kg, more than an order of magnitude lighter wrt. equivalent Wolter-I optics) making lobster-eye optics very attractive for small space borne instruments.

Coupled to the MOP is the MXT Camera (MCAM) which implements a focal plane assembly based on a pnCCD \citep{meidinger06} with its readout and control electronics (FEE). It also comprises a filter wheel, radiation shielding material and a thermo-electric cooling system \citep{meuris22}.
The detector has an active area of 256$\times$256 pixels of 75\,$\mu$m in side length, and a reduced frame store area with 75$\times$51\,$\mu$m pixels. Once transferred to the frame store area the charges are collected column-wise by two dedicated ASICs called CAMEX. 
This CCD is fully depleted (450\,$\mu$m depth), and its read-out rate is of 10 frames per second. The detector is actively cooled to $-$65$^{\circ}$C, in order to guarantee a low thermal noise and to reduce the radiation damage effects in flight. The filter wheel allows to put an $^{55}$Fe calibration source or a shutter in front of the detector when needed. The nominal MXT energy range is 0.2--10 keV.

The MDPU is responsible for the MXT thermal control, the calibration wheel control, the generation of the telemetry and the handling of the MXT telecommands. Its scientific partition deals with the processing of ‘‘dark frames'' for onboard offset and noise calculations, which are then used to select the pixels transmitted to the ground, only those whose deposited charge exceed a pre-defined threshold (see \citealt{schneider22} for more details). In addition the MDPU scientific partition is responsible for analysing the MXT data stream on board, by identifying valid X-ray patterns,
building sky images, and detecting and localizing afterglow candidates in near-real-time. The afterglow candidate positions are improved during the observation, as long as more data are accumulated, and regularly transmitted to ground (every $\sim$\,30 s) in order to allow ground based robotic telescopes to look for optical afterglows in a more efficient way. In fact, the MXT sky error areas will be on average ten times smaller than the ones produced by ECLAIRs, highly enhancing the chances for the VT and other telescopes to correctly identify the optical GRB counterparts. The optical identification and the subsequent spectroscopic distance measurement being a critical task in GRB science, the driving requirement for the MXT design was \textit{to be able to localize 90\,\% of the GRBs pointed to after a slew to better than 2 arc min (J2000)}.

\section{The MXT Scientific Performance}
\label{sec:perfo}
The MXT has been designed to fulfill the scientific performance presented in Tab. \ref{tab:perfo}.
   \begin{table}[ht]
    \centering
    \caption{MXT expected scientific performance.}
    \renewcommand{\arraystretch}{1.25}
        \begin{tabular}{ll}
            \hline \hline
            Energy range             & $0.2 - 10$~keV \\
            Field of View            & $58 \times 58$~arcmin \\
            Angular resolution       & 10~arcmin at 1.5~keV \\
            Source location accuracy & $< 120$~arcsec for 80\,\% GRBs \\
            Effective area           & $\sim 35$ cm$^2$ at 1.5~keV \\
            Sensitivity ($5 \sigma$) & 10~mCrab in 10~s \\
                                     & 150~$\mu$Crab in 10~ks \\
            Energy resolution        & $<80$ eV at 1.5~keV \\
            Time resolution          & 100~ms \\
            \hline
        \end{tabular}
    \label{tab:perfo}
    \end{table}
In order to test and validate this performance the MXT (proto-)flight model has been installed in its final configuration in the vacuum chamber of the MPE PANTER facility \citep[e.g.][]{panter} in Neuried near M\"unchen (Germany) and illuminated with a quasi-parallel flux of X-rays at different energies and positions inside and outside the FOV, between October 20th and November 5th 2021. The X-ray source in PANTER is located at $\sim$\,130\,m distance from the MXT instrument and is able to produce a uniform flux (5\,\% level) over about 1\,m of diameter. The X-rays, generated by Bremsstrahlung, illuminate different targets and so mono-energetic fluorescence beams are created.
\newline

\noindent The main goals of the calibration campaign were:
\begin{itemize}
    \item G1: to measure the MXT effective area as a function of energy and the vignetting over the FOV,
    \item G2: to characterize the spectral response of the MXT, including the relationship between ADU and keV (gain and offset), the Charge Transfer Efficiency (CTE), and the response non-linearities,
    \item G3: to validate the on board software localization performance.
\end{itemize}

The measure of the characteristics of the MXT PSF has been performed during a previous campaign dedicated to the stand alone MOP FM. However we measured again the optical properties of the integrated telescope, their temperature dependence, and the line of sight of the integrated telescope.  The details of this analysis are given in an accompanying paper by \citet{feldman22}, but here we recall the main results in \S \ref{sec:optics}.

\begin{figure*}[ht]
\centering
\includegraphics[width=8cm]{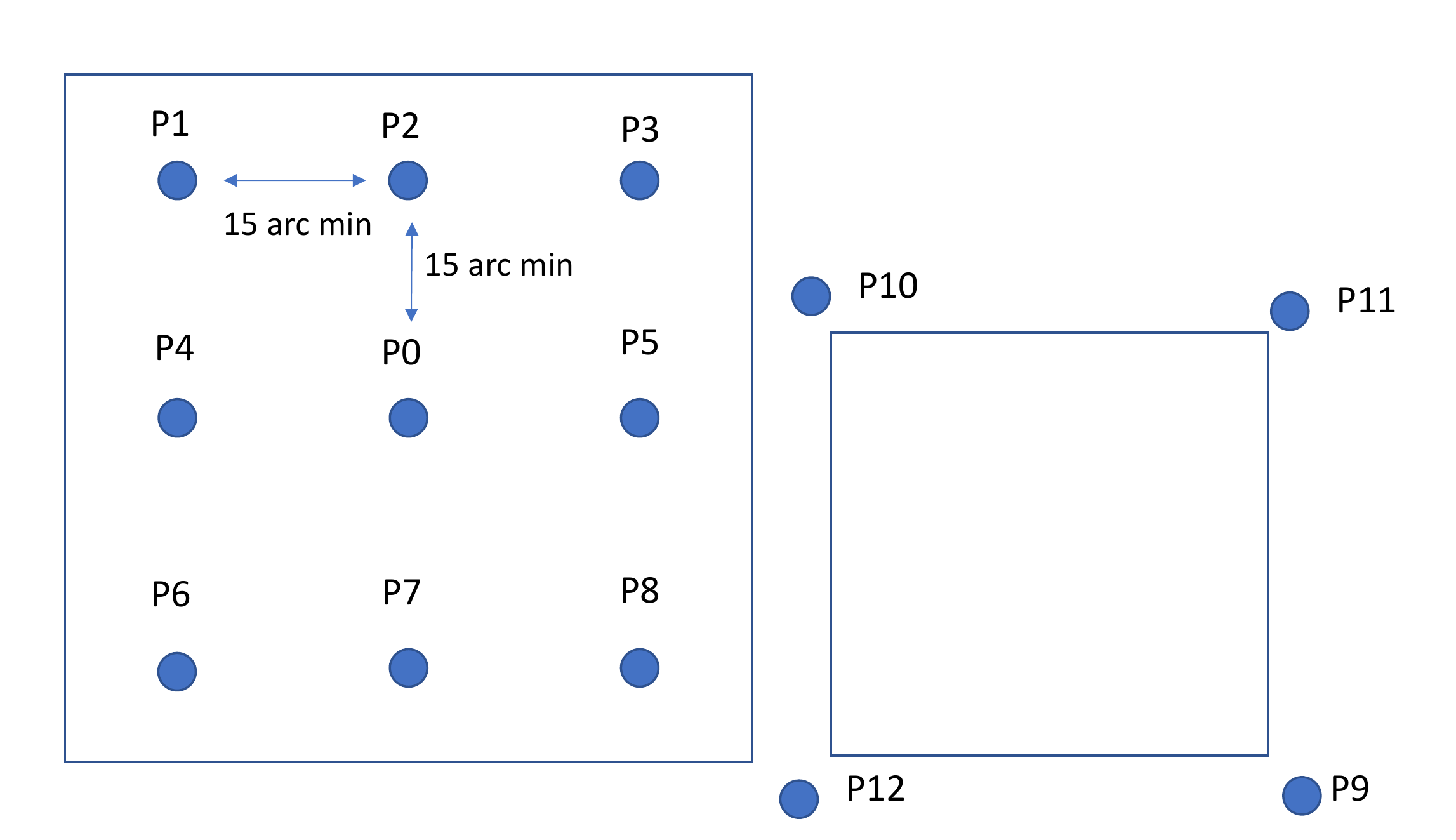}
\caption{Left: sketch of the nine in-FOV positions used for testing. P0 is the on-axis position. Right: sketch of the out-of-FOV positions used mainly for spectral calibration purposes.}
\label{fig:positions}
\end{figure*}

In order to cover in the most uniform way the entire energy range we chose to use the energies specified in table \ref{tab:energies}, and in order to spatially cover the entire field of view, we defined nine positions 15 arcmin apart from each other. The nine in-FOV positions have been complemented by four out-of-FOV positions (50 arcmin off-axis on both axes). The latter have been defined in order to exploit the straight-through flux (see \S \ref{sec:mxt}) and to cover the entire detector in a uniform manner for spectral characterization purposes (G2). The 12 positions are sketched in Fig.\,\ref{fig:positions}.

   \begin{table}
    \centering
    \caption{Summary of the main energy lines used for calibration.}
    \renewcommand{\arraystretch}{1.25}
        \begin{tabular}{ll}
            \hline \hline
            Name            & Energy (keV) \\
            \hline
            C-K	&	0.277 \\
            O-K	&	0.525 \\
            Cu-L&	0.93 \\
            Mg-K&	1.253 \\
            Al-K&	1.486\\
            Ti-K$_{\alpha}$&	4.508 \\
            Fe-K$_{\alpha}$&	6.398 \\
            Fe-K$_{\beta}$&	7.053 \\
            \hline
        \end{tabular}
    \label{tab:energies}
    \end{table}
  
\subsection{Optical performance}
\label{sec:optics}
The PSF profile cut along both axes can be fitted (see Fig.\,\ref{fig:psf}) with a Lorentzian profile offset by a constant of the form:
\begin{equation}
    f(x)=\frac{1}{1+\left(\frac{2x}{G}\right)^2}+c,
\end{equation}

\noindent while the global 2D PSF can be represented by the following function:

\begin{equation}
    F(x,y)=\frac{A(f_{1}(x)+f_{2}(x))(f_{1}(y)+f_{2}(y))}{(1+\eta)^2},
\end{equation}

\noindent where 

\begin{eqnarray}
f_{1}(x)=\frac{1}{1+\left(\frac{2x}{G}\right)^2} \hspace{1.5cm} f_{2}(x)=\eta\left(1-\left(\frac{x}{H}\right)^2\right)\\
f_{1}(y)=\frac{1}{1+\left(\frac{2y}{G}\right)^2} \hspace{1.5cm} f_{2}(y)=\eta\left(1-\left(\frac{y}{H}\right)^2\right)
\end{eqnarray}

\noindent where G represents the width of the Lorentzian, A is a normalization parameter, H=2d/L (with d the pore width and L the pore length) and $\eta$ is the relative strength of the cross arms and straight through flux w.r.t. the central spot.

\begin{figure*}[ht]
\centering
\includegraphics[width=8cm]{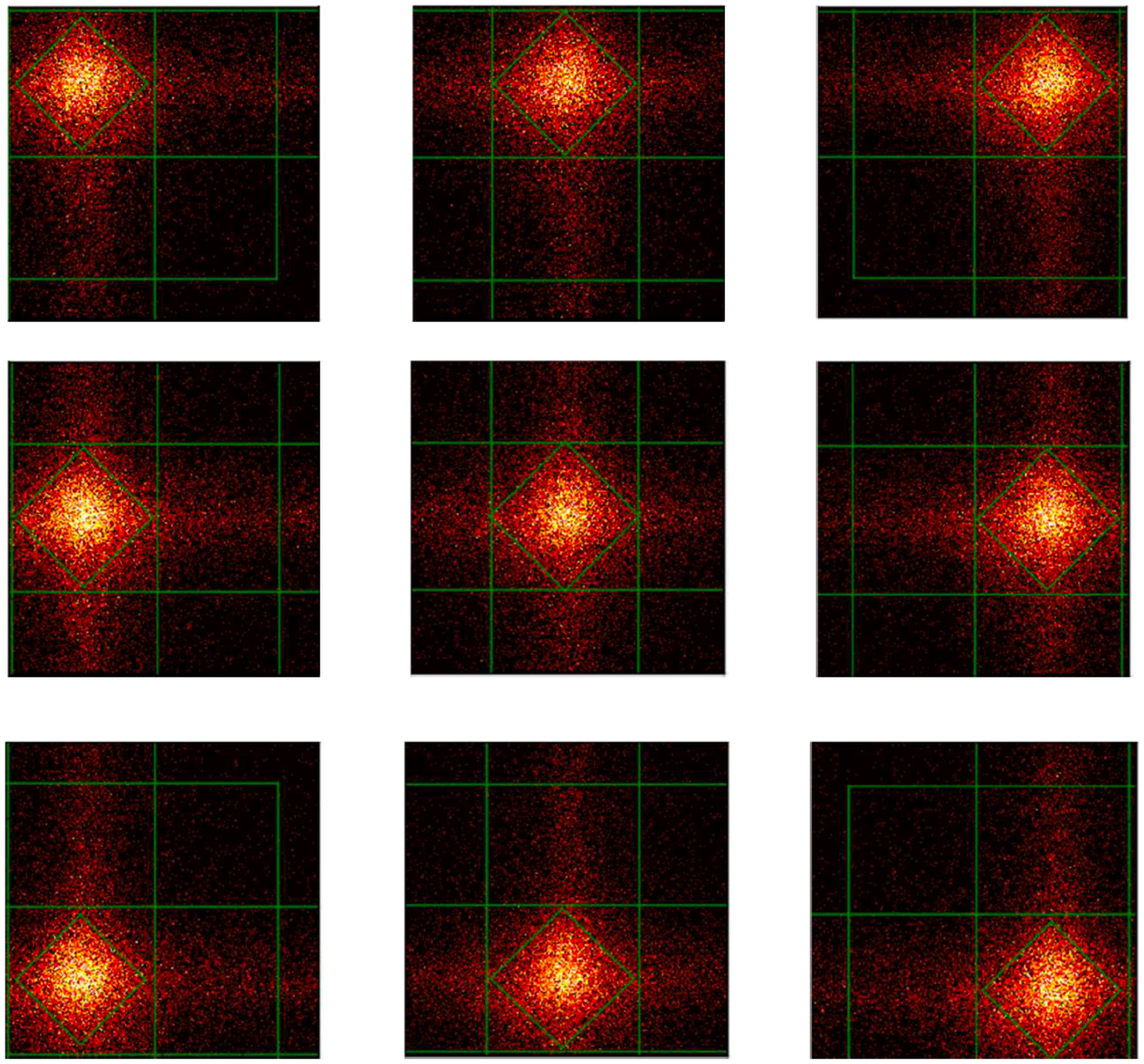}
\caption{Mosaic of the nine in-FOV positions for the Al-K source.}
\label{fig:mosaic}
\end{figure*}

By fitting the data of the nine in-FOV positions (see Fig.\,\ref{fig:mosaic}) for the Al-K source the MXT plate scale, and hence the in flight focal distance of 1137\,mm, could be measured.

The FWHM of the PSF at Al-K could also be accurately measured, turning out to be 11 arcmin, 10\,\% above the required value of 10\,arcmin. The reasons of this are the intrinsic defects in the MPOs (dominant component), alignment errors during MPOs integration on their frame, and mechanical errors on the support frame.  

\begin{figure*}[ht]
\centering
\includegraphics[width=6cm]{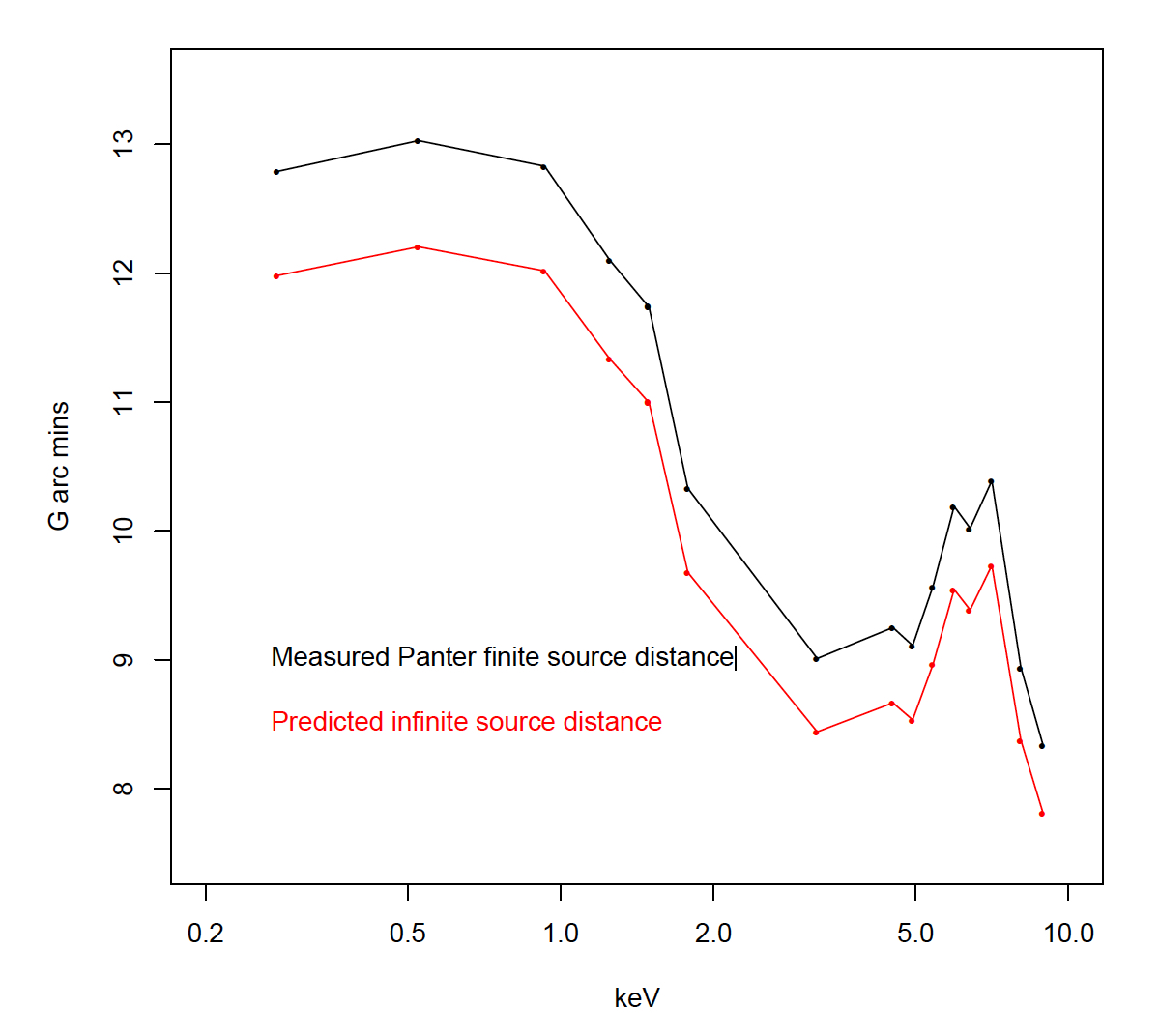}\includegraphics[width=6cm]{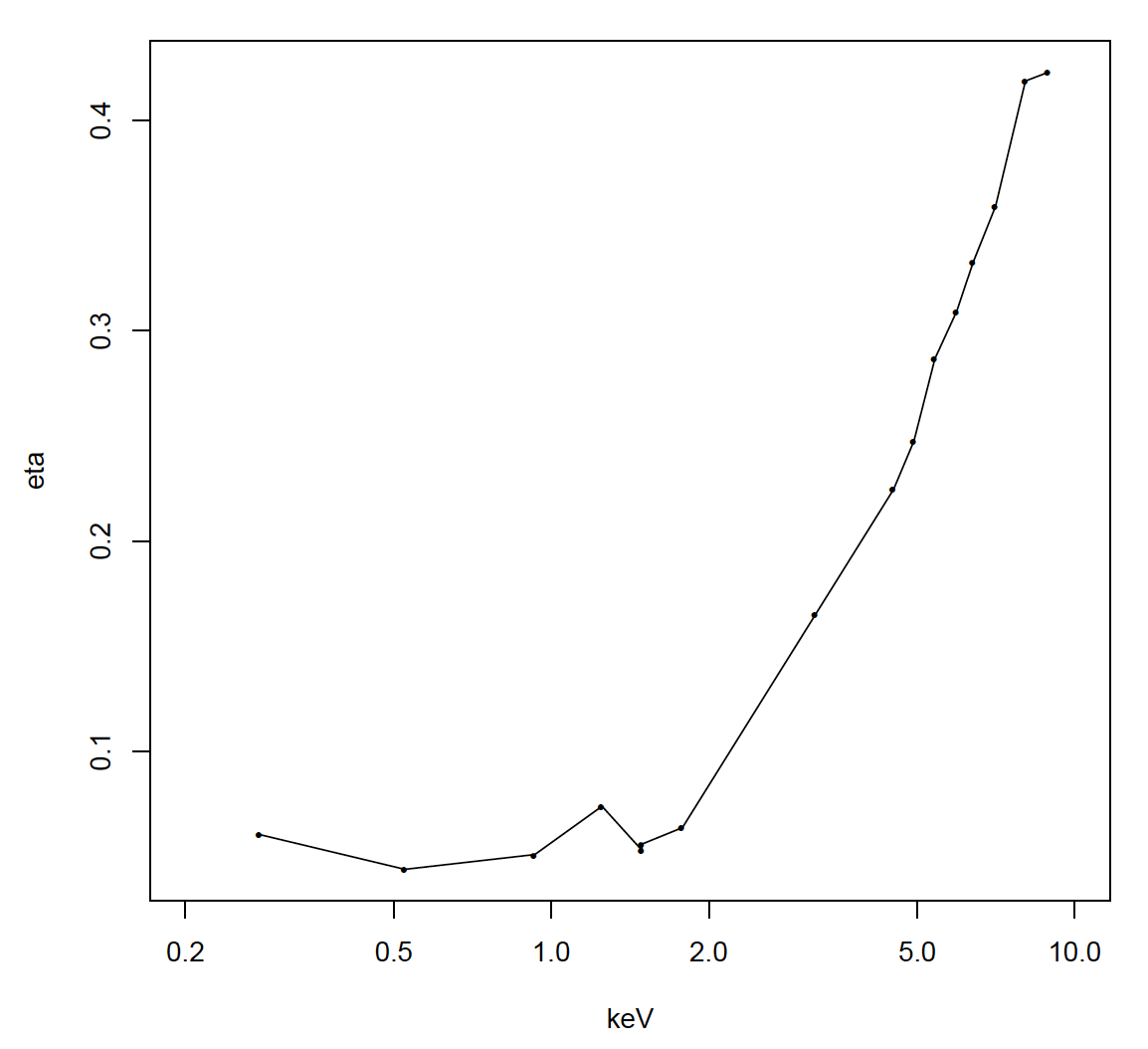}
\caption{Left: measured and predicted G value as a function of energy. Right: measured $\eta$ parameter versus energy.}
\label{fig:fwhmvsenergy}
\end{figure*}

The same measurement has been repeated at different energies and the resulting FWHM (G) as a function of energy is represented in Fig.\,\ref{fig:fwhmvsenergy}. As can be seen a significant variation of G with energy is present. This is due to the fact that different MPOs are contributing to imaging at different energies, and a consequence of the energy dependence of the limiting grazing reflection angle at X-ray energies: at high energies, only the central MPO is contributing and G is smaller due to the fact that alignment errors are negligible, while at lower energies the outer MPOs contribute significantly to the PSF creation, and in this case the different alignment errors between the plates add up. On the same Figure one can see the measured
$\eta$ value as a function of energy. While the relative intensity of the cross arms w.r.t. the central spot increases with energy, the size of the spot itself decreases with energy, allowing for fair imaging capabilities of the MXT at the high end of its energy range.

\subsection{Effective Area}
\label{sec:effarea}

One of the main goals of the PANTER campaign was to measure the telescope effective area as a function of energy. In order to obtain such a measurement, we decided to compare the total number of collected counts on the MXT detector with  the ones collected by an SDD detector intercepting the PANTER beam at a distance of 34.37\,m from the source. The entrance window of the SDD is circular with a radius of 2.33\,mm, implying a geometrical surface of 17.07\,mm$^{2}$. The MOP was placed at 131.4\,m from the source.
By taking into account the MOP geometric area, the MXT pnCCD quantum efficiency (QE), as calculated by \citet{meidinger06}, the SDD QE, as provided by the manufacturer, and the global 170\,nm Al-K filters (100 for the pnCCD and 70 for the MOP), one can derive the effective area of the MXT. In order to measure it as accurately as possible we decided to add additional line energies (see Tab.\,\ref{tab:adden}) and also the Bremsstrahlung continuum produced by the PANTER source up to about 4\,keV.

  \begin{table}
    \centering
    \caption{Summary of the additional energy lines used for effective area calibration.}
    \renewcommand{\arraystretch}{1.25}
        \begin{tabular}{ll}
            \hline \hline
            Name            & Energy (keV) \\
            \hline
            W-M & 1.78\\
            Ag-L&	3.20\\
            Cr-K&	5.40\\
            Cu-K&	8.04 \\
            Ge-K& 9.89 \\
            \hline
        \end{tabular}
    \label{tab:adden}
    \end{table}

\begin{figure*}[ht]
\centering
\includegraphics[width=6cm]{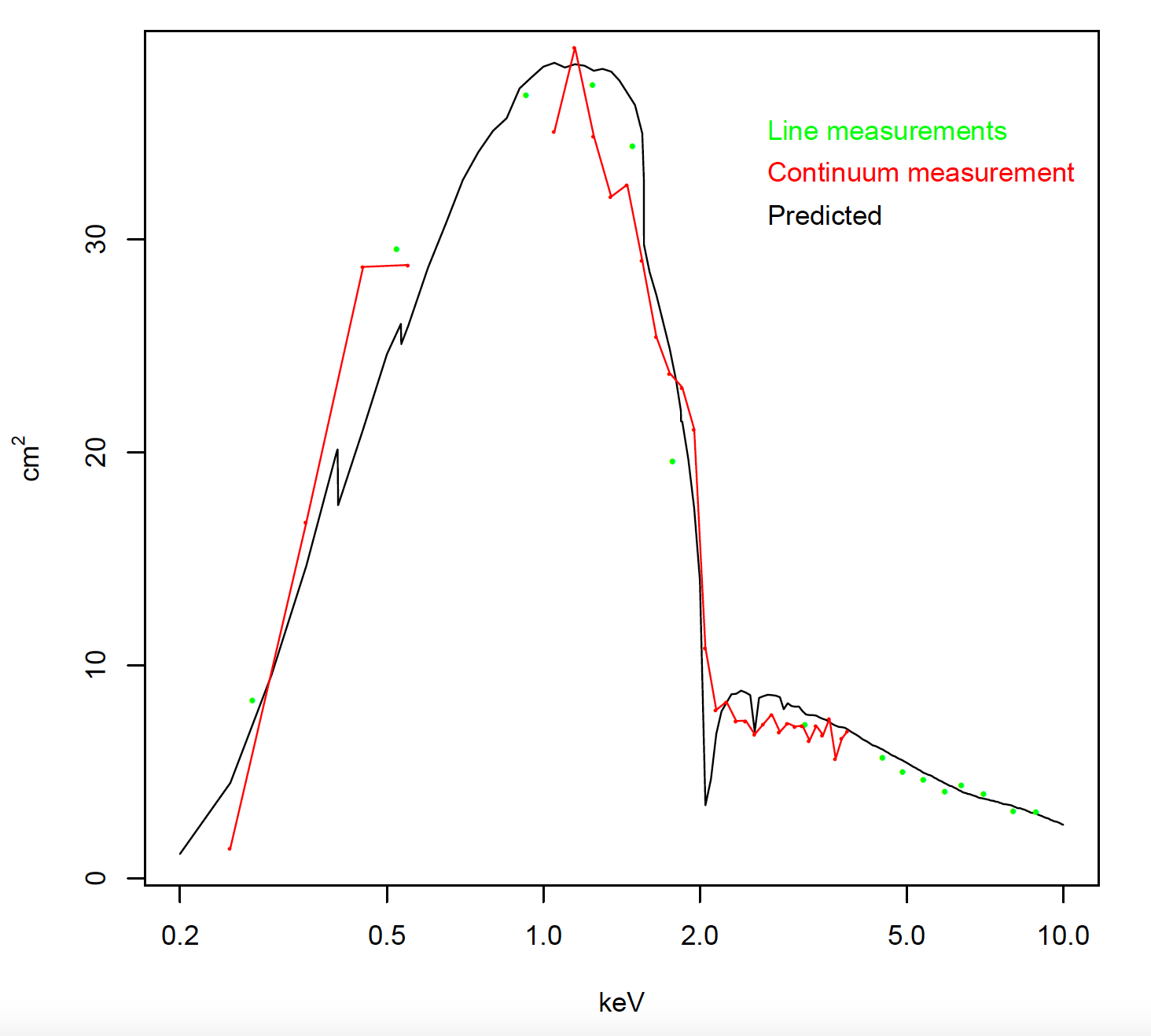}\includegraphics[width=6cm]{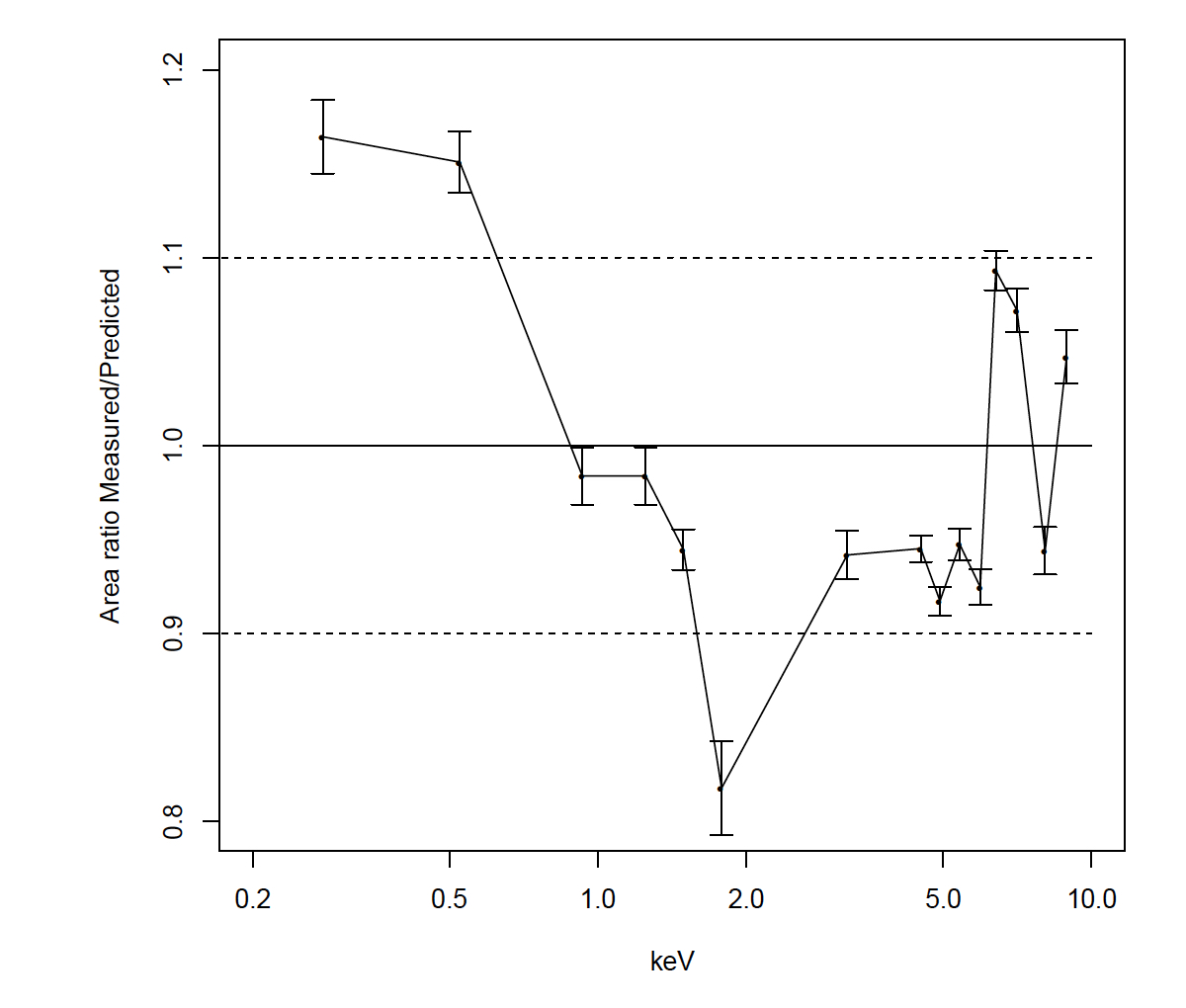}
\caption{Left: measured MXT effective area using line and continuum measurements, compared to the expected model. Right: ratio between the measured effective area and the model.}
\label{fig:aeff}
\end{figure*}

The results are shown on Fig.\,\ref{fig:aeff}, where we show the comparison between the MXT measured total (i.e. peak and cross arms) effective area, as derived from line and continuum measurements compared to the expected model, derived from the MOP FM stand alone measurement performed in PANTER in January 2021.

We note that the agreement is, in general, very good, and the ($<$\,20\,\%) discrepancies are probably due to the fact that the QE of the SDD and of the pnCCD have not been measured in a dedicated way. For the pnCCD this could not be done due to the project schedule constraints, but we foresee to absolutely calibrate the SDD in a metrology line in a synchrotron facility before the launch of SVOM. This will help in further reducing the uncertainties in the MXT effective area. In addition, once in flight, the effective area of the MXT will be tested against known calibration sources, such as the Crab nebula.

\subsection{Vignetting}

The effective area of the MXT at off-axis positions depends on the area vignetting function of the MOP and on the field of view of the MXT detector. As a source moves off-axis some portions of the PSF fall outside the MXT detector FOV and other portions enter the FOV. The full vignetting function of the MOP was measured at many off-axis positions and for a series of line energies in the FM MOP PANTER tests in January 2021. During the MXT PANTER tests in fall 2021 we could verify the vignetting properties of the integrated telescope at different energies. It turns out that in the central part of the FOV (i.e. $\pm$\,15\,arcmin over the two axes) the vignetting is less than $\sim$\,10\,\% at all energies. 

\subsection{Spectral Response}

\begin{figure*}[ht]
\centering
\includegraphics[width=10cm]{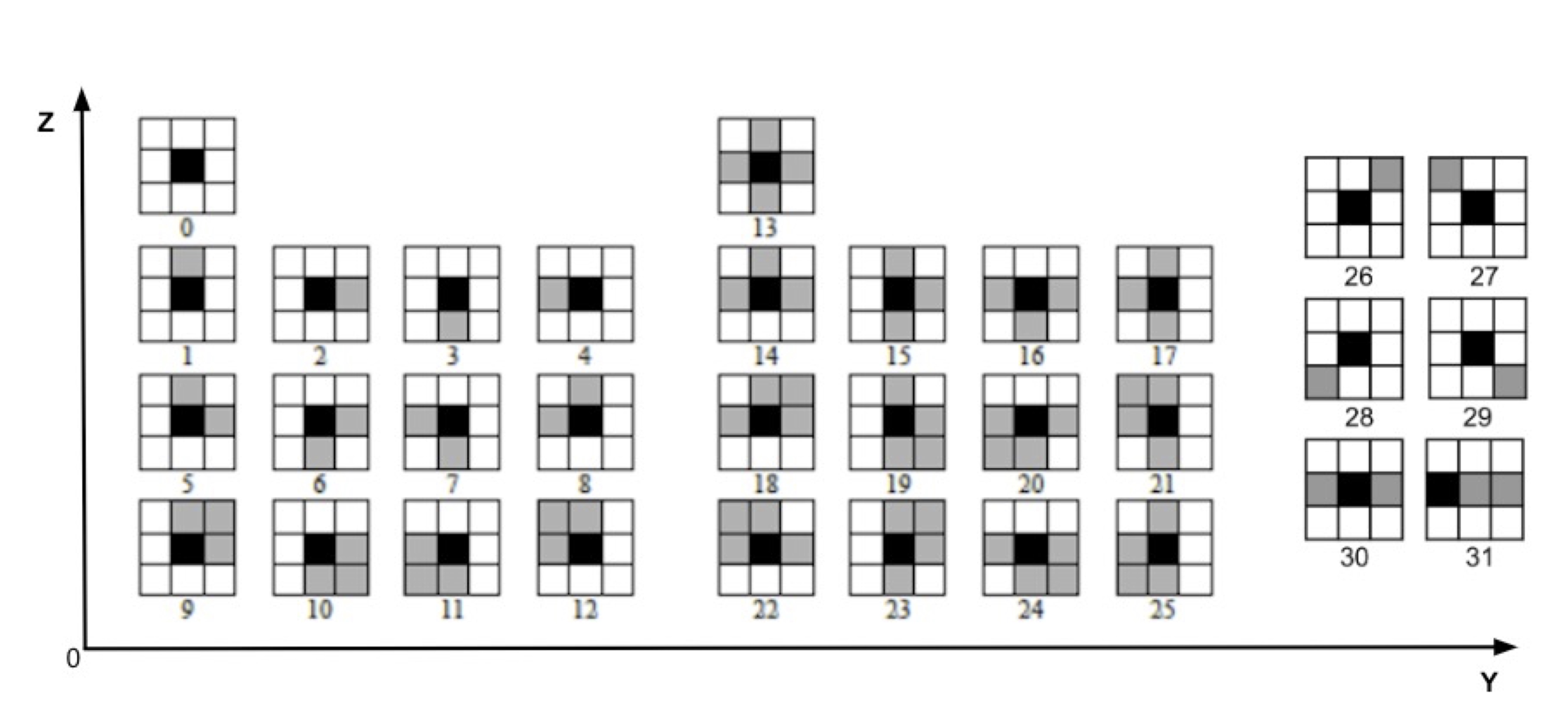}
\caption{ADU patterns as classified by the MXT recognition algorithm. Only patterns 0 to 12 are considered as valid photons. Patterns 26 to 29 could correspond to two single pixel real photons. However, due to the large oversampling of the PSF by the detector pixels, this occurrence has been estimated to be negligible ($<$\,1\,\%).}
\label{fig:patterns}
\end{figure*}

In this section we describe the ground processing of the MXT data, that has been developed in order to derive the spectral response of the pnCCD detector using PANTER data. This processing allowed us to derive the initial calibration values for the MXT and will be applied later on to on-flight data.
It can be divided into two main processes:
\begin{itemize}
    \item pattern recognition: it consists in identifying in each pnCCD frame the groups of adjacent pixels with a charge (ADU) value above the threshold in order to decide if they originate from a single X-ray photon or if they have another origin (most of the time an ionizing particle). The pixels are tagged accordingly and the deposited charges are co-added. Note that for the pnCCD valid X-ray patterns are considered as those involving up to a maximum of four pixels, see Fig.\,\ref{fig:patterns}
    \item camera calibration: this step allows the conversion of the ADU value(s) of the pixel(s) forming a valid X-ray photon into a single energy, that of the photon, in physical units (eV).
\end{itemize}

The second step is described in detail in the accompanying paper by Schneider et al. Here we just recall the three main steps that are applied to the data in order to obtain calibrated event lists:
\begin{enumerate}
    \item the correction for the Charge Transfer Inefficiency (CTI\footnote{CTI=1-CTE}). This accounts for the charge loss suffered by the initial electron packet when shifted from its creation position through the image area and the frame store to the anode. This is caused by crystalline defects (in very low concentration) in the bulk material acting as trapping centers for the electrons. It is modeled by a single value for a given energy, corresponding to the relative loss for a shift by one pixel. This value is about a few 10$^{-5}$ at launch and will probably increase during the mission with  the radiation level suffered by the detector;
    \item the “ADU to eV” conversion, also called the spectral calibration, which allows the conversion of the digital measurement (ADU) of the deposited charge (in a single pixel) into a physical energy (eV). This is a characteristic of the electronic read-out circuit, and there is thus one conversion function per electronic chain, i.e. one per column. Calibration results show that a linear function, with one offset and one gain (per column) can adequately represent this conversion; the dispersion of the gain values is quite small ($\sim$\,1\,\%) and the one of the offsets is of the order of ($\sim$\,10\,\%), showing that the spectral response of the pnCCD is quite uniform over the detector; 
    \item the correction for charge sharing. This accounts for the “non registered” part of the electron packet, i.e. for actual charges (in a single pixel) below the electronic threshold set for transmitting the pixel to the ground.
\end{enumerate}

\begin{figure*}[ht]
\centering
\includegraphics[width=10cm]{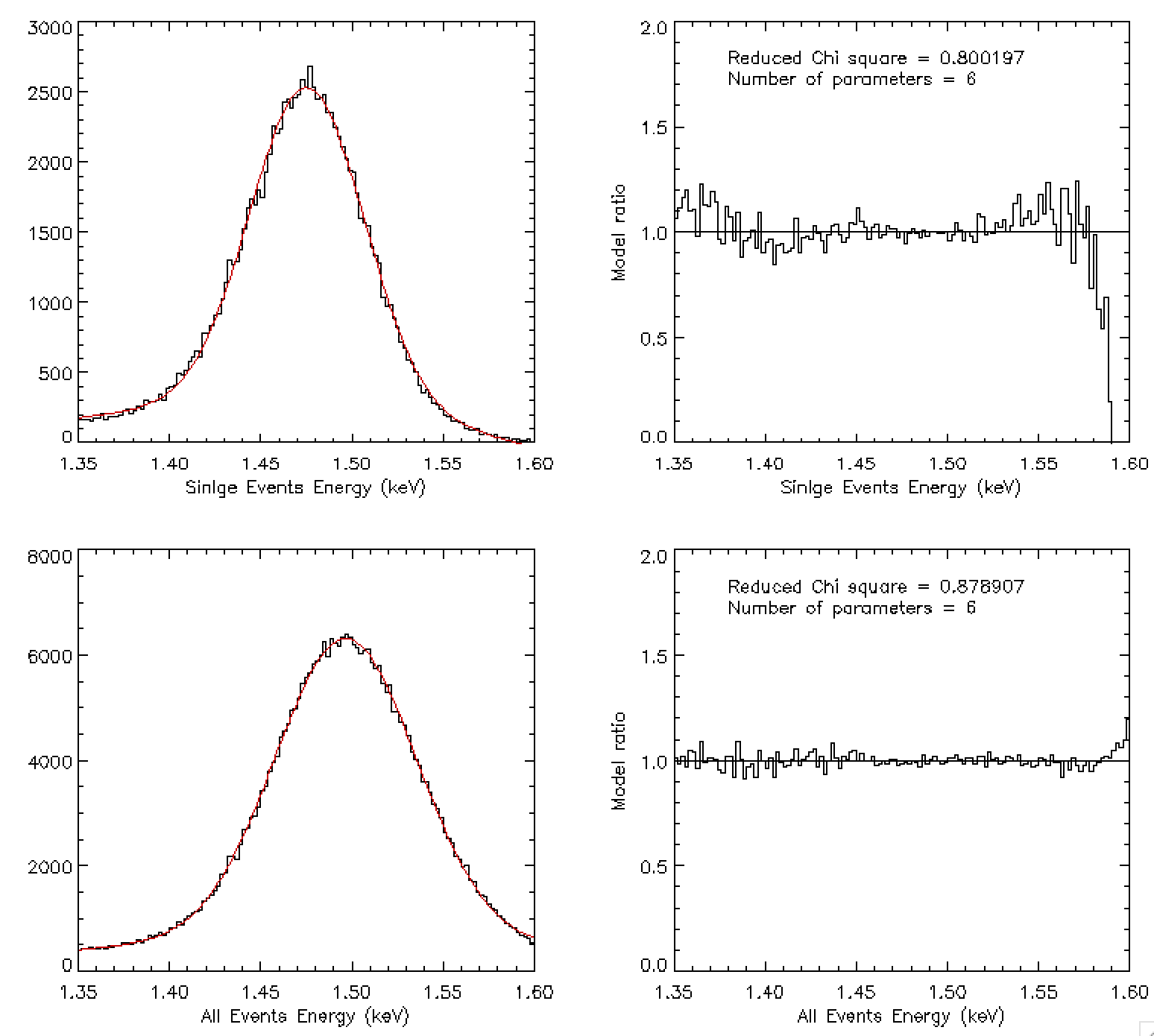}
\caption{Top Left: Al-K spectrum obtained using only single events. The black line represents the data, in 5\,eV bins, while the red line represents the fit using the function described in the text.  Top Right: The residuals of the fit in terms of ratio between the data and the model. The reduced $\chi^{2}$ is reported as well as the number of parameters used in the fit. Bottom Panels: same as top panels but using all events (singles to quadruples).}
\label{fig:fit}
\end{figure*}

Once all the calibration steps are applied, we combined the data of the four out-of-FOV data sets (see Fig. \ref{fig:positions} right) for each energy and we fitted resulting spectra using a combination of a Gaussian functions, and, where needed, a function (constant, linear or quadratic) in order to take into account the residual background, using up to six free parameters. One example of the fit and its residuals is given in Fig.\,\ref{fig:fit}.

The measured line widths as a function of energy are reported in Table~\ref{tab:spfwhm} for a number of representative line energies. As can be seen the mission requirement of 80\,eV at Al-K is met for single events, as well as for the totality of the events. In the same table one can also find the reconstructed energy after calibration, which shows that the energy scale is sufficiently accurate (less than 13\,eV) to meet the required mission accuracy of $\pm$\,20\,eV.

 \begin{table}[ht]
    \centering
    \caption{Summary of the MXT spectral performance. The upper part of the Table applies to single events, while the lower part to all the multiplicities.}
    \renewcommand{\arraystretch}{1.25}
        \begin{tabular}{lllll}
            \hline \hline
            Energy line (keV)  & Reconstructed  & Difference & FWHM (eV) & FWHM  \\
                                & energy (keV) & (eV) &     &   1 $\sigma$  error (eV)\\ 
            \hline
0.277	&0.280	& 3	& 48.1	&1.6	\\
0.525	&0.525	& 0	&60.0	&0.8	\\
0.930	&0.928	& 2	&65.3	&0.9	\\
1.253	&1.252	& 1	&69.2	&1.4	\\
1.486	&1.483	& 3	&72.7	&0.6	\\
4.508	&4.502	& 6	&115.5	&0.6	\\
5.405	&5.404	& 1 &127.3	&0.5	\\
6.398	&6.399	& 1	&136.8	&0.8	\\
            \hline
0.277	& 0.284 & 7	&49.5	&0.9\\
0.525	&0.531	& 6	&64.5	&0.6\\
0.930	&0.934	& 4	&70.4	&0.8\\
1.253	&1.261	& 8	&73.4	&0.8\\
1.486	&1.492	& 6	&79.3	&0.4\\
4.508	&4.513	& 5	&123.0	&0.4\\
5.405	&5.416	& 11&134.8	&0.2\\
6.398	&6.411	& 13&146.6	&0.3\\
\hline
        \end{tabular}
    \label{tab:spfwhm}
    \end{table}


\subsection{Background and Sensitivity}

The X-ray background on the MXT detector is dominated by the focalization of the CXB, see \S \ref{sec:loc} However there are additional components, due to the space radiation environment, that contribute to the induced X-ray background on the focal
plane. These components have been evaluated through GEANT4 Monte-Carlo simulations, and primary particles have been generated isotropically from the inner surface of a sphere, centered to the detector and of radius (typically 2 m) at least 10 times the size of the MXT mass model, range of emission angle (typically 2$^\circ$) restricted to intercept at most the model. As background sources (primary particles) the CXB (0.1 keV--100 GeV),
comic protons ($<$ 20 MeV), trapped protons ($<$ 400 MeV), and trapped electrons ($<$3.5 MeV) have been considered. The probability or each component to depose energy in the detector in the 0.2--10 keV energy range has been calculated.
In addition to primary interactions, the secondary radiation induced by the activation (especially after the South Atlantic Anomaly passages) of the materials  surrounding the the detector has been simulated with 
GEANT4. The sum of both components is expected to be of the order of 1.3$\times$10$^{-6}$ counts/pixel/s, which implies an expected contribution of $\sim$0.1 counts/s over the entire detector, confirming that the 
focused CXB is the main background component for the MXT (1 count/s).

Using the MXT effective area derived in \S \ref{sec:effarea}, one can estimate the expected counts for
an astrophysical source in the MXT and obtain its sensitivity by comparing this quantity with the expected
background. If, for example, we consider a Crab-like spectrum with a photon index $\Gamma$=2.1, a normalization
of 9 photons/cm$^{2}$/s/keV at 1 keV and an N$_{H}$ equivalent column density of 0.45$\times$10$^{22}$ atoms/cm$
^{2}$, we obtain an expcted count rate of 121 cts/s in the MXT detector over the entire energy range in 1 ks. The 5$\sigma$ MXT sensitivity is hence about 1 mCrab in this case, 13 mCrab in 10 s, and about 400 mCrab for a 10 ks observations.
These values can be considered consistent with the specified ones, see Tab. \ref{tab:perfo}, since they are based on the simple comparison of the expected counts over the entire the detector and they do not consider the advantage of the imaging properties of the MXT, where $>$ 50\% of the counts are concentrated in the center of the PSF. The latter is spread over an area of about 100$\times$100 pixels$^2$, and the expected background within this area is about a 0.15 fraction of the one expected on the whole detector. Taking this into account, the final sensitivity value is improved by a factor 30\%.

\subsection{Scientific On Board Software Performance}
\label{sec:obsw}
The MXT on-board scientific software was developed to localize X-ray sources in the field of view of the telescope. The method relies on a cross-correlation technique (performed in the Fourier domain), coupled to a barycenter method, which offers good localization performance, including for faint sources. In addition, several correction analyses were implemented to account for effects such as Fourier transform spectral leakage, or insensitive columns of camera pixels.

\begin{figure}   
  \centering
  \includegraphics[width=12cm]{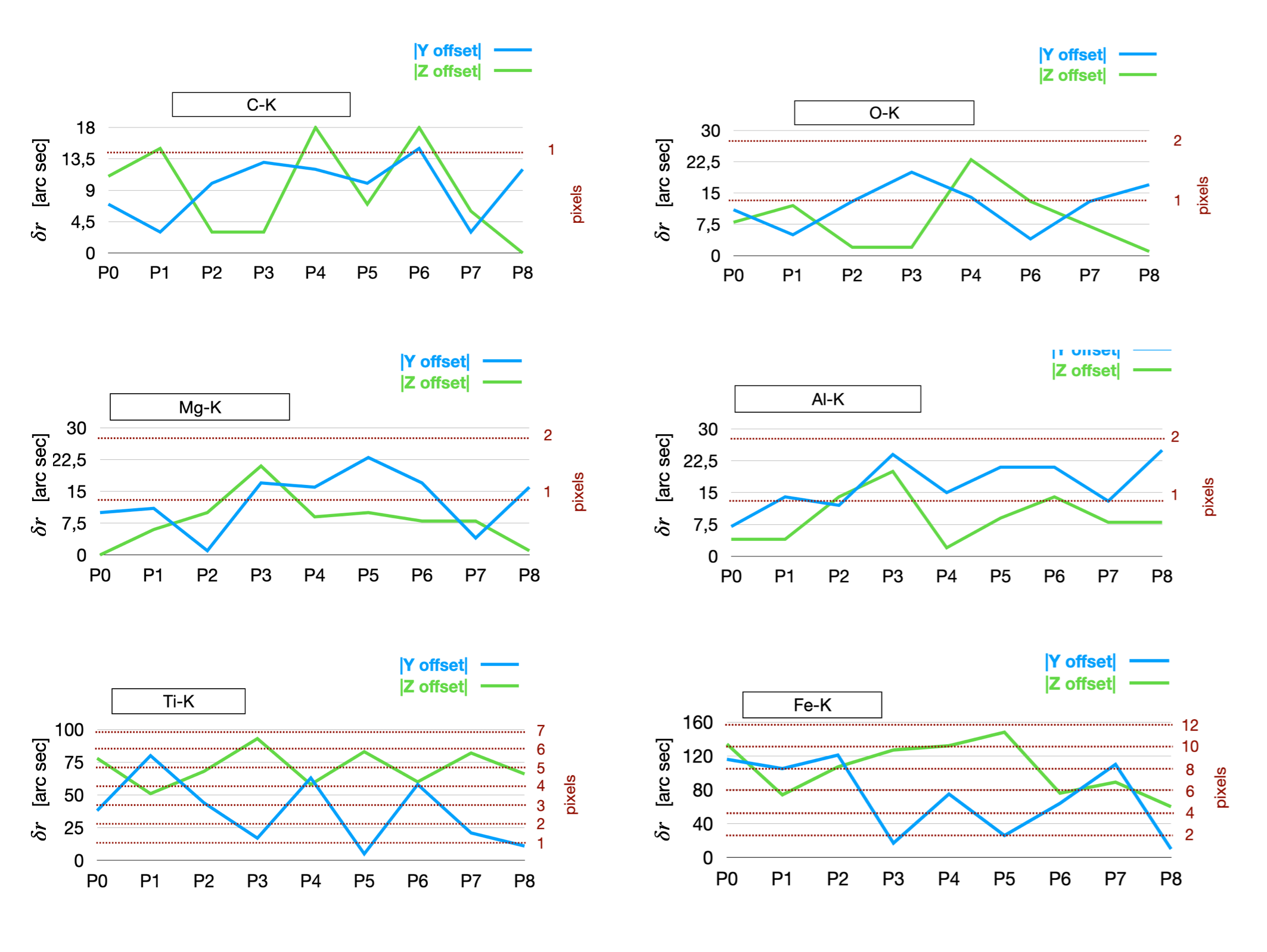}
  \caption{Localization performance for six photon energies: C-K (0.277~keV), O-K (0.525~keV), Mg-K (1.253~keV), Al-K (1.486~keV), Ti-K$_\alpha$ (4.508~keV), and Fe-K$_\alpha$ (6.398~keV). The angular (or pixel) difference between the measured position and the true position is evaluated in the $y$ direction (blue) and in the $z$ direction (green).}
  \label{fig:loc_perfo} 
\end{figure}

The localization algorithm was characterized at the PANTER facility using the X-ray sources listed in Tab.~\ref{tab:energies}, and for multiple beam positions represented in Fig.~\ref{fig:positions}. This is summarized in Fig.~\ref{fig:loc_perfo}. We measured the angular difference $\delta r$ between the true beam position and the reconstructed one in both the $y$ and $z$ directions. A sub-pixel resolution ($<15$~arcsec) is achieved for energies lower than 1~keV. For the highest energy source (Fe-K, 6.40~keV), the resolution is better than 120~arcsec. Moreover, the localization performance is uniform across the tested beam positions.

\begin{figure}
  \centering
  \includegraphics[width=10cm]{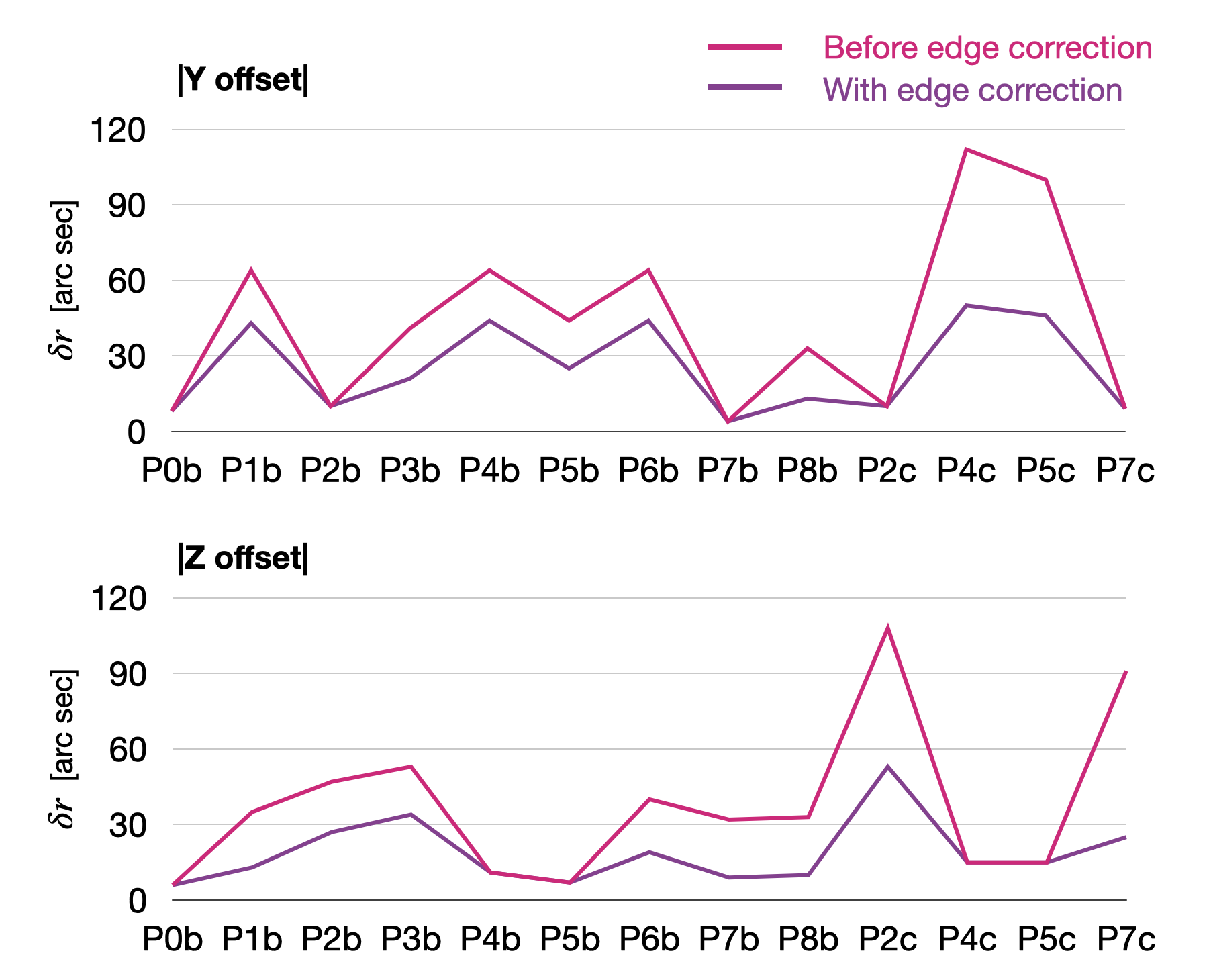}
  \caption{Localization performance tested for 13 near-the-edge positions and for an Al-K (1.486~keV) source. The angular difference between the measured position and the true position is evaluated in the $y$ direction (top) and $z$ direction (bottom).}
  \label{fig:loc_edge_bias}
\end{figure}

The cross-correlation method is biased when the source is located near the edges of the camera plane due to spectral leakage: recovered positions are shifted towards the center of the camera plane. This bias is corrected on-board to guarantee a uniform performance over the entire camera plane. The spectral leakage correction is characterized in Fig.~\ref{fig:loc_edge_bias}, where near-the-edge positions (labeled with a ``b'' for a 22~arcmin shift from the center and ``c'' for a 25~arcmin shift from the center) were tested with an Al-K (1.486~keV) source. For some beam positions, the localization accuracy is degraded up to a factor two. After applying the bias correction, the localization accuracy is almost back to nominal values (see also Fig.~\ref{fig:loc_perfo}/Al-K for a reference).

\begin{figure}
  \centering
  \includegraphics[width=10cm]{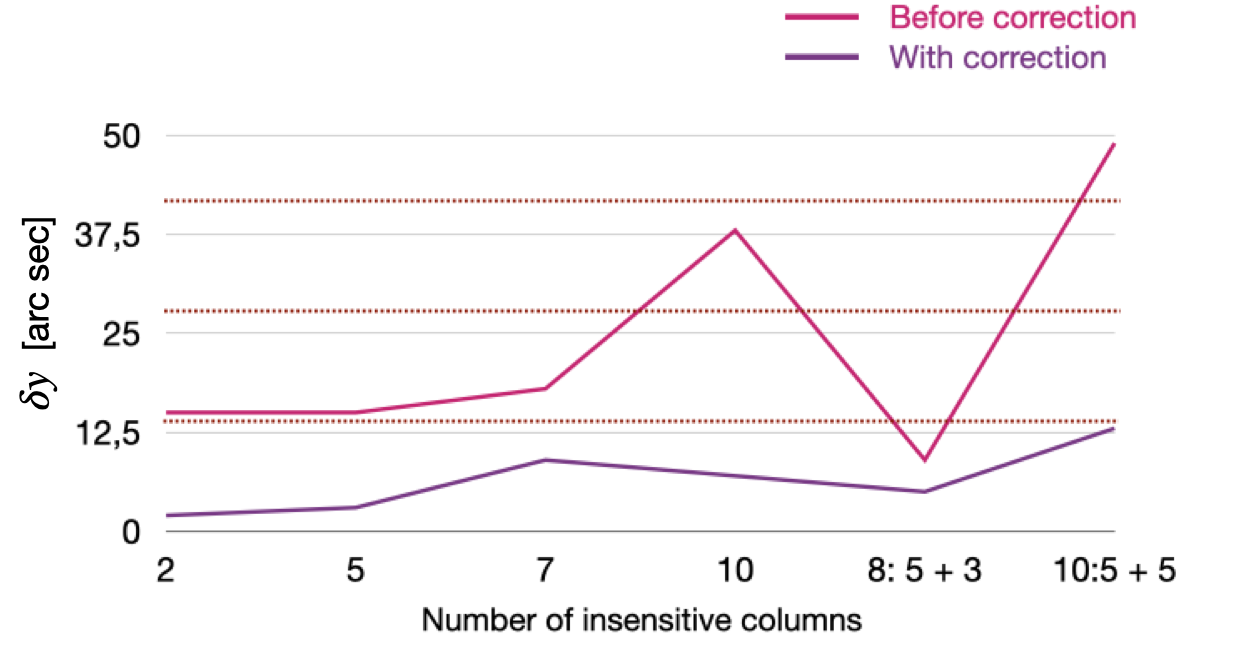}
  \caption{Localization performance tested with insensitive columns of camera pixels, for an Al-K (1.486~keV) source, and with a $P_0$ beam position. The angular difference between the measured position and the true position is evaluated before (pink curve) and after (purple curve) applying the correction function. The last two points were tested with two separated bunch of insensitive columns.}
  \label{fig:loc_insensitive}
\end{figure}

In order to test the robustness of the on board software, we decided to test it against a degraded scenario where one or more columns of the camera
would be insensitive.
 To correct for this effect, we apply a correction function which only depends on the position of the insensitive columns and the number of contiguous insensitive columns. At PANTER, a few columns were de-activated to simulate insensitive columns. This configuration was used to characterize the insensitive-column correction. Figure~\ref{fig:loc_insensitive} shows the effect of the correction for different numbers of insensitive columns. After correction, the localization performance is similar to a fully-active detector (see also Fig.~\ref{fig:loc_perfo} for the nominal performance).

The PANTER test campaign has shown that the on-board localization algorithm performance is compatible with the scientific requirements: an X-ray source, detected with a few hundreds of photons, can be localized with an accuracy better than 120 arcsec. The on-board scientific software also includes features to compensate hardware effects discovered during the development process like the presence of insensitive columns of pixels. If this effect appears during the mission, the onboard scientific software is able to recover the loss of localization accuracy almost completely.

\subsection{Localization Capabilities}
\label{sec:loc}

The data presented above have been used to estimate the in flight localization capabilities for the MXT. In order to obtain this result, we performed a simulation process that needs as ingredients the PSF shape (\S \ref{sec:optics}), the telescope effective area (\S \ref{sec:effarea}), the expected background in flight, and some hypotheses on the SVOM mission. Concerning the background, two main components need to be considered, the Cosmic X-ray Background (CXB) and the particle induced background. If we restrain ourselves to high galactic latitudes (which is compatible with the SVOM pointing law), there is a good agreement on the CXB measurement in the MXT energy range \citep[e.g.][]{cxb}. By folding the CXB spectrum with the MXT effective area one expects about 1 count/s over the entire detector for the CXB.
The non X-ray background has been estimated using GEANT4 simulations of the SVOM space
environment and the MXT mass model. The primary particles were simulated isotropically from the inner surface of a sphere, centered to the detector and of radius (typically 2 m) at least 10 times the size of the mass model. 
The incident particles in the keV-MeV range were the CXB as measured by \citet{moretti09} (in and out of the FOV), cosmic protons, SAA trapped protons, SAA trapped electrons. The average expected non-Xray background
is 0.023 counts/s/cm$^{2}$ (i.e. $\sim$\,0.1~counts/s over the entire detector) dominating above 2 keV.

\begin{figure*}[ht]
\centering
\includegraphics[width=6.cm]{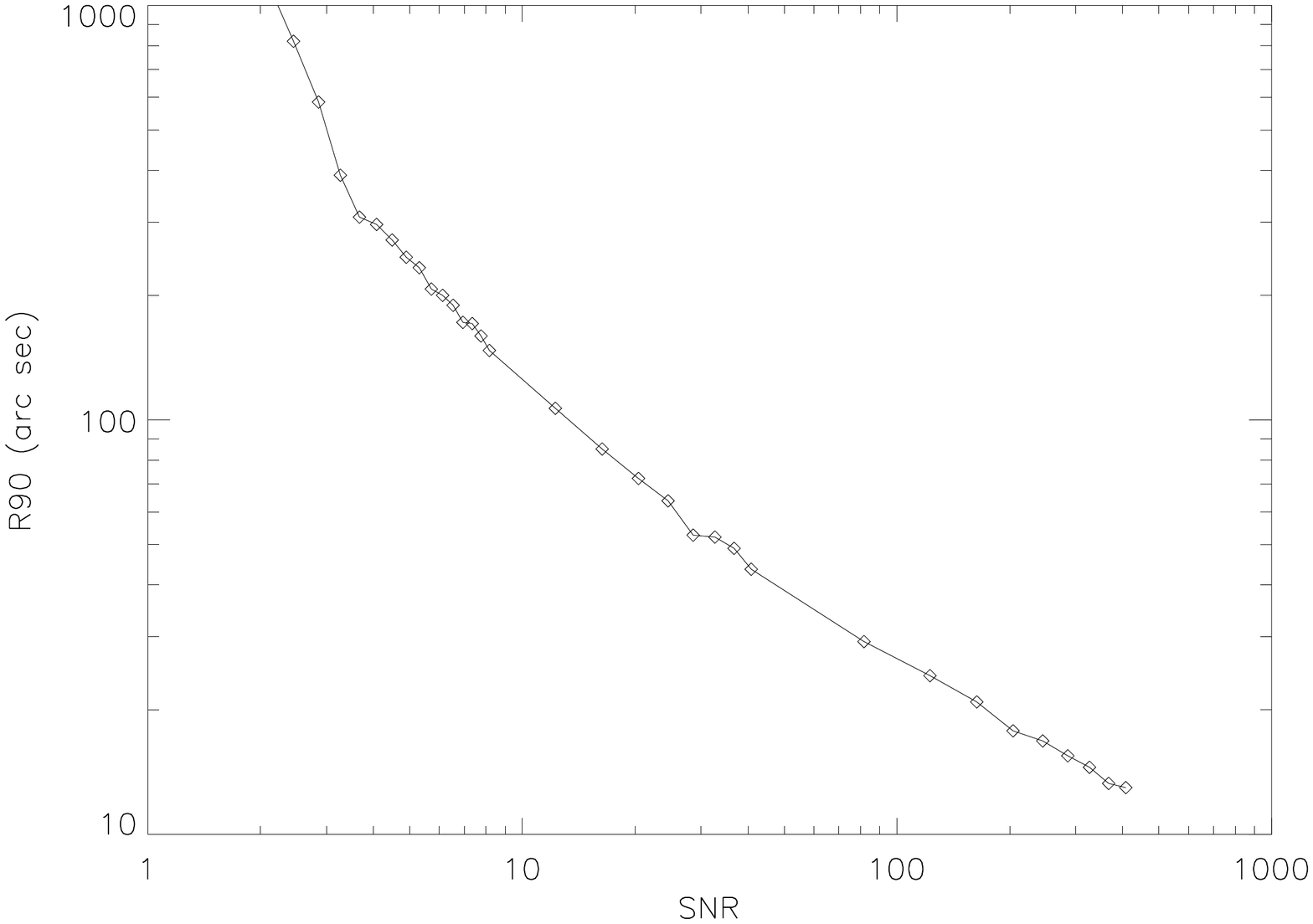}
\includegraphics[width=6.1cm]{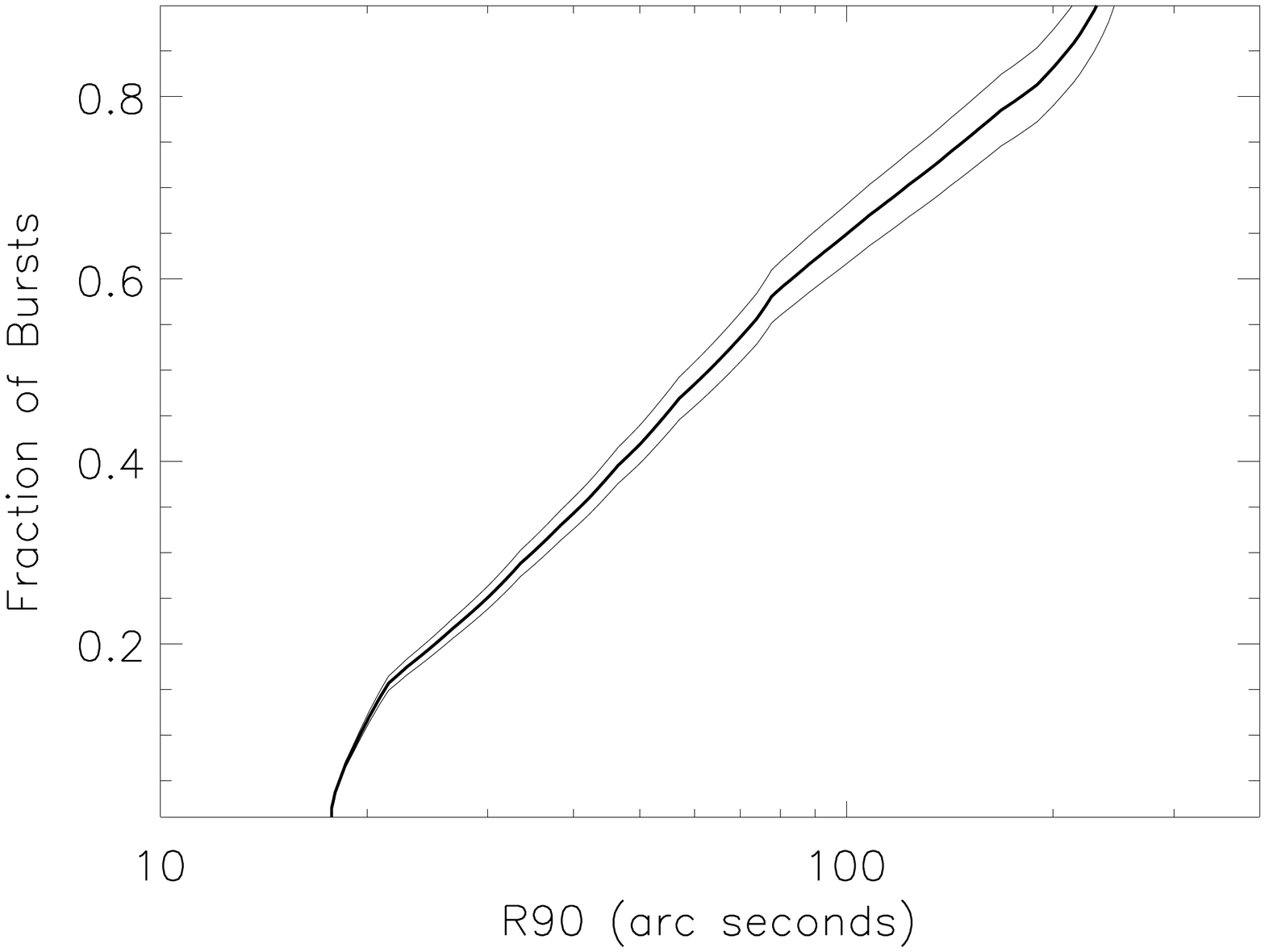}
\caption{Left: The R$_{90}$ vs SNR curve. Right: Expected location accuracy of MXT afterglows, after 10 minutes of observation of GRBs detected by ECLAIRs with more than 50\% probability. See text for more details. The central (thick line) represents the result of the simulation, while the lighter lines represents 5\% uncertainty regions.}
\label{fig:r90}
\end{figure*}

Using these background values we determined the R$_{90}$ vs Signal-to-Noise Ratio curve. What we wanted to test is the ability for MXT to deliver GRB positions with a given accuracy 10 minutes after the end of the slew. This corresponds  roughly to 600 counts of background integrated over the detector. Using the MXT PSF we simulated 1000 sources, distributed uniformly on the MXT FOV and with fluxes comprised between 10 and 10$^{4}$ counts, reconstructed their positions, and compared the latter to the injected positions. So we could define R$_{90}$ as the radius within which 90\,\% of the reconstructed positions fall. In other words, if we associate a given R$_{90}$ to an MXT reconstructed position we have 90\,\% of chances that the true source position is within this radius.
In order to derive the MXT position we implemented the following method: we cross correlate the data 
with the MXT PSF, we look for the maximum in the correlated image, we define an area of
$\pm$\,35 pixels around the maximum, we select the pixels within this area which have at least
90\,\% of the value of the maximum, and finally we barycenter those pixels\footnote{This is not exactly the same algorithm as the one used on board. The latter is described in \S \ref{sec:obsw}.}. The R$_{90}$ vs SNR curve is reported in Fig.\,\ref{fig:r90}.

Once we determined the relation between the localization accuracy and the MXT SNR, we downloaded from
the public \textit{Swift}/NGO archive the pre-processed XRT light curves \citep{evans09} for the GRBs detected between 2004 and 2017 (1128 GRBs). Among those GRBs we selected the ones for which BAT data and spectral parameters were available. This selection has been done because we needed to check which of the BAT GRBs, would have been detected by ECLAIRs: ECLAIRs is less sensitive than BAT and we wanted to restrain our simulations to those GRBs that have at least 50\,\% of probability to be detected by ECLAIRs at a SNR level that would trigger a SVOM slew. We then made the hypothesis that the SVOM and Swift slewing capabilities are similar, so that XRT and MXT are on source at the same time and that they experience the same Earth occultations (Swift and SVOM have similar orbital parameters).

The MXT has a smaller effective area than XRT. Hence, in order to be able to use the XRT light curves, we computed for each GRB the ratio of the MXT expected counts by using the time average spectral parameters measured by XRT and performing XSPEC \citep{arnaud99} simulations for both instruments with the same parameters. Once we determined the count rate conversion factor for all the GRBs afterglows, we integrated, for each light curve, the counts between the start of the observation and 10 minutes, and, using the R$_{90}$ vs SNR curve, we could determine the expected localization uncertainty for the whole GRB sample. 

The results can be summarized as follows: for about 30\,\% of the GRBs, the MXT will provide a localization accuracy better than 30 arcsec; about 50\,\% of them will be localized to better than 1 arc minute and the vast majority of them (>80\,\%) will be localized to better than 2 arc minutes, see Fig. \ref{fig:r90}.

\section{Conclusions}
The PANTER End-to-End tests performed in October/November 2021 are the endpoint of an activity lasting several years, which includes the integration of the so-called MXT Performance Model (PM) in 2019, which is a complete MXT model based on qualification sub-systems of the MXT, which were, however, representative of the scientific flight performance.
The MXT PM was tested in PANTER in February 2020, and those tests allowed us to rehearse the End-to-End tests, providing the necessary return of experience for a successful final calibration of the MXT telescope.

The End-to-End test campaign allowed us to measure the scientific performance of the MXT under various respects: imaging, effective area, spectral properties etc. The measured values are coherent with the ones derived for the PM model, which served as a basis for the simulations we developed in order to prepare the final test campaign.
The results presented here have allowed us to prove that the MXT design responds to the needs of the mission, and we
expect the first narrow-field “Lobster-Eye'' type telescope to be flown on a satellite to fully
accomplish its task, namely to detect, localize and characterize GRB afterglows and other X-ray transients at the dawn of modern multi-messenger astrophysics.

\section*{Acknowledgements}
We thank the PANTER team for the continuous and invaluable support through all the test campaign. 
We also warmly thank the CNES and CEA-Irfu technical staff for setting up the instrument at PANTER and for their fundamental support in operating the telescope. 
The work performed at the PANTER X-ray test facility has in part been supported by the European Union’s Horizon 2020 Program under the AHEAD2020 project (grant agreement n. 871158).

\section*{Declarations}
\subsection*{Authors' Contributions}
D.G\"otz defined the calibration campaign objectives, supported the operations at PANTER, analysed the data, coordinated the manuscript contributions. M.Boutelier, A.Fort supported the operations at PANTER. V. Burwitz coordinated the PANTER team, contributed to the manuscript revision. R. Chipaux performed background simulations. B. Cordier and F.Gonzalez contributed to the manuscript revision. P.Ferrando analysed the data and contributed to the manuscript revision. A. Gros, S. Hussein, F. Robinet analysed the data and contributed to the manuscript.
J.-M. Le Duigou contributed to the PANTER operations.
N. Meidinger contributed to the manuscript revision.
K. Mercier contributed to the manuscript revision, coordinated the technical operations at PANTER.
A. Meuris contributed to the test objectives definition and to the revision of the manuscript. J.Pearson contributed to the test objectives definition. N. Renault Tinacci, B. Schneider and R. Willingale contributed to the data analysis.

All authors read and approved the final manuscript.

\subsection*{Funding}
SVOM/MXT is a project funded by the French Space Agency (CNES),
with the contribution of MPE, CEA, CNRS and the University of Leicester. CNES took in charge the transportation and installation of the MXT telescope at Panter with the support of MPE. 
MPE provided the technical support to the operations. CNES, CEA, CNRS, and the University of Leicester funded the journeys of all the staff (scientific and technical) in order to achieved the data taking shifts.

The work performed at the PANTER X-ray test facility has in part been supported by the European Union’s Horizon 2020 Program under the AHEAD2020 project (grant agreement n. 871158).

\subsection*{Conflict of Interest}
The authors declare that they have nor financial nor non-financial interests in this work.

\bibliographystyle{aasjournal}
\bibliography{references}

\begin{thebibliography}{}
\expandafter\ifx\csname natexlab\endcsname\relax\def\natexlab#1{#1}\fi
\providecommand{\url}[1]{\href{#1}{#1}}
\providecommand{\dodoi}[1]{doi:~\href{http://doi.org/#1}{\nolinkurl{#1}}}
\providecommand{\doeprint}[1]{\href{http://ascl.net/#1}{\nolinkurl{http://ascl.net/#1}}}
\providecommand{\doarXiv}[1]{\href{https://arxiv.org/abs/#1}{\nolinkurl{https://arxiv.org/abs/#1}}}

\bibitem[{{Abbott} {et~al.}(2017{\natexlab{a}}){Abbott}, {Abbott}, {Abbott},
  {Acernese}, {Ackley}, {Adams}, {Adams}, {Addesso}, {Adhikari}, {Adya},
  {Affeldt}, {Afrough}, {Agarwal}, {Agathos}, {Agatsuma}, {Aggarwal}, {Aguiar},
  {Aiello}, {Ain}, {Ajith}, {Allen}, {Allen}, {Allocca}, {Aloy}, {Altin},
  {Amato}, {Ananyeva}, {Anderson}, {Anderson}, {Angelova}, {Antier}, {Appert},
  {Arai}, {Araya}, {Areeda}, {Arnaud}, {Arun}, {Ascenzi}, {Ashton}, {Ast},
  {Aston}, {Astone}, {Atallah}, {Aufmuth}, {Aulbert}, {AultONeal}, {Austin},
  {Avila-Alvarez}, {Babak}, {Bacon}, {Bader}, {Bae}, {Baker}, {Baldaccini},
  {Ballardin}, {Ballmer}, {Banagiri}, {Barayoga}, {Barclay}, {Barish},
  {Barker}, {Barkett}, {Barone}, {Barr}, {Barsotti}, {Barsuglia}, {Barta},
  {Bartlett}, {Bartos}, {Bassiri}, {Basti}, {Batch}, {Bawaj}, {Bayley},
  {Bazzan}, {B{\'e}csy}, {Beer}, {Bejger}, {Belahcene}, {Bell}, {Berger},
  {Bergmann}, {Bero}, {Berry}, {Bersanetti}, {Bertolini}, {Betzwieser},
  {Bhagwat}, {Bhandare}, {Bilenko}, {Billingsley}, {Billman}, {Birch},
  {Birney}, {Birnholtz}, {Biscans}, {Biscoveanu}, {Bisht}, {Bitossi}, {Biwer},
  {Bizouard}, {Blackburn}, {Blackman}, {Blair}, {Blair}, {Blair}, {Bloemen},
  {Bock}, {Bode}, {Boer}, {Bogaert}, {Bohe}, {Bondu}, {Bonilla}, {Bonnand},
  {Boom}, {Bork}, {Boschi}, {Bose}, {Bossie}, {Bouffanais}, {Bozzi},
  {Bradaschia}, {Brady}, {Branchesi}, {Brau}, {Briant}, {Brillet}, {Brinkmann},
  {Brisson}, {Brockill}, {Broida}, {Brooks}, {Brown}, {Brown}, {Brunett},
  {Buchanan}, {Buikema}, {Bulik}, {Bulten}, {Buonanno}, {Buskulic}, {Buy},
  {Byer}, {Cabero}, {Cadonati}, {Cagnoli}, {Cahillane}, {Calder{\'o}n
  Bustillo}, {Callister}, {Calloni}, {Camp}, {Canepa}, {Canizares}, {Cannon},
  {Cao}, {Cao}, {Capano}, {Capocasa}, {Carbognani}, {Caride}, {Carney},
  {Casanueva Diaz}, {Casentini}, {Caudill}, {Cavagli{\`a}}, {Cavalier},
  {Cavalieri}, {Cella}, {Cepeda}, {Cerd{\'a}-Dur{\'a}n}, {Cerretani},
  {Cesarini}, {Chamberlin}, {Chan}, {Chao}, {Charlton}, {Chase},
  {Chassande-Mottin}, {Chatterjee}, {Chatziioannou}, {Cheeseboro}, {Chen},
  {Chen}, {Chen}, {Cheng}, {Chia}, {Chincarini}, {Chiummo}, {Chmiel}, {Cho},
  {Cho}, {Chow}, {Christensen}, {Chu}, {Chua}, {Chua}, {Chung}, {Chung},
  {Ciani}, {Ciolfi}, {Cirelli}, {Cirone}, {Clara}, {Clark}, {Clearwater},
  {Cleva}, {Cocchieri}, {Coccia}, {Cohadon}, {Cohen}, {Colla}, {Collette},
  {Cominsky}, {Constancio}, {Conti}, {Cooper}, {Corban}, {Corbitt},
  {Cordero-Carri{\'o}n}, {Corley}, {Cornish}, {Corsi}, {Cortese}, {Costa},
  {Coughlin}, {Coughlin}, {Coulon}, {Countryman}, {Couvares}, {Covas}, {Cowan},
  {Coward}, {Cowart}, {Coyne}, {Coyne}, {Creighton}, {Creighton}, {Cripe},
  {Crowder}, {Cullen}, {Cumming}, {Cunningham}, {Cuoco}, {Dal Canton},
  {D{\'a}lya}, {Danilishin}, {D'Antonio}, {Danzmann}, {Dasgupta}, {Da Silva
  Costa}, {Dattilo}, {Dave}, {Davier}, {Davis}, {Daw}, {Day}, {De}, {DeBra},
  {Degallaix}, {De Laurentis}, {Del{\'e}glise}, {Del Pozzo}, {Demos}, {Denker},
  {Dent}, {De Pietri}, {Dergachev}, {De Rosa}, {DeRosa}, {De Rossi}, {DeSalvo},
  {de Varona}, {Devenson}, {Dhurandhar}, {D{\'\i}az}, {Di Fiore}, {Di
  Giovanni}, {Di Girolamo}, {Di Lieto}, {Di Pace}, {Di Palma}, {Di Renzo},
  {Doctor}, {Dolique}, {Donovan}, {Dooley}, {Doravari}, {Dorrington},
  {Douglas}, {Dovale {\'A}lvarez}, {Downes}, {Drago}, {Dreissigacker},
  {Driggers}, {Du}, {Ducrot}, {Dupej}, {Dwyer}, {Edo}, {Edwards}, {Effler},
  {Eggenstein}, {Ehrens}, {Eichholz}, {Eikenberry}, {Eisenstein}, {Essick},
  {Estevez}, {Etienne}, {Etzel}, {Evans}, {Evans}, {Factourovich}, {Fafone},
  {Fair}, {Fairhurst}, {Fan}, {Farinon}, {Farr}, {Farr}, {Fauchon-Jones},
  {Favata}, {Fays}, {Fee}, {Fehrmann}, {Feicht}, {Fejer}, {Fernandez-Galiana},
  {Ferrante}, {Ferreira}, {Ferrini}, {Fidecaro}, {Finstad}, {Fiori},
  {Fiorucci}, {Fishbach}, {Fisher}, {Fitz-Axen}, {Flaminio}, {Fletcher},
  {Fong}, {Font}, {Forsyth}, {Forsyth}, {Fournier}, {Frasca}, {Frasconi},
  {Frei}, {Freise}, {Frey}, {Frey}, {Fries}, {Fritschel}, {Frolov}, {Fulda},
  {Fyffe}, {Gabbard}, {Gadre}, {Gaebel}, {Gair}, {Gammaitoni}, {Ganija},
  {Gaonkar}, {Garcia-Quiros}, {Garufi}, {Gateley}, {Gaudio}, {Gaur},
  {Gayathri}, {Gehrels}, {Gemme}, {Genin}, {Gennai}, {George}, {George},
  {Gergely}, {Germain}, {Ghonge}, {Ghosh}, {Ghosh}, {Ghosh}, {Giaime},
  {Giardina}, {Giazotto}, {Gill}, {Glover}, {Goetz}, {Goetz}, {Gomes},
  {Goncharov}, {Gonz{\'a}lez}, {Gonzalez Castro}, {Gopakumar}, {Gorodetsky},
  {Gossan}, {Gosselin}, {Gouaty}, {Grado}, {Graef}, {Granata}, {Grant}, {Gras},
  {Gray}, {Greco}, {Green}, {Gretarsson}, {Groot}, {Grote}, {Grunewald},
  {Gruning}, {Guidi}, {Guo}, {Gupta}, {Gupta}, {Gushwa}, {Gustafson},
  {Gustafson}, {Halim}, {Hall}, {Hall}, {Hamilton}, {Hammond}, {Haney},
  {Hanke}, {Hanks}, {Hanna}, {Hannam}, {Hannuksela}, {Hanson}, {Hardwick},
  {Harms}, {Harry}, {Harry}, {Hart}, {Haster}, {Haughian}, {Healy}, {Heidmann},
  {Heintze}, {Heitmann}, {Hello}, {Hemming}, {Hendry}, {Heng}, {Hennig},
  {Heptonstall}, {Heurs}, {Hild}, {Hinderer}, {Hoak}, {Hofman}, {Holt}, {Holz},
  {Hopkins}, {Horst}, {Hough}, {Houston}, {Howell}, {Hreibi}, {Hu}, {Huerta},
  {Huet}, {Hughey}, {Husa}, {Huttner}, {Huynh-Dinh}, {Indik}, {Inta}, {Intini},
  {Isa}, {Isac}, {Isi}, {Iyer}, {Izumi}, {Jacqmin}, {Jani}, {Jaranowski},
  {Jawahar}, {Jim{\'e}nez-Forteza}, {Johnson}, {Johnson-McDaniel}, {Jones},
  {Jones}, {Jonker}, {Ju}, {Junker}, {Kalaghatgi}, {Kalogera}, {Kamai},
  {Kandhasamy}, {Kang}, {Kanner}, {Kapadia}, {Karki}, {Karvinen}, {Kasprzack},
  {Kastaun}, {Katolik}, {Katsavounidis}, {Katzman}, {Kaufer}, {Kawabe},
  {K{\'e}f{\'e}lian}, {Keitel}, {Kemball}, {Kennedy}, {Kent}, {Key}, {Khalili},
  {Khan}, {Khan}, {Khan}, {Khazanov}, {Kijbunchoo}, {Kim}, {Kim}, {Kim}, {Kim},
  {Kim}, {Kim}, {Kimbrell}, {King}, {King}, {Kinley-Hanlon}, {Kirchhoff},
  {Kissel}, {Kleybolte}, {Klimenko}, {Knowles}, {Koch}, {Koehlenbeck}, {Koley},
  {Kondrashov}, {Kontos}, {Korobko}, {Korth}, {Kowalska}, {Kozak},
  {Kr{\"a}mer}, {Kringel}, {Krishnan}, {Kr{\'o}lak}, {Kuehn}, {Kumar}, {Kumar},
  {Kumar}, {Kuo}, {Kutynia}, {Kwang}, {Lackey}, {Lai}, {Landry}, {Lang},
  {Lange}, {Lantz}, {Lanza}, {Lartaux-Vollard}, {Lasky}, {Laxen}, {Lazzarini},
  {Lazzaro}, {Leaci}, {Leavey}, {Lee}, {Lee}, {Lee}, {Lee}, {Lee}, {Lehmann},
  {Lenon}, {Leonardi}, {Leroy}, {Letendre}, {Levin}, {Li}, {Linker},
  {Littenberg}, {Liu}, {Lo}, {Lockerbie}, {London}, {Lord}, {Lorenzini},
  {Loriette}, {Lormand}, {Losurdo}, {Lough}, {Lousto}, {Lovelace}, {L{\"u}ck},
  {Lumaca}, {Lundgren}, {Lynch}, {Ma}, {Macas}, {Macfoy}, {Machenschalk},
  {MacInnis}, {Macleod}, {Maga{\~n}a Hernandez}, {Maga{\~n}a-Sandoval},
  {Maga{\~n}a Zertuche}, {Magee}, {Majorana}, {Maksimovic}, {Man}, {Mandic},
  {Mangano}, {Mansell}, {Manske}, {Mantovani}, {Marchesoni}, {Marion},
  {M{\'a}rka}, {M{\'a}rka}, {Markakis}, {Markosyan}, {Markowitz}, {Maros},
  {Marquina}, {Martelli}, {Martellini}, {Martin}, {Martin}, {Martynov},
  {Mason}, {Massera}, {Masserot}, {Massinger}, {Masso-Reid}, {Mastrogiovanni},
  {Matas}, {Matichard}, {Matone}, {Mavalvala}, {Mazumder}, {McCarthy},
  {McClelland}, {McCormick}, {McCuller}, {McGuire}, {McIntyre}, {McIver},
  {McManus}, {McNeill}, {McRae}, {McWilliams}, {Meacher}, {Meadors}, {Mehmet},
  {Meidam}, {Mejuto-Villa}, {Melatos}, {Mendell}, {Mercer}, {Merilh},
  {Merzougui}, {Meshkov}, {Messenger}, {Messick}, {Metzdorff}, {Meyers},
  {Miao}, {Michel}, {Middleton}, {Mikhailov}, {Milano}, {Miller}, {Miller},
  {Miller}, {Millhouse}, {Milovich-Goff}, {Minazzoli}, {Minenkov}, {Ming},
  {Mishra}, {Mitra}, {Mitrofanov}, {Mitselmakher}, {Mittleman}, {Moffa},
  {Moggi}, {Mogushi}, {Mohan}, {Mohapatra}, {Montani}, {Moore}, {Moraru},
  {Moreno}, {Morriss}, {Mours}, {Mow-Lowry}, {Mueller}, {Muir}, {Mukherjee},
  {Mukherjee}, {Mukherjee}, {Mukund}, {Mullavey}, {Munch}, {Mu{\~n}iz},
  {Muratore}, {Murray}, {Napier}, {Nardecchia}, {Naticchioni}, {Nayak},
  {Neilson}, {Nelemans}, {Nelson}, {Nery}, {Neunzert}, {Nevin}, {Newport},
  {Newton}, {Ng}, {Nguyen}, {Nichols}, {Nielsen}, {Nissanke}, {Nitz}, {Noack},
  {Nocera}, {Nolting}, {North}, {Nuttall}, {Oberling}, {O'Dea}, {Ogin}, {Oh},
  {Oh}, {Ohme}, {Okada}, {Oliver}, {Oppermann}, {Oram}, {O'Reilly}, {Ormiston},
  {Ortega}, {O'Shaughnessy}, {Ossokine}, {Ottaway}, {Overmier}, {Owen}, {Pace},
  {Page}, {Page}, {Pai}, {Pai}, {Palamos}, {Palashov}, {Palomba}, {Pal-Singh},
  {Pan}, {Pan}, {Pang}, {Pang}, {Pankow}, {Pannarale}, {Pant}, {Paoletti},
  {Paoli}, {Papa}, {Parida}, {Parker}, {Pascucci}, {Pasqualetti},
  {Passaquieti}, {Passuello}, {Patil}, {Patricelli}, {Pearlstone}, {Pedraza},
  {Pedurand}, {Pekowsky}, {Pele}, {Penn}, {Perez}, {Perreca}, {Perri},
  {Pfeiffer}, {Phelps}, {Piccinni}, {Pichot}, {Piergiovanni}, {Pierro},
  {Pillant}, {Pinard}, {Pinto}, {Pirello}, {Pitkin}, {Poe}, {Poggiani},
  {Popolizio}, {Porter}, {Post}, {Powell}, {Prasad}, {Pratt}, {Pratten},
  {Predoi}, {Prestegard}, {Prijatelj}, {Principe}, {Privitera}, {Prodi},
  {Prokhorov}, {Puncken}, {Punturo}, {Puppo}, {P{\"u}rrer}, {Qi}, {Quetschke},
  {Quintero}, {Quitzow-James}, {Raab}, {Rabeling}, {Radkins}, {Raffai}, {Raja},
  {Rajan}, {Rajbhandari}, {Rakhmanov}, {Ramirez}, {Ramos-Buades}, {Rapagnani},
  {Raymond}, {Razzano}, {Read}, {Regimbau}, {Rei}, {Reid}, {Reitze}, {Ren},
  {Reyes}, {Ricci}, {Ricker}, {Rieger}, {Riles}, {Rizzo}, {Robertson}, {Robie},
  {Robinet}, {Rocchi}, {Rolland}, {Rollins}, {Roma}, {Romano}, {Romel},
  {Romie}, {Rosi{\'n}ska}, {Ross}, {Rowan}, {R{\"u}diger}, {Ruggi}, {Rutins},
  {Ryan}, {Sachdev}, {Sadecki}, {Sadeghian}, {Sakellariadou}, {Salconi},
  {Saleem}, {Salemi}, {Samajdar}, {Sammut}, {Sampson}, {Sanchez}, {Sanchez},
  {Sanchis-Gual}, {Sandberg}, {Sanders}, {Sassolas}, {Sathyaprakash},
  {Saulson}, {Sauter}, {Savage}, {Sawadsky}, {Schale}, {Scheel}, {Scheuer},
  {Schmidt}, {Schmidt}, {Schnabel}, {Schofield}, {Sch{\"o}nbeck}, {Schreiber},
  {Schuette}, {Schulte}, {Schutz}, {Schwalbe}, {Scott}, {Scott}, {Seidel},
  {Sellers}, {Sengupta}, {Sentenac}, {Sequino}, {Sergeev}, {Shaddock},
  {Shaffer}, {Shah}, {Shahriar}, {Shaner}, {Shao}, {Shapiro}, {Shawhan},
  {Sheperd}, {Shoemaker}, {Shoemaker}, {Siellez}, {Siemens}, {Sieniawska},
  {Sigg}, {Silva}, {Singer}, {Singh}, {Singhal}, {Sintes}, {Slagmolen},
  {Smith}, {Smith}, {Smith}, {Somala}, {Son}, {Sonnenberg}, {Sorazu},
  {Sorrentino}, {Souradeep}, {Spencer}, {Srivastava}, {Staats}, {Staley},
  {Steinke}, {Steinlechner}, {Steinlechner}, {Steinmeyer}, {Stevenson},
  {Stone}, {Stops}, {Strain}, {Stratta}, {Strigin}, {Strunk}, {Sturani},
  {Stuver}, {Summerscales}, {Sun}, {Sunil}, {Suresh}, {Sutton}, {Swinkels},
  {Szczepa{\'n}czyk}, {Tacca}, {Tait}, {Talbot}, {Talukder}, {Tanner},
  {T{\'a}pai}, {Taracchini}, {Tasson}, {Taylor}, {Taylor}, {Tewari}, {Theeg},
  {Thies}, {Thomas}, {Thomas}, {Thomas}, {Thorne}, {Thorne}, {Thrane},
  {Tiwari}, {Tiwari}, {Tokmakov}, {Toland}, {Tonelli}, {Tornasi},
  {Torres-Forn{\'e}}, {Torrie}, {T{\"o}yr{\"a}}, {Travasso}, {Traylor},
  {Trinastic}, {Tringali}, {Trozzo}, {Tsang}, {Tse}, {Tso}, {Tsukada}, {Tsuna},
  {Tuyenbayev}, {Ueno}, {Ugolini}, {Unnikrishnan}, {Urban}, {Usman},
  {Vahlbruch}, {Vajente}, {Valdes}, {van Bakel}, {van Beuzekom}, {van den
  Brand}, {Van Den Broeck}, {Vander-Hyde}, {van der Schaaf}, {van Heijningen},
  {van Veggel}, {Vardaro}, {Varma}, {Vass}, {Vas{\'u}th}, {Vecchio},
  {Vedovato}, {Veitch}, {Veitch}, {Venkateswara}, {Venugopalan}, {Verkindt},
  {Vetrano}, {Vicer{\'e}}, {Viets}, {Vinciguerra}, {Vine}, {Vinet}, {Vitale},
  {Vo}, {Vocca}, {Vorvick}, {Vyatchanin}, {Wade}, {Wade}, {Wade}, {Walet},
  {Walker}, {Wallace}, {Walsh}, {Wang}, {Wang}, {Wang}, {Wang}, {Wang}, {Ward},
  {Warner}, {Was}, {Watchi}, {Weaver}, {Wei}, {Weinert}, {Weinstein}, {Weiss},
  {Wen}, {Wessel}, {We{\ss}els}, {Westerweck}, {Westphal}, {Wette}, {Whelan},
  {Whitcomb}, {Whiting}, {Whittle}, {Wilken}, {Williams}, {Williams},
  {Williamson}, {Willis}, {Willke}, {Wimmer}, {Winkler}, {Wipf}, {Wittel},
  {Woan}, {Woehler}, {Wofford}, {Wong}, {Worden}, {Wright}, {Wu}, {Wysocki},
  {Xiao}, {Yamamoto}, {Yancey}, {Yang}, {Yap}, {Yazback}, {Yu}, {Yu}, {Yvert},
  {Zadro{\.z}ny}, {Zanolin}, {Zelenova}, {Zendri}, {Zevin}, {Zhang}, {Zhang},
  {Zhang}, {Zhang}, {Zhao}, {Zhou}, {Zhou}, {Zhu}, {Zhu}, {Zimmerman},
  {Zucker}, {Zweizig}, {(LIGO Scientific Collaboration}, {Virgo Collaboration},
  {Burns}, {Veres}, {Kocevski}, {Racusin}, {Goldstein}, {Connaughton},
  {Briggs}, {Blackburn}, {Hamburg}, {Hui}, {von Kienlin}, {McEnery}, {Preece},
  {Wilson-Hodge}, {Bissaldi}, {Cleveland}, {Gibby}, {Giles}, {Kippen},
  {McBreen}, {Meegan}, {Paciesas}, {Poolakkil}, {Roberts}, {Stanbro},
  {Gamma-ray Burst Monitor}, {Savchenko}, {Ferrigno}, {Kuulkers}, {Bazzano},
  {Bozzo}, {Brandt}, {Chenevez}, {Courvoisier}, {Diehl}, {Domingo}, {Hanlon},
  {Jourdain}, {Laurent}, {Lebrun}, {Lutovinov}, {Mereghetti}, {Natalucci},
  {Rodi}, {Roques}, {Sunyaev}, {Ubertini}, \& {(INTEGRAL}}]{abbott17a}
{Abbott}, B.~P., {Abbott}, R., {Abbott}, T.~D., {et~al.} 2017{\natexlab{a}},
  \apjl, 848, L13, \dodoi{10.3847/2041-8213/aa920c}

\bibitem[{{Abbott} {et~al.}(2017{\natexlab{b}}){Abbott}, {Abbott}, {Abbott},
  {Acernese}, {Ackley}, {Adams}, {Adams}, {Addesso}, {Adhikari}, {Adya},
  {Affeldt}, {Afrough}, {Agarwal}, {Agathos}, {Agatsuma}, {Aggarwal}, {Aguiar},
  {Aiello}, {Ain}, {Ajith}, {Allen}, {Allen}, {Allocca}, {Altin}, {Amato},
  {Ananyeva}, {Anderson}, {Anderson}, {Angelova}, {Antier}, {Appert}, {Arai},
  {Araya}, {Areeda}, {Arnaud}, {Arun}, {Ascenzi}, {Ashton}, {Ast}, {Aston},
  {Astone}, {Atallah}, {Aufmuth}, {Aulbert}, {AultONeal}, {Austin},
  {Avila-Alvarez}, {Babak}, {Bacon}, {Bader}, {Bae}, {Baker}, {Baldaccini},
  {Ballardin}, {Ballmer}, {Banagiri}, {Barayoga}, {Barclay}, {Barish},
  {Barker}, {Barkett}, {Barone}, {Barr}, {Barsotti}, {Barsuglia}, {Barta},
  {Barthelmy}, {Bartlett}, {Bartos}, {Bassiri}, {Basti}, {Batch}, {Bawaj},
  {Bayley}, {Bazzan}, {B{\'e}csy}, {Beer}, {Bejger}, {Belahcene}, {Bell},
  {Berger}, {Bergmann}, {Bero}, {Berry}, {Bersanetti}, {Bertolini},
  {Betzwieser}, {Bhagwat}, {Bhandare}, {Bilenko}, {Billingsley}, {Billman},
  {Birch}, {Birney}, {Birnholtz}, {Biscans}, {Biscoveanu}, {Bisht}, {Bitossi},
  {Biwer}, {Bizouard}, {Blackburn}, {Blackman}, {Blair}, {Blair}, {Blair},
  {Bloemen}, {Bock}, {Bode}, {Boer}, {Bogaert}, {Bohe}, {Bondu}, {Bonilla},
  {Bonnand}, {Boom}, {Bork}, {Boschi}, {Bose}, {Bossie}, {Bouffanais}, {Bozzi},
  {Bradaschia}, {Brady}, {Branchesi}, {Brau}, {Briant}, {Brillet}, {Brinkmann},
  {Brisson}, {Brockill}, {Broida}, {Brooks}, {Brown}, {Brown}, {Brunett},
  {Buchanan}, {Buikema}, {Bulik}, {Bulten}, {Buonanno}, {Buskulic}, {Buy},
  {Byer}, {Cabero}, {Cadonati}, {Cagnoli}, {Cahillane}, {Calder{\'o}n
  Bustillo}, {Callister}, {Calloni}, {Camp}, {Canepa}, {Canizares}, {Cannon},
  {Cao}, {Cao}, {Capano}, {Capocasa}, {Carbognani}, {Caride}, {Carney},
  {Casanueva Diaz}, {Casentini}, {Caudill}, {Cavagli{\`a}}, {Cavalier},
  {Cavalieri}, {Cella}, {Cepeda}, {Cerd{\'a}-Dur{\'a}n}, {Cerretani},
  {Cesarini}, {Chamberlin}, {Chan}, {Chao}, {Charlton}, {Chase},
  {Chassande-Mottin}, {Chatterjee}, {Chatziioannou}, {Cheeseboro}, {Chen},
  {Chen}, {Chen}, {Cheng}, {Chia}, {Chincarini}, {Chiummo}, {Chmiel}, {Cho},
  {Cho}, {Chow}, {Christensen}, {Chu}, {Chua}, {Chua}, {Chung}, {Chung},
  {Ciani}, {Ciolfi}, {Cirelli}, {Cirone}, {Clara}, {Clark}, {Clearwater},
  {Cleva}, {Cocchieri}, {Coccia}, {Cohadon}, {Cohen}, {Colla}, {Collette},
  {Cominsky}, {Constancio}, {Conti}, {Cooper}, {Corban}, {Corbitt},
  {Cordero-Carri{\'o}n}, {Corley}, {Cornish}, {Corsi}, {Cortese}, {Costa},
  {Coughlin}, {Coughlin}, {Coulon}, {Countryman}, {Couvares}, {Covas}, {Cowan},
  {Coward}, {Cowart}, {Coyne}, {Coyne}, {Creighton}, {Creighton}, {Cripe},
  {Crowder}, {Cullen}, {Cumming}, {Cunningham}, {Cuoco}, {Dal Canton},
  {D{\'a}lya}, {Danilishin}, {D'Antonio}, {Danzmann}, {Dasgupta}, {Da Silva
  Costa}, {Dattilo}, {Dave}, {Davier}, {Davis}, {Daw}, {Day}, {De}, {DeBra},
  {Degallaix}, {De Laurentis}, {Del{\'e}glise}, {Del Pozzo}, {Demos}, {Denker},
  {Dent}, {De Pietri}, {Dergachev}, {De Rosa}, {DeRosa}, {De Rossi}, {DeSalvo},
  {de Varona}, {Devenson}, {Dhurandhar}, {D{\'\i}az}, {Di Fiore}, {Di
  Giovanni}, {Di Girolamo}, {Di Lieto}, {Di Pace}, {Di Palma}, {Di Renzo},
  {Doctor}, {Dolique}, {Donovan}, {Dooley}, {Doravari}, {Dorrington},
  {Douglas}, {Dovale {\'A}lvarez}, {Downes}, {Drago}, {Dreissigacker},
  {Driggers}, {Du}, {Ducrot}, {Dupej}, {Dwyer}, {Edo}, {Edwards}, {Effler},
  {Ehrens}, {Eichholz}, {Eikenberry}, {Eisenstein}, {Essick}, {Estevez},
  {Etienne}, {Etzel}, {Evans}, {Evans}, {Factourovich}, {Fafone}, {Fair},
  {Fairhurst}, {Fan}, {Farinon}, {Farr}, {Farr}, {Fauchon-Jones}, {Favata},
  {Fays}, {Fee}, {Fehrmann}, {Feicht}, {Fejer}, {Fernandez-Galiana},
  {Ferrante}, {Ferreira}, {Ferrini}, {Fidecaro}, {Finstad}, {Fiori},
  {Fiorucci}, {Fishbach}, {Fisher}, {Fitz-Axen}, {Flaminio}, {Fletcher},
  {Fong}, {Font}, {Forsyth}, {Forsyth}, {Fournier}, {Frasca}, {Frasconi},
  {Frei}, {Freise}, {Frey}, {Frey}, {Fries}, {Fritschel}, {Frolov}, {Fulda},
  {Fyffe}, {Gabbard}, {Gadre}, {Gaebel}, {Gair}, {Gammaitoni}, {Ganija},
  {Gaonkar}, {Garcia-Quiros}, {Garufi}, {Gateley}, {Gaudio}, {Gaur},
  {Gayathri}, {Gehrels}, {Gemme}, {Genin}, {Gennai}, {George}, {George},
  {Gergely}, {Germain}, {Ghonge}, {Ghosh}, {Ghosh}, {Ghosh}, {Giaime},
  {Giardina}, {Giazotto}, {Gill}, {Glover}, {Goetz}, {Goetz}, {Gomes},
  {Goncharov}, {Gonz{\'a}lez}, {Gonzalez Castro}, {Gopakumar}, {Gorodetsky},
  {Gossan}, {Gosselin}, {Gouaty}, {Grado}, {Graef}, {Granata}, {Grant}, {Gras},
  {Gray}, {Greco}, {Green}, {Gretarsson}, {Griswold}, {Groot}, {Grote},
  {Grunewald}, {Gruning}, {Guidi}, {Guo}, {Gupta}, {Gupta}, {Gushwa},
  {Gustafson}, {Gustafson}, {Halim}, {Hall}, {Hall}, {Hamilton}, {Hammond},
  {Haney}, {Hanke}, {Hanks}, {Hanna}, {Hannam}, {Hannuksela}, {Hanson},
  {Hardwick}, {Harms}, {Harry}, {Harry}, {Hart}, {Haster}, {Haughian}, {Healy},
  {Heidmann}, {Heintze}, {Heitmann}, {Hello}, {Hemming}, {Hendry}, {Heng},
  {Hennig}, {Heptonstall}, {Heurs}, {Hild}, {Hinderer}, {Hoak}, {Hofman},
  {Holt}, {Holz}, {Hopkins}, {Horst}, {Hough}, {Houston}, {Howell}, {Hreibi},
  {Hu}, {Huerta}, {Huet}, {Hughey}, {Husa}, {Huttner}, {Huynh-Dinh}, {Indik},
  {Inta}, {Intini}, {Isa}, {Isac}, {Isi}, {Iyer}, {Izumi}, {Jacqmin}, {Jani},
  {Jaranowski}, {Jawahar}, {Jim{\'e}nez-Forteza}, {Johnson}, {Jones}, {Jones},
  {Jonker}, {Ju}, {Junker}, {Kalaghatgi}, {Kalogera}, {Kamai}, {Kandhasamy},
  {Kang}, {Kanner}, {Kapadia}, {Karki}, {Karvinen}, {Kasprzack}, {Katolik},
  {Katsavounidis}, {Katzman}, {Kaufer}, {Kawabe}, {K{\'e}f{\'e}lian}, {Keitel},
  {Kemball}, {Kennedy}, {Kent}, {Key}, {Khalili}, {Khan}, {Khan}, {Khan},
  {Khazanov}, {Kijbunchoo}, {Kim}, {Kim}, {Kim}, {Kim}, {Kim}, {Kim},
  {Kimbrell}, {King}, {King}, {Kinley-Hanlon}, {Kirchhoff}, {Kissel},
  {Kleybolte}, {Klimenko}, {Knowles}, {Koch}, {Koehlenbeck}, {Koley},
  {Kondrashov}, {Kontos}, {Korobko}, {Korth}, {Kowalska}, {Kozak},
  {Kr{\"a}mer}, {Kringel}, {Krishnan}, {Kr{\'o}lak}, {Kuehn}, {Kumar}, {Kumar},
  {Kumar}, {Kuo}, {Kutynia}, {Kwang}, {Lackey}, {Lai}, {Landry}, {Lang},
  {Lange}, {Lantz}, {Lanza}, {Larson}, {Lartaux-Vollard}, {Lasky}, {Laxen},
  {Lazzarini}, {Lazzaro}, {Leaci}, {Leavey}, {Lee}, {Lee}, {Lee}, {Lee}, {Lee},
  {Lehmann}, {Lenon}, {Leonardi}, {Leroy}, {Letendre}, {Levin}, {Li}, {Linker},
  {Littenberg}, {Liu}, {Lo}, {Lockerbie}, {London}, {Lord}, {Lorenzini},
  {Loriette}, {Lormand}, {Losurdo}, {Lough}, {Lousto}, {Lovelace}, {L{\"u}ck},
  {Lumaca}, {Lundgren}, {Lynch}, {Ma}, {Macas}, {Macfoy}, {Machenschalk},
  {MacInnis}, {Macleod}, {Maga{\~n}a Hernandez}, {Maga{\~n}a-Sandoval},
  {Maga{\~n}a Zertuche}, {Magee}, {Majorana}, {Maksimovic}, {Man}, {Mandic},
  {Mangano}, {Mansell}, {Manske}, {Mantovani}, {Marchesoni}, {Marion},
  {M{\'a}rka}, {M{\'a}rka}, {Markakis}, {Markosyan}, {Markowitz}, {Maros},
  {Marquina}, {Marsh}, {Martelli}, {Martellini}, {Martin}, {Martin},
  {Martynov}, {Mason}, {Massera}, {Masserot}, {Massinger}, {Masso-Reid},
  {Mastrogiovanni}, {Matas}, {Matichard}, {Matone}, {Mavalvala}, {Mazumder},
  {McCarthy}, {McClelland}, {McCormick}, {McCuller}, {McGuire}, {McIntyre},
  {McIver}, {McManus}, {McNeill}, {McRae}, {McWilliams}, {Meacher}, {Meadors},
  {Mehmet}, {Meidam}, {Mejuto-Villa}, {Melatos}, {Mendell}, {Mercer}, {Merilh},
  {Merzougui}, {Meshkov}, {Messenger}, {Messick}, {Metzdorff}, {Meyers},
  {Miao}, {Michel}, {Middleton}, {Mikhailov}, {Milano}, {Miller}, {Miller},
  {Miller}, {Millhouse}, {Milovich-Goff}, {Minazzoli}, {Minenkov}, {Ming},
  {Mishra}, {Mitra}, {Mitrofanov}, {Mitselmakher}, {Mittleman}, {Moffa},
  {Moggi}, {Mogushi}, {Mohan}, {Mohapatra}, {Montani}, {Moore}, {Moraru},
  {Moreno}, {Morriss}, {Mours}, {Mow-Lowry}, {Mueller}, {Muir}, {Mukherjee},
  {Mukherjee}, {Mukherjee}, {Mukund}, {Mullavey}, {Munch}, {Mu{\~n}iz},
  {Muratore}, {Murray}, {Napier}, {Nardecchia}, {Naticchioni}, {Nayak},
  {Neilson}, {Nelemans}, {Nelson}, {Nery}, {Neunzert}, {Nevin}, {Newport},
  {Newton}, {Ng}, {Nguyen}, {Nguyen}, {Nichols}, {Nielsen}, {Nissanke}, {Nitz},
  {Noack}, {Nocera}, {Nolting}, {North}, {Nuttall}, {Oberling}, {O'Dea},
  {Ogin}, {Oh}, {Oh}, {Ohme}, {Okada}, {Oliver}, {Oppermann}, {Oram},
  {O'Reilly}, {Ormiston}, {Ortega}, {O'Shaughnessy}, {Ossokine}, {Ottaway},
  {Overmier}, {Owen}, {Pace}, {Page}, {Page}, {Pai}, {Pai}, {Palamos},
  {Palashov}, {Palomba}, {Pal-Singh}, {Pan}, {Pan}, {Pang}, {Pang}, {Pankow},
  {Pannarale}, {Pant}, {Paoletti}, {Paoli}, {Papa}, {Parida}, {Parker},
  {Pascucci}, {Pasqualetti}, {Passaquieti}, {Passuello}, {Patil}, {Patricelli},
  {Pearlstone}, {Pedraza}, {Pedurand}, {Pekowsky}, {Pele}, {Penn}, {Perez},
  {Perreca}, {Perri}, {Pfeiffer}, {Phelps}, {Piccinni}, {Pichot},
  {Piergiovanni}, {Pierro}, {Pillant}, {Pinard}, {Pinto}, {Pirello}, {Pitkin},
  {Poe}, {Poggiani}, {Popolizio}, {Porter}, {Post}, {Powell}, {Prasad},
  {Pratt}, {Pratten}, {Predoi}, {Prestegard}, {Price}, {Prijatelj}, {Principe},
  {Privitera}, {Prodi}, {Prokhorov}, {Puncken}, {Punturo}, {Puppo},
  {P{\"u}rrer}, {Qi}, {Quetschke}, {Quintero}, {Quitzow-James}, {Raab},
  {Rabeling}, {Radkins}, {Raffai}, {Raja}, {Rajan}, {Rajbhandari}, {Rakhmanov},
  {Ramirez}, {Ramos-Buades}, {Rapagnani}, {Raymond}, {Razzano}, {Read},
  {Regimbau}, {Rei}, {Reid}, {Reitze}, {Ren}, {Reyes}, {Ricci}, {Ricker},
  {Rieger}, {Riles}, {Rizzo}, {Robertson}, {Robie}, {Robinet}, {Rocchi},
  {Rolland}, {Rollins}, {Roma}, {Romano}, {Romel}, {Romie}, {Rosi{\'n}ska},
  {Ross}, {Rowan}, {R{\"u}diger}, {Ruggi}, {Rutins}, {Ryan}, {Sachdev},
  {Sadecki}, {Sadeghian}, {Sakellariadou}, {Salconi}, {Saleem}, {Salemi},
  {Samajdar}, {Sammut}, {Sampson}, {Sanchez}, {Sanchez}, {Sanchis-Gual},
  {Sandberg}, {Sanders}, {Sassolas}, {Sathyaprakash}, {Saulson}, {Sauter},
  {Savage}, {Sawadsky}, {Schale}, {Scheel}, {Scheuer}, {Schmidt}, {Schmidt},
  {Schnabel}, {Schofield}, {Sch{\"o}nbeck}, {Schreiber}, {Schuette}, {Schulte},
  {Schutz}, {Schwalbe}, {Scott}, {Scott}, {Seidel}, {Sellers}, {Sengupta},
  {Sentenac}, {Sequino}, {Sergeev}, {Shaddock}, {Shaffer}, {Shah}, {Shahriar},
  {Shaner}, {Shao}, {Shapiro}, {Shawhan}, {Sheperd}, {Shoemaker}, {Shoemaker},
  {Siellez}, {Siemens}, {Sieniawska}, {Sigg}, {Silva}, {Singer}, {Singh},
  {Singhal}, {Sintes}, {Slagmolen}, {Smith}, {Smith}, {Smith}, {Somala}, {Son},
  {Sonnenberg}, {Sorazu}, {Sorrentino}, {Souradeep}, {Spencer}, {Srivastava},
  {Staats}, {Staley}, {Steinke}, {Steinlechner}, {Steinlechner}, {Steinmeyer},
  {Stevenson}, {Stone}, {Stops}, {Strain}, {Stratta}, {Strigin}, {Strunk},
  {Sturani}, {Stuver}, {Summerscales}, {Sun}, {Sunil}, {Suresh}, {Sutton},
  {Swinkels}, {Szczepa{\'n}czyk}, {Tacca}, {Tait}, {Talbot}, {Talukder},
  {Tanner}, {T{\'a}pai}, {Taracchini}, {Tasson}, {Taylor}, {Taylor}, {Tewari},
  {Theeg}, {Thies}, {Thomas}, {Thomas}, {Thomas}, {Thorne}, {Thorne}, {Thrane},
  {Tiwari}, {Tiwari}, {Tokmakov}, {Toland}, {Tonelli}, {Tornasi},
  {Torres-Forn{\'e}}, {Torrie}, {T{\"o}yr{\"a}}, {Travasso}, {Traylor},
  {Trinastic}, {Tringali}, {Trozzo}, {Tsang}, {Tse}, {Tso}, {Tsukada}, {Tsuna},
  {Tuyenbayev}, {Ueno}, {Ugolini}, {Unnikrishnan}, {Urban}, {Usman},
  {Vahlbruch}, {Vajente}, {Valdes}, {van Bakel}, {van Beuzekom}, {van den
  Brand}, {Van Den Broeck}, {Vander-Hyde}, {van der Schaaf}, {van Heijningen},
  {van Veggel}, {Vardaro}, {Varma}, {Vass}, {Vas{\'u}th}, {Vecchio},
  {Vedovato}, {Veitch}, {Veitch}, {Venkateswara}, {Venugopalan}, {Verkindt},
  {Vetrano}, {Vicer{\'e}}, {Viets}, {Vinciguerra}, {Vine}, {Vinet}, {Vitale},
  {Vo}, {Vocca}, {Vorvick}, {Vyatchanin}, {Wade}, {Wade}, {Wade}, {Walet},
  {Walker}, {Wallace}, {Walsh}, {Wang}, {Wang}, {Wang}, {Wang}, {Wang}, {Ward},
  {Warner}, {Was}, {Watchi}, {Weaver}, {Wei}, {Weinert}, {Weinstein}, {Weiss},
  {Wen}, {Wessel}, {Wessels}, {Westerweck}, {Westphal}, {Wette}, {Whelan},
  {Whitcomb}, {Whiting}, {Whittle}, {Wilken}, {Williams}, {Williams},
  {Williamson}, {Willis}, {Willke}, {Wimmer}, {Winkler}, {Wipf}, {Wittel},
  {Woan}, {Woehler}, {Wofford}, {Wong}, {Worden}, {Wright}, {Wu}, {Wysocki},
  {Xiao}, {Yamamoto}, {Yancey}, {Yang}, {Yap}, {Yazback}, {Yu}, {Yu}, {Yvert},
  {Zadro{\.z}ny}, {Zanolin}, {Zelenova}, {Zendri}, {Zevin}, {Zhang}, {Zhang},
  {Zhang}, {Zhang}, {Zhao}, {Zhou}, {Zhou}, {Zhu}, {Zhu}, {Zimmerman},
  {Zucker}, {Zweizig}, {LIGO Scientific Collaboration}, {Virgo Collaboration},
  {Wilson-Hodge}, {Bissaldi}, {Blackburn}, {Briggs}, {Burns}, {Cleveland},
  {Connaughton}, {Gibby}, {Giles}, {Goldstein}, {Hamburg}, {Jenke}, {Hui},
  {Kippen}, {Kocevski}, {McBreen}, {Meegan}, {Paciesas}, {Poolakkil}, {Preece},
  {Racusin}, {Roberts}, {Stanbro}, {Veres}, {von Kienlin}, {GBM}, {Savchenko},
  {Ferrigno}, {Kuulkers}, {Bazzano}, {Bozzo}, {Brandt}, {Chenevez},
  {Courvoisier}, {Diehl}, {Domingo}, {Hanlon}, {Jourdain}, {Laurent}, {Lebrun},
  {Lutovinov}, {Martin-Carrillo}, {Mereghetti}, {Natalucci}, {Rodi}, {Roques},
  {Sunyaev}, {Ubertini}, {INTEGRAL}, {Aartsen}, {Ackermann}, {Adams},
  {Aguilar}, {Ahlers}, {Ahrens}, {Samarai}, {Altmann}, {Andeen}, {Anderson},
  {Ansseau}, {Anton}, {Arg{\"u}elles}, {Auffenberg}, {Axani}, {Bagherpour},
  {Bai}, {Barron}, {Barwick}, {Baum}, {Bay}, {Beatty}, {Becker Tjus},
  {Bernardini}, {Besson}, {Binder}, {Bindig}, {Blaufuss}, {Blot}, {Bohm},
  {B{\"o}rner}, {Bos}, {Bose}, {B{\"o}ser}, {Botner}, {Bourbeau}, {Bourbeau},
  {Bradascio}, {Braun}, {Brayeur}, {Brenzke}, {Bretz}, {Bron},
  {Brostean-Kaiser}, {Burgman}, {Carver}, {Casey}, {Casier}, {Cheung},
  {Chirkin}, {Christov}, {Clark}, {Classen}, {Coenders}, {Collin}, {Conrad},
  {Cowen}, {Cross}, {Day}, {de Andr{\'e}}, {De Clercq}, {DeLaunay},
  {Dembinski}, {De Ridder}, {Desiati}, {de Vries}, {de Wasseige}, {de With},
  {DeYoung}, {D{\'\i}az-V{\'e}lez}, {di Lorenzo}, {Dujmovic}, {Dumm},
  {Dunkman}, {Dvorak}, {Eberhardt}, {Ehrhardt}, {Eichmann}, {Eller}, {Evenson},
  {Fahey}, {Fazely}, {Felde}, {Filimonov}, {Finley}, {Flis}, {Franckowiak},
  {Friedman}, {Fuchs}, {Gaisser}, {Gallagher}, {Gerhardt}, {Ghorbani}, {Giang},
  {Glauch}, {Gl{\"u}senkamp}, {Goldschmidt}, {Gonzalez}, {Grant}, {Griffith},
  {Haack}, {Hallgren}, {Halzen}, {Hanson}, {Hebecker}, {Heereman}, {Helbing},
  {Hellauer}, {Hickford}, {Hignight}, {Hill}, {Hoffman}, {Hoffmann},
  {Hokanson-Fasig}, {Hoshina}, {Huang}, {Huber}, {Hultqvist}, {H{\"u}nnefeld},
  {In}, {Ishihara}, {Jacobi}, {Japaridze}, {Jeong}, {Jero}, {Jones},
  {Kalaczynski}, {Kang}, {Kappes}, {Karg}, {Karle}, {Kauer}, {Keivani},
  {Kelley}, {Kheirandish}, {Kim}, {Kim}, {Kintscher}, {Kiryluk}, {Kittler},
  {Klein}, {Kohnen}, {Koirala}, {Kolanoski}, {K{\"o}pke}, {Kopper}, {Kopper},
  {Koschinsky}, {Koskinen}, {Kowalski}, {Krings}, {Kroll}, {Kr{\"u}ckl},
  {Kunnen}, {Kunwar}, {Kurahashi}, {Kuwabara}, {Kyriacou}, {Labare},
  {Lanfranchi}, {Larson}, {Lauber}, {Lesiak-Bzdak}, {Leuermann}, {Liu}, {Lu},
  {L{\"u}nemann}, {Luszczak}, {Madsen}, {Maggi}, {Mahn}, {Mancina}, {Maruyama},
  {Mase}, {Maunu}, {McNally}, {Meagher}, {Medici}, {Meier}, {Menne}, {Merino},
  {Meures}, {Miarecki}, {Micallef}, {Moment{\'e}}, {Montaruli}, {Moore},
  {Moulai}, {Nahnhauer}, {Nakarmi}, {Naumann}, {Neer}, {Niederhausen},
  {Nowicki}, {Nygren}, {Obertacke Pollmann}, {Olivas}, {O'Murchadha},
  {Palczewski}, {Pandya}, {Pankova}, {Peiffer}, {Pepper}, {P{\'e}rez de los
  Heros}, {Pieloth}, {Pinat}, {Price}, {Przybylski}, {Raab}, {R{\"a}del},
  {Rameez}, {Rawlins}, {Rea}, {Reimann}, {Relethford}, {Relich}, {Resconi},
  {Rhode}, {Richman}, {Robertson}, {Rongen}, {Rott}, {Ruhe}, {Ryckbosch},
  {Rysewyk}, {S{\"a}lzer}, {Sanchez Herrera}, {Sandrock}, {Sandroos},
  {Santander}, {Sarkar}, {Sarkar}, {Satalecka}, {Schlunder}, {Schmidt},
  {Schneider}, {Schoenen}, {Sch{\"o}neberg}, {Schumacher}, {Seckel},
  {Seunarine}, {Soedingrekso}, {Soldin}, {Song}, {Spiczak}, {Spiering},
  {Stachurska}, {Stamatikos}, {Stanev}, {Stasik}, {Stettner}, {Steuer},
  {Stezelberger}, {Stokstad}, {St{\"o}ssl}, {Strotjohann}, {Stuttard},
  {Sullivan}, {Sutherland}, {Taboada}, {Tatar}, {Tenholt}, {Ter-Antonyan},
  {Terliuk}, {Te{\v{s}}i{\'c}}, {Tilav}, {Toale}, {Tobin}, {Toscano}, {Tosi},
  {Tselengidou}, {Tung}, {Turcati}, {Turley}, {Ty}, {Unger}, {Usner},
  {Vandenbroucke}, {Van Driessche}, {van Eijndhoven}, {Vanheule}, {van Santen},
  {Vehring}, {Vogel}, {Vraeghe}, {Walck}, {Wallace}, {Wallraff}, {Wandler},
  {Wandkowsky}, {Waza}, {Weaver}, {Weiss}, {Wendt}, {Werthebach}, {Whelan},
  {Wiebe}, {Wiebusch}, {Wille}, {Williams}, {Wills}, {Wolf}, {Wood}, {Woolsey},
  {Woschnagg}, {Xu}, {Xu}, {Xu}, {Yanez}, {Yodh}, {Yoshida}, {Yuan}, {Zoll},
  {IceCube Collaboration}, {Balasubramanian}, {Mate}, {Bhalerao},
  {Bhattacharya}, {Vibhute}, {Dewangan}, {Rao}, {Vadawale}, {AstroSat Cadmium
  Zinc Telluride Imager Team}, {Svinkin}, {Hurley}, {Aptekar}, {Frederiks},
  {Golenetskii}, {Kozlova}, {Lysenko}, {Oleynik}, {Tsvetkova}, {Ulanov},
  {Cline}, {IPN Collaboration}, {Li}, {Xiong}, {Zhang}, {Lu}, {Song}, {Cao},
  {Chang}, {Chen}, {Chen}, {Chen}, {Chen}, {Chen}, {Chen}, {Cui}, {Cui},
  {Deng}, {Dong}, {Du}, {Fu}, {Gao}, {Gao}, {Gao}, {Ge}, {Gu}, {Guan}, {Guo},
  {Han}, {Hu}, {Huang}, {Huo}, {Jia}, {Jiang}, {Jiang}, {Jin}, {Jin}, {Li},
  {Li}, {Li}, {Li}, {Li}, {Li}, {Li}, {Li}, {Li}, {Li}, {Li}, {Liang}, {Liao},
  {Liu}, {Liu}, {Liu}, {Liu}, {Liu}, {Liu}, {Liu}, {Lu}, {Lu}, {Luo}, {Ma},
  {Meng}, {Nang}, {Nie}, {Ou}, {Qu}, {Sai}, {Sun}, {Tan}, {Tao}, {Tao}, {Tuo},
  {Wang}, {Wang}, {Wang}, {Wang}, {Wang}, {Wen}, {Wu}, {Wu}, {Xiao}, {Xu},
  {Xu}, {Yan}, {Yang}, {Yang}, {Yang}, {Zhang}, {Zhang}, {Zhang}, {Zhang},
  {Zhang}, {Zhang}, {Zhang}, {Zhang}, {Zhang}, {Zhang}, {Zhang}, {Zhang},
  {Zhang}, {Zhang}, {Zhang}, {Zhang}, {Zhang}, {Zhang}, {Zhao}, {Zhao}, {Zhao},
  {Zheng}, {Zhu}, {Zhu}, {Zou}, {Insight-HXMT Collaboration}, {Albert},
  {Andr{\'e}}, {Anghinolfi}, {Ardid}, {Aubert}, {Aublin}, {Avgitas}, {Baret},
  {Barrios-Mart{\'\i}}, {Basa}, {Belhorma}, {Bertin}, {Biagi}, {Bormuth},
  {Bourret}, {Bouwhuis}, {Br{\^a}nza{\c{s}}}, {Bruijn}, {Brunner}, {Busto},
  {Capone}, {Caramete}, {Carr}, {Celli}, {Cherkaoui El Moursli}, {Chiarusi},
  {Circella}, {Coelho}, {Coleiro}, {Coniglione}, {Costantini}, {Coyle},
  {Creusot}, {D{\'\i}az}, {Deschamps}, {De Bonis}, {Distefano}, {Di Palma},
  {Domi}, {Donzaud}, {Dornic}, {Drouhin}, {Eberl}, {El Bojaddaini}, {El
  Khayati}, {Els{\"a}sser}, {Enzenh{\"o}fer}, {Ettahiri}, {Fassi}, {Felis},
  {Fusco}, {Gay}, {Giordano}, {Glotin}, {Gr{\'e}goire}, {Ruiz}, {Graf},
  {Hallmann}, {van Haren}, {Heijboer}, {Hello}, {Hern{\'a}ndez-Rey},
  {H{\"o}ssl}, {Hofest{\"a}dt}, {Hugon}, {Illuminati}, {James}, {de Jong},
  {Jongen}, {Kadler}, {Kalekin}, {Katz}, {Kiessling}, {Kouchner}, {Kreter},
  {Kreykenbohm}, {Kulikovskiy}, {Lachaud}, {Lahmann}, {Lef{\`e}vre}, {Leonora},
  {Lotze}, {Loucatos}, {Marcelin}, {Margiotta}, {Marinelli},
  {Mart{\'\i}nez-Mora}, {Mele}, {Melis}, {Michael}, {Migliozzi}, {Moussa},
  {Navas}, {Nezri}, {Organokov}, {P{\u{a}}v{\u{a}}la{\c{s}}}, {Pellegrino},
  {Perrina}, {Piattelli}, {Popa}, {Pradier}, {Quinn}, {Racca}, {Riccobene},
  {S{\'a}nchez-Losa}, {Salda{\~n}a}, {Salvadori}, {Samtleben}, {Sanguineti},
  {Sapienza}, {Sieger}, {Spurio}, {Stolarczyk}, {Taiuti}, {Tayalati},
  {Trovato}, {Turpin}, {T{\"o}nnis}, {Vallage}, {Van Elewyck}, {Versari},
  {Vivolo}, {Vizzoca}, {Wilms}, {Zornoza}, {Z{\'u}{\~n}iga}, {ANTARES
  Collaboration}, {Beardmore}, {Breeveld}, {Burrows}, {Cenko}, {Cusumano},
  {D'A{\`\i}}, {de Pasquale}, {Emery}, {Evans}, {Giommi}, {Gronwall}, {Kennea},
  {Krimm}, {Kuin}, {Lien}, {Marshall}, {Melandri}, {Nousek}, {Oates},
  {Osborne}, {Pagani}, {Page}, {Palmer}, {Perri}, {Siegel}, {Sbarufatti},
  {Tagliaferri}, {Tohuvavohu}, {Swift Collaboration}, {Tavani}, {Verrecchia},
  {Bulgarelli}, {Evangelista}, {Pacciani}, {Feroci}, {Pittori}, {Giuliani},
  {Del Monte}, {Donnarumma}, {Argan}, {Trois}, {Ursi}, {Cardillo}, {Piano},
  {Longo}, {Lucarelli}, {Munar-Adrover}, {Fuschino}, {Labanti}, {Marisaldi},
  {Minervini}, {Fioretti}, {Parmiggiani}, {Gianotti}, {Trifoglio}, {Di Persio},
  {Antonelli}, {Barbiellini}, {Caraveo}, {Cattaneo}, {Costa}, {Colafrancesco},
  {D'Amico}, {Ferrari}, {Morselli}, {Paoletti}, {Picozza}, {Pilia}, {Rappoldi},
  {Soffitta}, {Vercellone}, {AGILE Team}, {Foley}, {Coulter}, {Kilpatrick},
  {Drout}, {Piro}, {Shappee}, {Siebert}, {Simon}, {Ulloa}, {Kasen}, {Madore},
  {Murguia-Berthier}, {Pan}, {Prochaska}, {Ramirez-Ruiz}, {Rest},
  {Rojas-Bravo}, {1M2H Team}, {Berger}, {Soares-Santos}, {Annis}, {Alexander},
  {Allam}, {Balbinot}, {Blanchard}, {Brout}, {Butler}, {Chornock}, {Cook},
  {Cowperthwaite}, {Diehl}, {Drlica-Wagner}, {Drout}, {Durret}, {Eftekhari},
  {Finley}, {Fong}, {Frieman}, {Fryer}, {Garc{\'\i}a-Bellido}, {Gruendl},
  {Hartley}, {Herner}, {Kessler}, {Lin}, {Lopes}, {Louren{\c{c}}o}, {Margutti},
  {Marshall}, {Matheson}, {Medina}, {Metzger}, {Mu{\~n}oz}, {Muir}, {Nicholl},
  {Nugent}, {Palmese}, {Paz-Chinch{\'o}n}, {Quataert}, {Sako}, {Sauseda},
  {Schlegel}, {Scolnic}, {Secco}, {Smith}, {Sobreira}, {Villar}, {Vivas},
  {Wester}, {Williams}, {Yanny}, {Zenteno}, {Zhang}, {Abbott}, {Banerji},
  {Bechtol}, {Benoit-L{\'e}vy}, {Bertin}, {Brooks}, {Buckley-Geer}, {Burke},
  {Capozzi}, {Carnero Rosell}, {Carrasco Kind}, {Castander}, {Crocce}, {Cunha},
  {D'Andrea}, {da Costa}, {Davis}, {DePoy}, {Desai}, {Dietrich}, {Eifler},
  {Fernandez}, {Flaugher}, {Fosalba}, {Gaztanaga}, {Gerdes}, {Giannantonio},
  {Goldstein}, {Gruen}, {Gschwend}, {Gutierrez}, {Honscheid}, {James},
  {Jeltema}, {Johnson}, {Johnson}, {Kent}, {Krause}, {Kron}, {Kuehn}, {Lahav},
  {Lima}, {Maia}, {March}, {Martini}, {McMahon}, {Menanteau}, {Miller},
  {Miquel}, {Mohr}, {Nichol}, {Ogando}, {Plazas}, {Romer}, {Roodman}, {Rykoff},
  {Sanchez}, {Scarpine}, {Schindler}, {Schubnell}, {Sevilla-Noarbe}, {Sheldon},
  {Smith}, {Smith}, {Stebbins}, {Suchyta}, {Swanson}, {Tarle}, {Thomas},
  {Troxel}, {Tucker}, {Vikram}, {Walker}, {Wechsler}, {Weller}, {Carlin},
  {Gill}, {Li}, {Marriner}, {Neilsen}, {Dark Energy Camera GW-EM
  Collaboration}, {DES Collaboration}, {Haislip}, {Kouprianov}, {Reichart},
  {Sand}, {Tartaglia}, {Valenti}, {Yang}, {DLT40 Collaboration}, {Benetti},
  {Brocato}, {Campana}, {Cappellaro}, {Covino}, {D'Avanzo}, {D'Elia}, {Getman},
  {Ghirlanda}, {Ghisellini}, {Limatola}, {Nicastro}, {Palazzi}, {Pian},
  {Piranomonte}, {Possenti}, {Rossi}, {Salafia}, {Tomasella}, {Amati},
  {Antonelli}, {Bernardini}, {Bufano}, {Capaccioli}, {Casella}, {Dadina}, {De
  Cesare}, {Di Paola}, {Giuffrida}, {Giunta}, {Israel}, {Lisi}, {Maiorano},
  {Mapelli}, {Masetti}, {Pescalli}, {Pulone}, {Salvaterra}, {Schipani},
  {Spera}, {Stamerra}, {Stella}, {Testa}, {Turatto}, {Vergani}, {Aresu},
  {Bachetti}, {Buffa}, {Burgay}, {Buttu}, {Caria}, {Carretti}, {Casasola},
  {Castangia}, {Carboni}, {Casu}, {Concu}, {Corongiu}, {Deiana}, {Egron},
  {Fara}, {Gaudiomonte}, {Gusai}, {Ladu}, {Loru}, {Leurini}, {Marongiu},
  {Melis}, {Melis}, {Migoni}, {Milia}, {Navarrini}, {Orlati}, {Ortu}, {Palmas},
  {Pellizzoni}, {Perrodin}, {Pisanu}, {Poppi}, {Righini}, {Saba}, {Serra},
  {Serrau}, {Stagni}, {Surcis}, {Vacca}, {Vargiu}, {Hunt}, {Jin}, {Klose},
  {Kouveliotou}, {Mazzali}, {M{\o}ller}, {Nava}, {Piran}, {Selsing}, {Vergani},
  {Wiersema}, {Toma}, {Higgins}, {Mundell}, {di Serego Alighieri}, {G{\'o}tz},
  {Gao}, {Gomboc}, {Kaper}, {Kobayashi}, {Kopac}, {Mao}, {Starling}, {Steele},
  {van der Horst}, {GRAWITA: GRAvitational Wave Inaf TeAm}, {Acero}, {Atwood},
  {Baldini}, {Barbiellini}, {Bastieri}, {Berenji}, {Bellazzini}, {Bissaldi},
  {Blandford}, {Bloom}, {Bonino}, {Bottacini}, {Bregeon}, {Buehler}, {Buson},
  {Cameron}, {Caputo}, {Caraveo}, {Cavazzuti}, {Chekhtman}, {Cheung}, {Chiang},
  {Ciprini}, {Cohen-Tanugi}, {Cominsky}, {Costantin}, {Cuoco}, {D'Ammando}, {de
  Palma}, {Digel}, {Di Lalla}, {Di Mauro}, {Di Venere}, {Dubois}, {Fegan},
  {Focke}, {Franckowiak}, {Fukazawa}, {Funk}, {Fusco}, {Gargano}, {Gasparrini},
  {Giglietto}, {Giordano}, {Giroletti}, {Glanzman}, {Green}, {Grondin},
  {Guillemot}, {Guiriec}, {Harding}, {Horan}, {J{\'o}hannesson}, {Kamae},
  {Kensei}, {Kuss}, {La Mura}, {Latronico}, {Lemoine-Goumard}, {Longo},
  {Loparco}, {Lovellette}, {Lubrano}, {Magill}, {Maldera}, {Manfreda},
  {Mazziotta}, {McEnery}, {Meyer}, {Michelson}, {Mirabal}, {Monzani},
  {Moretti}, {Morselli}, {Moskalenko}, {Negro}, {Nuss}, {Ojha}, {Omodei},
  {Orienti}, {Orlando}, {Palatiello}, {Paliya}, {Paneque}, {Pesce-Rollins},
  {Piron}, {Porter}, {Principe}, {Rain{\`o}}, {Rando}, {Razzano}, {Razzaque},
  {Reimer}, {Reimer}, {Reposeur}, {Rochester}, {Saz Parkinson}, {Sgr{\`o}},
  {Siskind}, {Spada}, {Spandre}, {Suson}, {Takahashi}, {Tanaka}, {Thayer},
  {Thayer}, {Thompson}, {Tibaldo}, {Torres}, {Torresi}, {Troja}, {Venters},
  {Vianello}, {Zaharijas}, {Fermi Large Area Telescope Collaboration},
  {Allison}, {Bannister}, {Dobie}, {Kaplan}, {Lenc}, {Lynch}, {Murphy},
  {Sadler}, {Australia Telescope Compact Array}, {Hotan}, {James}, {Oslowski},
  {Raja}, {Shannon}, {Whiting}, {Australian SKA Pathfinder}, {Arcavi},
  {Howell}, {McCully}, {Hosseinzadeh}, {Hiramatsu}, {Poznanski}, {Barnes},
  {Zaltzman}, {Vasylyev}, {Maoz}, {Las Cumbres Observatory Group}, {Cooke},
  {Bailes}, {Wolf}, {Deller}, {Lidman}, {Wang}, {Gendre}, {Andreoni}, {Ackley},
  {Pritchard}, {Bessell}, {Chang}, {M{\"o}ller}, {Onken}, {Scalzo},
  {Ridden-Harper}, {Sharp}, {Tucker}, {Farrell}, {Elmer}, {Johnston},
  {Venkatraman Krishnan}, {Keane}, {Green}, {Jameson}, {Hu}, {Ma}, {Sun}, {Wu},
  {Wang}, {Shang}, {Hu}, {Ashley}, {Yuan}, {Li}, {Tao}, {Zhu}, {Zhang},
  {Suntzeff}, {Zhou}, {Yang}, {Orange}, {Morris}, {Cucchiara}, {Giblin},
  {Klotz}, {Staff}, {Thierry}, {Schmidt}, {OzGrav}, {(Deeper}, {Wider},
  {program}, {AST3}, {CAASTRO Collaborations}, {Tanvir}, {Levan}, {Cano}, {de
  Ugarte-Postigo}, {Gonz{\'a}lez-Fern{\'a}ndez}, {Greiner}, {Hjorth}, {Irwin},
  {Kr{\"u}hler}, {Mandel}, {Milvang-Jensen}, {O'Brien}, {Rol}, {Rosetti},
  {Rosswog}, {Rowlinson}, {Steeghs}, {Th{\"o}ne}, {Ulaczyk}, {Watson}, {Bruun},
  {Cutter}, {Figuera Jaimes}, {Fujii}, {Fruchter}, {Gompertz}, {Jakobsson},
  {Hodosan}, {J{\`e}rgensen}, {Kangas}, {Kann}, {Rabus}, {Schr{\o}der},
  {Stanway}, {Wijers}, {VINROUGE Collaboration}, {Lipunov}, {Gorbovskoy},
  {Kornilov}, {Tyurina}, {Balanutsa}, {Kuznetsov}, {Vlasenko}, {Podesta},
  {Lopez}, {Podesta}, {Levato}, {Saffe}, {Mallamaci}, {Budnev}, {Gress},
  {Kuvshinov}, {Gorbunov}, {Vladimirov}, {Zimnukhov}, {Gabovich}, {Yurkov},
  {Sergienko}, {Rebolo}, {Serra-Ricart}, {Tlatov}, {Ishmuhametova}, {MASTER
  Collaboration}, {Abe}, {Aoki}, {Aoki}, {Asakura}, {Baar}, {Barway}, {Bond},
  {Doi}, {Finet}, {Fujiyoshi}, {Furusawa}, {Honda}, {Itoh}, {Kanda},
  {Kawabata}, {Kawabata}, {Kim}, {Koshida}, {Kuroda}, {Lee}, {Liu},
  {Matsubayashi}, {Miyazaki}, {Morihana}, {Morokuma}, {Motohara}, {Murata},
  {Nagai}, {Nagashima}, {Nagayama}, {Nakaoka}, {Nakata}, {Ohsawa}, {Ohshima},
  {Ohta}, {Okita}, {Saito}, {Saito}, {Sako}, {Sekiguchi}, {Sumi}, {Tajitsu},
  {Takahashi}, {Takayama}, {Tamura}, {Tanaka}, {Tanaka}, {Terai}, {Tominaga},
  {Tristram}, {Uemura}, {Utsumi}, {Yamaguchi}, {Yasuda}, {Yoshida}, {Zenko},
  {J-GEM}, {Adams}, {Anupama}, {Bally}, {Barway}, {Bellm}, {Blagorodnova},
  {Cannella}, {Chandra}, {Chatterjee}, {Clarke}, {Cobb}, {Cook}, {Copperwheat},
  {De}, {Emery}, {Feindt}, {Foster}, {Fox}, {Frail}, {Fremling}, {Frohmaier},
  {Garcia}, {Ghosh}, {Giacintucci}, {Goobar}, {Gottlieb}, {Grefenstette},
  {Hallinan}, {Harrison}, {Heida}, {Helou}, {Ho}, {Horesh}, {Hotokezaka}, {Ip},
  {Itoh}, {Jacobs}, {Jencson}, {Kasen}, {Kasliwal}, {Kassim}, {Kim}, {Kiran},
  {Kuin}, {Kulkarni}, {Kupfer}, {Lau}, {Madsen}, {Mazzali}, {Miller},
  {Miyasaka}, {Mooley}, {Myers}, {Nakar}, {Ngeow}, {Nugent}, {Ofek},
  {Palliyaguru}, {Pavana}, {Perley}, {Peters}, {Pike}, {Piran}, {Qi}, {Quimby},
  {Rana}, {Rosswog}, {Rusu}, {Sadler}, {Van Sistine}, {Sollerman}, {Xu}, {Yan},
  {Yatsu}, {Yu}, {Zhang}, {Zhao}, {GROWTH}, {JAGWAR}, {Caltech-NRAO},
  {TTU-NRAO}, {NuSTAR Collaborations}, {Chambers}, {Huber}, {Schultz},
  {Bulger}, {Flewelling}, {Magnier}, {Lowe}, {Wainscoat}, {Waters}, {Willman},
  {Pan-STARRS}, {Ebisawa}, {Hanyu}, {Harita}, {Hashimoto}, {Hidaka}, {Hori},
  {Ishikawa}, {Isobe}, {Iwakiri}, {Kawai}, {Kawai}, {Kawamuro}, {Kawase},
  {Kitaoka}, {Makishima}, {Matsuoka}, {Mihara}, {Morita}, {Morita}, {Nakahira},
  {Nakajima}, {Nakamura}, {Negoro}, {Oda}, {Sakamaki}, {Sasaki}, {Serino},
  {Shidatsu}, {Shimomukai}, {Sugawara}, {Sugita}, {Sugizaki}, {Tachibana},
  {Takao}, {Tanimoto}, {Tomida}, {Tsuboi}, {Tsunemi}, {Ueda}, {Ueno}, {Yamada},
  {Yamaoka}, {Yamauchi}, {Yatabe}, {Yoneyama}, {Yoshii}, {MAXI Team}, {Coward},
  {Crisp}, {Macpherson}, {Andreoni}, {Laugier}, {Noysena}, {Klotz}, {Gendre},
  {Thierry}, {Turpin}, {Consortium}, {Im}, {Choi}, {Kim}, {Yoon}, {Lim}, {Lee},
  {Lee}, {Kim}, {Ko}, {Joe}, {Kwon}, {Kim}, {Lim}, {Choi}, {KU Collaboration},
  {Fynbo}, {Malesani}, {Xu}, {Optical Telescope}, {Smartt}, {Jerkstrand},
  {Kankare}, {Sim}, {Fraser}, {Inserra}, {Maguire}, {Leloudas}, {Magee},
  {Shingles}, {Smith}, {Young}, {Kotak}, {Gal-Yam}, {Lyman}, {Homan},
  {Agliozzo}, {Anderson}, {Angus}, {Ashall}, {Barbarino}, {Bauer}, {Berton},
  {Botticella}, {Bulla}, {Cannizzaro}, {Cartier}, {Cikota}, {Clark}, {De Cia},
  {Della Valle}, {Dennefeld}, {Dessart}, {Dimitriadis}, {Elias-Rosa}, {Firth},
  {Fl{\"o}rs}, {Frohmaier}, {Galbany}, {Gonz{\'a}lez-Gait{\'a}n}, {Gromadzki},
  {Guti{\'e}rrez}, {Hamanowicz}, {Harmanen}, {Heintz}, {Hernandez}, {Hodgkin},
  {Hook}, {Izzo}, {James}, {Jonker}, {Kerzendorf}, {Kostrzewa-Rutkowska},
  {Kromer}, {Kuncarayakti}, {Lawrence}, {Manulis}, {Mattila}, {McBrien},
  {M{\"u}ller}, {Nordin}, {O'Neill}, {Onori}, {Palmerio}, {Pastorello},
  {Patat}, {Pignata}, {Podsiadlowski}, {Razza}, {Reynolds}, {Roy}, {Ruiter},
  {Rybicki}, {Salmon}, {Pumo}, {Prentice}, {Seitenzahl}, {Smith}, {Sollerman},
  {Sullivan}, {Szegedi}, {Taddia}, {Taubenberger}, {Terreran}, {Van Soelen},
  {Vos}, {Walton}, {Wright}, {Wyrzykowski}, {Yaron}, {pre=''(''>ePESSTO},
  {Chen}, {Kr{\"u}hler}, {Schady}, {Wiseman}, {Greiner}, {Rau}, {Schweyer},
  {Klose}, {Nicuesa Guelbenzu}, {GROND}, {Palliyaguru}, {Tech University},
  {Shara}, {Williams}, {Vaisanen}, {Potter}, {Romero Colmenero}, {Crawford},
  {Buckley}, {Mao}, {SALT Group}, {D{\'\i}az}, {Macri}, {Garc{\'\i}a Lambas},
  {Mendes de Oliveira}, {Nilo Castell{\'o}n}, {Ribeiro}, {S{\'a}nchez},
  {Schoenell}, {Abramo}, {Akras}, {Alcaniz}, {Artola}, {Beroiz}, {Bonoli},
  {Cabral}, {Camuccio}, {Chavushyan}, {Coelho}, {Colazo}, {Costa-Duarte},
  {Cuevas Larenas}, {Dom{\'\i}nguez Romero}, {Dultzin}, {Fern{\'a}ndez},
  {Garc{\'\i}a}, {Girardini}, {Gon{\c{c}}alves}, {Gon{\c{c}}alves}, {Gurovich},
  {Jim{\'e}nez-Teja}, {Kanaan}, {Lares}, {Lopes de Oliveira}, {L{\'o}pez-Cruz},
  {Melia}, {Molino}, {Padilla}, {Pe{\~n}uela}, {Placco}, {Qui{\~n}ones},
  {Ram{\'\i}rez Rivera}, {Renzi}, {Riguccini}, {R{\'\i}os-L{\'o}pez},
  {Rodriguez}, {Sampedro}, {Schneiter}, {Sodr{\'e}}, {Starck}, {Torres-Flores},
  {Tornatore}, {Zadro{\.z}ny}, {Castillo}, {TOROS: Transient Robotic
  Observatory of South Collaboration}, {Castro-Tirado}, {Tello}, {Hu}, {Zhang},
  {Cunniffe}, {Castell{\'o}n}, {Hiriart}, {Caballero-Garc{\'\i}a},
  {Jel{\'\i}nek}, {Kub{\'a}nek}, {P{\'e}rez del Pulgar}, {Park}, {Jeong},
  {Castro Cer{\'o}n}, {Pandey}, {Yock}, {Querel}, {Fan}, {Wang}, {BOOTES
  Collaboration}, {Beardsley}, {Brown}, {Crosse}, {Emrich}, {Franzen},
  {Gaensler}, {Horsley}, {Johnston-Hollitt}, {Kenney}, {Morales}, {Pallot},
  {Sokolowski}, {Steele}, {Tingay}, {Trott}, {Walker}, {Wayth}, {Williams},
  {Wu}, {Murchison Widefield Array}, {Yoshida}, {Sakamoto}, {Kawakubo},
  {Yamaoka}, {Takahashi}, {Asaoka}, {Ozawa}, {Torii}, {Shimizu}, {Tamura},
  {Ishizaki}, {Cherry}, {Ricciarini}, {Penacchioni}, {Marrocchesi}, {CALET
  Collaboration}, {Pozanenko}, {Volnova}, {Mazaeva}, {Minaev}, {Krugov},
  {Kusakin}, {Reva}, {Moskvitin}, {Rumyantsev}, {Inasaridze}, {Klunko},
  {Tungalag}, {Schmalz}, {Burhonov}, {IKI-GW Follow-up Collaboration},
  {Abdalla}, {Abramowski}, {Aharonian}, {Ait Benkhali}, {Ang{\"u}ner},
  {Arakawa}, {Arrieta}, {Aubert}, {Backes}, {Balzer}, {Barnard}, {Becherini},
  {Becker Tjus}, {Berge}, {Bernhard}, {Bernl{\"o}hr}, {Blackwell},
  {B{\"o}ttcher}, {Boisson}, {Bolmont}, {Bonnefoy}, {Bordas}, {Bregeon},
  {Brun}, {Brun}, {Bryan}, {B{\"u}chele}, {Bulik}, {Capasso}, {Caroff},
  {Carosi}, {Casanova}, {Cerruti}, {Chakraborty}, {Chaves}, {Chen},
  {Chevalier}, {Colafrancesco}, {Condon}, {Conrad}, {Davids}, {Decock}, {Deil},
  {Devin}, {deWilt}, {Dirson}, {Djannati-Ata{\"\i}}, {Donath}, {O'C. Drury},
  {Dutson}, {Dyks}, {Edwards}, {Egberts}, {Emery}, {Ernenwein}, {Eschbach},
  {Farnier}, {Fegan}, {Fernandes}, {Fiasson}, {Fontaine}, {Funk},
  {F{\"u}ssling}, {Gabici}, {Gallant}, {Garrigoux}, {Gat{\'e}}, {Giavitto},
  {Giebels}, {Glawion}, {Glicenstein}, {Gottschall}, {Grondin}, {Hahn},
  {Haupt}, {Hawkes}, {Heinzelmann}, {Henri}, {Hermann}, {Hinton}, {Hofmann},
  {Hoischen}, {Holch}, {Holler}, {Horns}, {Ivascenko}, {Iwasaki},
  {Jacholkowska}, {Jamrozy}, {Jankowsky}, {Jankowsky}, {Jingo}, {Jouvin},
  {Jung-Richardt}, {Kastendieck}, {Katarzy{\'n}ski}, {Katsuragawa},
  {Kerszberg}, {Khangulyan}, {Kh{\'e}lifi}, {King}, {Klepser}, {Klochkov},
  {Klu{\'z}niak}, {Komin}, {Kosack}, {Krakau}, {Kraus}, {Kr{\"u}ger}, {Laffon},
  {Lamanna}, {Lau}, {Lees}, {Lefaucheur}, {Lemi{\`e}re}, {Lemoine-Goumard},
  {Lenain}, {Leser}, {Lohse}, {Lorentz}, {Liu}, {Lypova}, {Malyshev},
  {Marandon}, {Marcowith}, {Mariaud}, {Marx}, {Maurin}, {Maxted}, {Mayer},
  {Meintjes}, {Meyer}, {Mitchell}, {Moderski}, {Mohamed}, {Mohrmann},
  {Mor{\r{a}}}, {Moulin}, {Murach}, {Nakashima}, {de Naurois}, {Ndiyavala},
  {Niederwanger}, {Niemiec}, {Oakes}, {O'Brien}, {Odaka}, {Ohm}, {Ostrowski},
  {Oya}, {Padovani}, {Panter}, {Parsons}, {Pekeur}, {Pelletier}, {Perennes},
  {Petrucci}, {Peyaud}, {Piel}, {Pita}, {Poireau}, {Poon}, {Prokhorov},
  {Prokoph}, {P{\"u}hlhofer}, {Punch}, {Quirrenbach}, {Raab}, {Rauth},
  {Reimer}, {Reimer}, {Renaud}, {de los Reyes}, {Rieger}, {Rinchiuso},
  {Romoli}, {Rowell}, {Rudak}, {Rulten}, {Sahakian}, {Saito}, {Sanchez},
  {Santangelo}, {Sasaki}, {Schlickeiser}, {Sch{\"u}ssler}, {Schulz},
  {Schwanke}, {Schwemmer}, {Seglar-Arroyo}, {Settimo}, {Seyffert}, {Shafi},
  {Shilon}, {Shiningayamwe}, {Simoni}, {Sol}, {Spanier}, {Spir-Jacob},
  {Stawarz}, {Steenkamp}, {Stegmann}, {Steppa}, {Sushch}, {Takahashi},
  {Tavernet}, {Tavernier}, {Taylor}, {Terrier}, {Tibaldo}, {Tiziani},
  {Tluczykont}, {Trichard}, {Tsirou}, {Tsuji}, {Tuffs}, {Uchiyama}, {van der
  Walt}, {van Eldik}, {van Rensburg}, {van Soelen}, {Vasileiadis}, {Veh},
  {Venter}, {Viana}, {Vincent}, {Vink}, {Voisin}, {V{\"o}lk}, {Vuillaume},
  {Wadiasingh}, {Wagner}, {Wagner}, {Wagner}, {White}, {Wierzcholska},
  {Willmann}, {W{\"o}rnlein}, {Wouters}, {Yang}, {Zaborov}, {Zacharias},
  {Zanin}, {Zdziarski}, {Zech}, {Zefi}, {Ziegler}, {Zorn}, {{\.Z}ywucka},
  {H.~E.~S.~S. Collaboration}, {Fender}, {Broderick}, {Rowlinson}, {Wijers},
  {Stewart}, {ter Veen}, {Shulevski}, {LOFAR Collaboration}, {Kavic},
  {Simonetti}, {League}, {Tsai}, {Obenberger}, {Nathaniel}, {Taylor}, {Dowell},
  {Liebling}, {Estes}, {Lippert}, {Sharma}, {Vincent}, {Farella}, {Wavelength
  Array}, {Abeysekara}, {Albert}, {Alfaro}, {Alvarez}, {Arceo},
  {Arteaga-Vel{\'a}zquez}, {Avila Rojas}, {Ayala Solares}, {Barber}, {Becerra
  Gonzalez}, {Becerril}, {Belmont-Moreno}, {BenZvi}, {Berley}, {Bernal},
  {Braun}, {Brisbois}, {Caballero-Mora}, {Capistr{\'a}n}, {Carrami{\~n}ana},
  {Casanova}, {Castillo}, {Cotti}, {Cotzomi}, {Couti{\~n}o de Le{\'o}n}, {De
  Le{\'o}n}, {De la Fuente}, {Diaz Hernandez}, {Dichiara}, {Dingus},
  {DuVernois}, {D{\'\i}az-V{\'e}lez}, {Ellsworth}, {Engel},
  {Enr{\'\i}quez-Rivera}, {Fiorino}, {Fleischhack}, {Fraija},
  {Garc{\'\i}a-Gonz{\'a}lez}, {Garfias}, {Gerhardt}, {Gonz{\~o}lez Mu{\~n}oz},
  {Gonz{\'a}lez}, {Goodman}, {Hampel-Arias}, {Harding}, {Hernandez},
  {Hernandez-Almada}, {Hona}, {H{\"u}ntemeyer}, {Iriarte}, {Jardin-Blicq},
  {Joshi}, {Kaufmann}, {Kieda}, {Lara}, {Lauer}, {Lennarz}, {Le{\'o}n Vargas},
  {Linnemann}, {Longinotti}, {Raya}, {Luna-Garc{\'\i}a}, {L{\'o}pez-Coto},
  {Malone}, {Marinelli}, {Martinez}, {Martinez-Castellanos},
  {Mart{\'\i}nez-Castro}, {Mart{\'\i}nez-Huerta}, {Matthews},
  {Miranda-Romagnoli}, {Moreno}, {Mostaf{\'a}}, {Nellen}, {Newbold}, {Nisa},
  {Noriega-Papaqui}, {Pelayo}, {Pretz}, {P{\'e}rez-P{\'e}rez}, {Ren}, {Rho},
  {Rivi{\`e}re}, {Rosa-Gonz{\'a}lez}, {Rosenberg}, {Ruiz-Velasco}, {Salazar},
  {Salesa Greus}, {Sandoval}, {Schneider}, {Schoorlemmer}, {Sinnis}, {Smith},
  {Springer}, {Surajbali}, {Tibolla}, {Tollefson}, {Torres}, {Ukwatta},
  {Weisgarber}, {Westerhoff}, {Wisher}, {Wood}, {Yapici}, {Yodh}, {Younk},
  {Zhou}, {{\'A}lvarez}, {HAWC Collaboration}, {Aab}, {Abreu}, {Aglietta},
  {Albuquerque}, {Albury}, {Allekotte}, {Almela}, {Alvarez Castillo},
  {Alvarez-Mu{\~n}iz}, {Anastasi}, {Anchordoqui}, {Andrada}, {Andringa},
  {Aramo}, {Arsene}, {Asorey}, {Assis}, {Avila}, {Badescu}, {Balaceanu},
  {Barbato}, {Barreira Luz}, {Becker}, {Bellido}, {Berat}, {Bertaina},
  {Bertou}, {Biermann}, {Biteau}, {Blaess}, {Blanco}, {Blazek}, {Bleve},
  {Boh{\'a}{\v{c}}ov{\'a}}, {Bonifazi}, {Borodai}, {Botti}, {Brack}, {Brancus},
  {Bretz}, {Bridgeman}, {Briechle}, {Buchholz}, {Bueno}, {Buitink}, {Buscemi},
  {Caballero-Mora}, {Caccianiga}, {Cancio}, {Canfora}, {Caruso}, {Castellina},
  {Catalani}, {Cataldi}, {Cazon}, {Chavez}, {Chinellato}, {Chudoba}, {Clay},
  {Cobos Cerutti}, {Colalillo}, {Coleman}, {Collica}, {Coluccia},
  {Concei{\c{c}}{\~a}o}, {Consolati}, {Contreras}, {Cooper}, {Coutu},
  {Covault}, {Cronin}, {D'Amico}, {Daniel}, {Dasso}, {Daumiller}, {Dawson},
  {Day}, {de Almeida}, {de Jong}, {De Mauro}, {de Mello Neto}, {De Mitri}, {de
  Oliveira}, {de Souza}, {Debatin}, {Deligny}, {D{\'\i}az Castro}, {Diogo},
  {Dobrigkeit}, {D'Olivo}, {Dorosti}, {Dos Anjos}, {Dova}, {Dundovic}, {Ebr},
  {Engel}, {Erdmann}, {Erfani}, {Escobar}, {Espadanal}, {Etchegoyen}, {Falcke},
  {Farmer}, {Farrar}, {Fauth}, {Fazzini}, {Feldbusch}, {Fenu}, {Fick},
  {Figueira}, {Filip{\v{c}}i{\v{c}}}, {Freire}, {Fujii}, {Fuster},
  {Ga{\"\i}or}, {Garc{\'\i}a}, {Gat{\'e}}, {Gemmeke}, {Gherghel-Lascu}, {Ghia},
  {Giaccari}, {Giammarchi}, {Giller}, {G{\l}as}, {Glaser}, {Golup}, {G{\'o}mez
  Berisso}, {G{\'o}mez Vitale}, {Gonz{\'a}lez}, {Gorgi}, {Gottowik}, {Grillo},
  {Grubb}, {Guarino}, {Guedes}, {Halliday}, {Hampel}, {Hansen}, {Harari},
  {Harrison}, {Harvey}, {Haungs}, {Hebbeker}, {Heck}, {Heimann}, {Herve},
  {Hill}, {Hojvat}, {Holt}, {Homola}, {H{\"o}randel}, {Horvath},
  {Hrabovsk{\'y}}, {Huege}, {Hulsman}, {Insolia}, {Isar}, {Jandt}, {Johnsen},
  {Josebachuili}, {Jurysek}, {K{\"a}{\"a}p{\"a}}, {Kampert}, {Keilhauer},
  {Kemmerich}, {Kemp}, {Kieckhafer}, {Klages}, {Kleifges}, {Kleinfeller},
  {Krause}, {Krohm}, {Kuempel}, {Kukec Mezek}, {Kunka}, {Kuotb Awad}, {Lago},
  {LaHurd}, {Lang}, {Lauscher}, {Legumina}, {Leigui de Oliveira},
  {Letessier-Selvon}, {Lhenry-Yvon}, {Link}, {Lo Presti}, {Lopes}, {L{\'o}pez},
  {L{\'o}pez Casado}, {Lorek}, {Luce}, {Lucero}, {Malacari}, {Mallamaci},
  {Mandat}, {Mantsch}, {Mariazzi}, {Maris}, {Marsella}, {Martello}, {Martinez},
  {Mart{\'\i}nez Bravo}, {Mas{\'\i}as Meza}, {Mathes}, {Mathys}, {Matthews},
  {Matthiae}, {Mayotte}, {Mazur}, {Medina}, {Medina-Tanco}, {Melo},
  {Menshikov}, {Merenda}, {Michal}, {Micheletti}, {Middendorf}, {Miramonti},
  {Mitrica}, {Mockler}, {Mollerach}, {Montanet}, {Morello}, {Morlino},
  {M{\"u}ller}, {M{\"u}ller}, {Muller}, {M{\"u}ller}, {Mussa}, {Naranjo},
  {Nguyen}, {Niculescu-Oglinzanu}, {Niechciol}, {Niemietz}, {Niggemann},
  {Nitz}, {Nosek}, {Novotny}, {No{\v{z}}ka}, {N{\'u}{\~n}ez}, {Oikonomou},
  {Olinto}, {Palatka}, {Pallotta}, {Papenbreer}, {Parente}, {Parra}, {Paul},
  {Pech}, {Pedreira}, {P{\c{e}}kala}, {Pe{\~n}a-Rodriguez}, {Pereira},
  {Perlin}, {Perrone}, {Peters}, {Petrera}, {Phuntsok}, {Pierog}, {Pimenta},
  {Pirronello}, {Platino}, {Plum}, {Poh}, {Porowski}, {Prado}, {Privitera},
  {Prouza}, {Quel}, {Querchfeld}, {Quinn}, {Ramos-Pollan}, {Rautenberg},
  {Ravignani}, {Ridky}, {Riehn}, {Risse}, {Ristori}, {Rizi}, {Rodrigues de
  Carvalho}, {Rodriguez Fernandez}, {Rodriguez Rojo}, {Roncoroni}, {Roth},
  {Roulet}, {Rovero}, {Ruehl}, {Saffi}, {Saftoiu}, {Salamida}, {Salazar},
  {Saleh}, {Salina}, {S{\'a}nchez}, {Sanchez-Lucas}, {Santos}, {Santos},
  {Sarazin}, {Sarmento}, {Sarmiento-Cano}, {Sato}, {Schauer}, {Scherini},
  {Schieler}, {Schimp}, {Schmidt}, {Scholten}, {Schov{\'a}nek}, {Schr{\"o}der},
  {Schr{\"o}der}, {Schulz}, {Schumacher}, {Sciutto}, {Segreto}, {Shadkam},
  {Shellard}, {Sigl}, {Silli}, {{\v{S}}m{\'\i}da}, {Snow}, {Sommers},
  {Sonntag}, {Soriano}, {Squartini}, {Stanca}, {Stani{\v{c}}}, {Stasielak},
  {Stassi}, {Stolpovskiy}, {Strafella}, {Streich}, {Suarez},
  {Suarez-Dur{\'a}n}, {Sudholz}, {Suomij{\"a}rvi}, {Supanitsky},
  {{\v{S}}up{\'\i}k}, {Swain}, {Szadkowski}, {Taboada}, {Taborda},
  {Timmermans}, {Todero Peixoto}, {Tomankova}, {Tom{\'e}}, {Torralba Elipe},
  {Travnicek}, {Trini}, {Tueros}, {Ulrich}, {Unger}, {Urban}, {Vald{\'e}s
  Galicia}, {Vali{\~n}o}, {Valore}, {van Aar}, {van Bodegom}, {van den Berg},
  {van Vliet}, {Varela}, {Vargas C{\'a}rdenas}, {V{\'a}zquez}, {Veberi{\v{c}}},
  {Ventura}, {Vergara Quispe}, {Verzi}, {Vicha}, {Villase{\~n}or}, {Vorobiov},
  {Wahlberg}, {Wainberg}, {Walz}, {Watson}, {Weber}, {Weindl}, {Wiede{\'n}ski},
  {Wiencke}, {Wilczy{\'n}ski}, {Wirtz}, {Wittkowski}, {Wundheiler}, {Yang},
  {Yushkov}, {Zas}, {Zavrtanik}, {Zavrtanik}, {Zepeda}, {Zimmermann},
  {Ziolkowski}, {Zong}, {Zuccarello}, {Pierre Auger Collaboration}, {Kim},
  {Schulze}, {Bauer}, {Corral-Santana}, {de Gregorio-Monsalvo},
  {Gonz{\'a}lez-L{\'o}pez}, {Hartmann}, {Ishwara-Chandra}, {Mart{\'\i}n},
  {Mehner}, {Misra}, {Micha{\l}owski}, {Resmi}, {ALMA Collaboration}, {Paragi},
  {Agudo}, {An}, {Beswick}, {Casadio}, {Frey}, {Jonker}, {Kettenis}, {Marcote},
  {Moldon}, {Szomoru}, {van Langevelde}, {Yang}, {Euro VLBI Team}, {Cwiek},
  {Cwiok}, {Czyrkowski}, {Dabrowski}, {Kasprowicz}, {Mankiewicz}, {Nawrocki},
  {Opiela}, {Piotrowski}, {Wrochna}, {Zaremba}, {{\.Z}arnecki}, {Pi of Sky
  Collaboration}, {Haggard}, {Nynka}, {Ruan}, {Chandra Team at McGill
  University}, {Bland}, {Booler}, {Devillepoix}, {de Gois}, {Hancock}, {Howie},
  {Paxman}, {Sansom}, {Towner}, {Desert Fireball Network}, {Tonry}, {Coughlin},
  {Stubbs}, {Denneau}, {Heinze}, {Stalder}, {Weiland}, {ATLAS}, {Eatough},
  {Kramer}, {Kraus}, {Time Resolution Universe Survey}, {Troja}, {Piro},
  {Becerra Gonz{\'a}lez}, {Butler}, {Fox}, {Khandrika}, {Kutyrev}, {Lee},
  {Ricci}, {Ryan}, {S{\'a}nchez-Ram{\'\i}rez}, {Veilleux}, {Watson},
  {Wieringa}, {Burgess}, {van Eerten}, {Fontes}, {Fryer}, {Korobkin},
  {Wollaeger}, {RIMAS}, {RATIR}, {Camilo}, {Foley}, {Goedhart}, {Makhathini},
  {Oozeer}, {Smirnov}, {Fender}, {Woudt}, \& {South
  Africa/MeerKAT}}]{abbott17b}
---. 2017{\natexlab{b}}, \apjl, 848, L12, \dodoi{10.3847/2041-8213/aa91c9}

\bibitem[{{Angel}(1979)}]{angel79}
{Angel}, J.~R.~P. 1979, \apj, 233, 364, \dodoi{10.1086/157397}

\bibitem[{{Arnaud} {et~al.}(1999){Arnaud}, {Dorman}, \& {Gordon}}]{arnaud99}
{Arnaud}, K., {Dorman}, B., \& {Gordon}, C. 1999, {XSPEC: An X-ray spectral
  fitting package}, Astrophysics Source Code Library, record ascl:9910.005.
\newblock \doeprint{9910.005}

\bibitem[{{Atteia} {et~al.}(2022){Atteia}, {Cordier}, \& {Wei}}]{svom}
{Atteia}, J.~L., {Cordier}, B., \& {Wei}, J. 2022, International Journal of
  Modern Physics D, 31, 2230008, \dodoi{10.1142/S0218271822300087}

\bibitem[{{Bradshaw} {et~al.}(2019){Bradshaw}, {Burwitz}, {Hartner},
  {Pelliciari}, {Langmeier}, {Liao}, {Friedrich}, {Valsecchi}, {Barri{\`e}re},
  {Collon}, \& {Vacanti}}]{panter}
{Bradshaw}, M., {Burwitz}, V., {Hartner}, G., {et~al.} 2019, in Society of
  Photo-Optical Instrumentation Engineers (SPIE) Conference Series, Vol. 11119,
  Optics for EUV, X-Ray, and Gamma-Ray Astronomy IX, 1111916,
  \dodoi{10.1117/12.2531709}

\bibitem[{{Chary} {et~al.}(2016){Chary}, {Petitjean}, {Robertson}, {Trenti}, \&
  {Vangioni}}]{chary16}
{Chary}, R., {Petitjean}, P., {Robertson}, B., {Trenti}, M., \& {Vangioni}, E.
  2016, \ssr, 202, 181, \dodoi{10.1007/s11214-016-0288-6}

\bibitem[{{Costa} {et~al.}(1997){Costa}, {Frontera}, {Heise}, {Feroci}, {in't
  Zand}, {Fiore}, {Cinti}, {Dal Fiume}, {Nicastro}, {Orlandini}, {Palazzi},
  {Rapisarda\#}, {Zavattini}, {Jager}, {Parmar}, {Owens}, {Molendi},
  {Cusumano}, {Maccarone}, {Giarrusso}, {Coletta}, {Antonelli}, {Giommi},
  {Muller}, {Piro}, \& {Butler}}]{costa97}
{Costa}, E., {Frontera}, F., {Heise}, J., {et~al.} 1997, \nat, 387, 783,
  \dodoi{10.1038/42885}

\bibitem[{{Cucchiara} {et~al.}(2011){Cucchiara}, {Levan}, {Fox}, {Tanvir},
  {Ukwatta}, {Berger}, {Kr{\"u}hler}, {K{\"u}pc{\"u} Yolda{\c{s}}}, {Wu},
  {Toma}, {Greiner}, {Olivares}, {Rowlinson}, {Amati}, {Sakamoto}, {Roth},
  {Stephens}, {Fritz}, {Fynbo}, {Hjorth}, {Malesani}, {Jakobsson}, {Wiersema},
  {O'Brien}, {Soderberg}, {Foley}, {Fruchter}, {Rhoads}, {Rutledge}, {Schmidt},
  {Dopita}, {Podsiadlowski}, {Willingale}, {Wolf}, {Kulkarni}, \&
  {D'Avanzo}}]{cucchiara11}
{Cucchiara}, A., {Levan}, A.~J., {Fox}, D.~B., {et~al.} 2011, \apj, 736, 7,
  \dodoi{10.1088/0004-637X/736/1/7}

\bibitem[{{Evans} {et~al.}(2009){Evans}, {Beardmore}, {Page}, {Osborne},
  {O'Brien}, {Willingale}, {Starling}, {Burrows}, {Godet}, {Vetere}, {Racusin},
  {Goad}, {Wiersema}, {Angelini}, {Capalbi}, {Chincarini}, {Gehrels}, {Kennea},
  {Margutti}, {Morris}, {Mountford}, {Pagani}, {Perri}, {Romano}, \&
  {Tanvir}}]{evans09}
{Evans}, P.~A., {Beardmore}, A.~P., {Page}, K.~L., {et~al.} 2009, \mnras, 397,
  1177, \dodoi{10.1111/j.1365-2966.2009.14913.x}

\bibitem[{{Feldman} \& {et al.}(2022)}]{feldman22}
{Feldman}, C., \& {et al.} 2022, Experimental Astronomy, in preparation

\bibitem[{{Frail} {et~al.}(1997){Frail}, {Kulkarni}, {Nicastro}, {Feroci}, \&
  {Taylor}}]{frail97}
{Frail}, D.~A., {Kulkarni}, S.~R., {Nicastro}, L., {Feroci}, M., \& {Taylor},
  G.~B. 1997, \nat, 389, 261, \dodoi{10.1038/38451}

\bibitem[{{Gehrels} {et~al.}(2004){Gehrels}, {Chincarini}, {Giommi}, {Mason},
  {Nousek}, {Wells}, {White}, {Barthelmy}, {Burrows}, {Cominsky}, {Hurley},
  {Marshall}, {M{\'e}sz{\'a}ros}, {Roming}, {Angelini}, {Barbier}, {Belloni},
  {Campana}, {Caraveo}, {Chester}, {Citterio}, {Cline}, {Cropper}, {Cummings},
  {Dean}, {Feigelson}, {Fenimore}, {Frail}, {Fruchter}, {Garmire}, {Gendreau},
  {Ghisellini}, {Greiner}, {Hill}, {Hunsberger}, {Krimm}, {Kulkarni}, {Kumar},
  {Lebrun}, {Lloyd-Ronning}, {Markwardt}, {Mattson}, {Mushotzky}, {Norris},
  {Osborne}, {Paczynski}, {Palmer}, {Park}, {Parsons}, {Paul}, {Rees},
  {Reynolds}, {Rhoads}, {Sasseen}, {Schaefer}, {Short}, {Smale}, {Smith},
  {Stella}, {Tagliaferri}, {Takahashi}, {Tashiro}, {Townsley}, {Tueller},
  {Turner}, {Vietri}, {Voges}, {Ward}, {Willingale}, {Zerbi}, \&
  {Zhang}}]{swift}
{Gehrels}, N., {Chincarini}, G., {Giommi}, P., {et~al.} 2004, \apj, 611, 1005,
  \dodoi{10.1086/422091}

\bibitem[{{Klebesadel} {et~al.}(1973){Klebesadel}, {Strong}, \&
  {Olson}}]{klebesadel73}
{Klebesadel}, R.~W., {Strong}, I.~B., \& {Olson}, R.~A. 1973, \apjl, 182, L85,
  \dodoi{10.1086/181225}

\bibitem[{{Liang} {et~al.}(2008){Liang}, {Racusin}, {Zhang}, {Zhang}, \&
  {Burrows}}]{liang08}
{Liang}, E.-W., {Racusin}, J.~L., {Zhang}, B., {Zhang}, B.-B., \& {Burrows},
  D.~N. 2008, \apj, 675, 528, \dodoi{10.1086/524701}

\bibitem[{{Meidinger} {et~al.}(2006){Meidinger}, {Andritschke}, {H{\"a}lker},
  {Hartmann}, {Herrmann}, {Holl}, {Lutz}, {Kimmel}, {Schaller}, {Schnecke},
  {Schopper}, {Soltau}, \& {Str{\"u}der}}]{meidinger06}
{Meidinger}, N., {Andritschke}, R., {H{\"a}lker}, O., {et~al.} 2006, Nuclear
  Instruments and Methods in Physics Research A, 568, 141,
  \dodoi{10.1016/j.nima.2006.05.268}

\bibitem[{{Mercier} {et~al.}(2022){Mercier}, {Gonzalez}, {G{\"o}tz},
  {Boutelier}, {Boufracha}, {Burwitz}, {Charmeau}, {Drumm}, {Feldman}, {Gomes},
  {Le Duigou}, {Meidinger}, {Meuris}, {O'Brien}, {Osborne}, {Pasqier},
  {Perraud}, {Pearson}, {Pinsard}, {Raynal}, \& {Willingale}}]{mercier22}
{Mercier}, K., {Gonzalez}, F., {G{\"o}tz}, D., {et~al.} 2022, in Society of
  Photo-Optical Instrumentation Engineers (SPIE) Conference Series, Vol. 12181,
  Space Telescopes and Instrumentation 2022: Ultraviolet to Gamma Ray, ed.
  J.-W.~A. {den Herder}, S.~{Nikzad}, \& K.~{Nakazawa}

\bibitem[{{Metzger} {et~al.}(1997){Metzger}, {Djorgovski}, {Kulkarni},
  {Steidel}, {Adelberger}, {Frail}, {Costa}, \& {Frontera}}]{metzger97}
{Metzger}, M.~R., {Djorgovski}, S.~G., {Kulkarni}, S.~R., {et~al.} 1997, \nat,
  387, 878, \dodoi{10.1038/43132}

\bibitem[{{Meuris} \& {et al.}(2022)}]{meuris22}
{Meuris}, A., \& {et al.} 2022, Nuclear Instruments and Methods in Physics
  Resaerch A, submitted

\bibitem[{{Moretti} {et~al.}(2009){Moretti}, {Pagani}, {Cusumano}, {Campana},
  {Perri}, {Abbey}, {Ajello}, {Beardmore}, {Burrows}, {Chincarini}, {Godet},
  {Guidorzi}, {Hill}, {Kennea}, {Nousek}, {Osborne}, \&
  {Tagliaferri}}]{moretti09}
{Moretti}, A., {Pagani}, C., {Cusumano}, G., {et~al.} 2009, \aap, 493, 501,
  \dodoi{10.1051/0004-6361:200811197}

\bibitem[{{Parmar} {et~al.}(1999){Parmar}, {Guainazzi}, {Oosterbroek}, {Orr},
  {Favata}, {Lumb}, \& {Malizia}}]{cxb}
{Parmar}, A.~N., {Guainazzi}, M., {Oosterbroek}, T., {et~al.} 1999, \aap, 345,
  611.
\newblock \doarXiv{astro-ph/9903109}

\bibitem[{{Perley} {et~al.}(2016){Perley}, {Niino}, {Tanvir}, {Vergani}, \&
  {Fynbo}}]{perley16}
{Perley}, D.~A., {Niino}, Y., {Tanvir}, N.~R., {Vergani}, S.~D., \& {Fynbo}, J.
  P.~U. 2016, \ssr, 202, 111, \dodoi{10.1007/s11214-016-0237-4}

\bibitem[{{Pian} {et~al.}(2006){Pian}, {Mazzali}, {Masetti}, {Ferrero},
  {Klose}, {Palazzi}, {Ramirez-Ruiz}, {Woosley}, {Kouveliotou}, {Deng},
  {Filippenko}, {Foley}, {Fynbo}, {Kann}, {Li}, {Hjorth}, {Nomoto}, {Patat},
  {Sauer}, {Sollerman}, {Vreeswijk}, {Guenther}, {Levan}, {O'Brien}, {Tanvir},
  {Wijers}, {Dumas}, {Hainaut}, {Wong}, {Baade}, {Wang}, {Amati}, {Cappellaro},
  {Castro-Tirado}, {Ellison}, {Frontera}, {Fruchter}, {Greiner}, {Kawabata},
  {Ledoux}, {Maeda}, {M{\o}ller}, {Nicastro}, {Rol}, \& {Starling}}]{pian06}
{Pian}, E., {Mazzali}, P.~A., {Masetti}, N., {et~al.} 2006, \nat, 442, 1011,
  \dodoi{10.1038/nature05082}

\bibitem[{{Schneider} \& {et al.}(2022)}]{schneider22}
{Schneider}, B., \& {et al.} 2022, Experimental Astronomy, submitted

\bibitem[{{Toma} {et~al.}(2016){Toma}, {Yoon}, \& {Bromm}}]{toma16}
{Toma}, K., {Yoon}, S.-C., \& {Bromm}, V. 2016, \ssr, 202, 159,
  \dodoi{10.1007/s11214-016-0250-7}

\bibitem[{{van Paradijs} {et~al.}(1997){van Paradijs}, {Groot}, {Galama},
  {Kouveliotou}, {Strom}, {Telting}, {Rutten}, {Fishman}, {Meegan}, {Pettini},
  {Tanvir}, {Bloom}, {Pedersen}, {N{\o}rdgaard-Nielsen}, {Linden-V{\o}rnle},
  {Melnick}, {Van der Steene}, {Bremer}, {Naber}, {Heise}, {in't Zand},
  {Costa}, {Feroci}, {Piro}, {Frontera}, {Zavattini}, {Nicastro}, {Palazzi},
  {Bennett}, {Hanlon}, \& {Parmar}}]{vanparadijs97}
{van Paradijs}, J., {Groot}, P.~J., {Galama}, T., {et~al.} 1997, \nat, 386,
  686, \dodoi{10.1038/386686a0}

\end{thebibliography}
\end{document}